\documentclass[12pt,preprint]{aastex}

\slugcomment{Submitted to Astronomical Journal}

\shorttitle{Dynamical Zodiacal Cloud Models}
\shortauthors{Ipatov et al.}

\begin{document}


\title{Dynamical Zodiacal Cloud Models Constrained \\ by High Resolution Spectroscopy of the Zodiacal Light}


\author{S. I. Ipatov}
\affil{Astronomy Department, University of Maryland,
 College Park, MD 20742, USA; Space Research Institute, Moscow, Russia} \email{siipatov@hotmail.com}

\author{A. S. Kutyrev} 
\affil{NASA/GSFC, Greenbelt, MD 20771} 

\author{G. J. Madsen$^1$} 
\affil{Anglo-Australian Observatory, P.O. Box 296, Epping, NSW 1710, Australia} 
\altaffiltext{1}{NSF Distinguished International Postdoctoral Research 
Fellow}

\author{J. C. Mather, S. H. Moseley} 
\affil{NASA/GSFC, Greenbelt, MD 20771} 

\and

\author{R. J. Reynolds} 
\affil{Department of Astronomy, 475 North Charter st., University of Wisconsin at Madison, Madison, WI 53706} 



\begin{abstract}

We present simulated observations of the Doppler shifts of the solar Mg~I 
Fraunhofer line scattered by asteroidal, cometary, and trans-Neptunian 
dust particles.  The studies are based on the results of 
integrations of orbital evolution of particles under the gravitational influence of 
planets, the Poynting-Robertson drag, radiation pressure, and solar wind 
drag. The derived shifts in the centroid and profile of the line with 
solar elongation are different for different sources of dust. A comparison 
of the velocities of zodiacal dust particles based on these numerical 
integrations with the velocities obtained from WHAM observations shows 
that the fraction of cometary dust particles among zodiacal dust particles 
is significant and can be dominant.
A considerable fraction of trans-Neptunian dust particles among zodiacal dust
particles also fits different observations.
The mean eccentricity of zodiacal dust particles is estimated to be about 0.5.

\end{abstract}





\section{Introduction}

Ground-based spectroscopic observations allow one to study Doppler shifts 
of scattered solar Fraunhofer lines in the zodiacal light (e.g., James 
1969, Hicks et al. 1974, Fried 1978, and East \& Reay 1984).  Analysis of 
these shifts thus provide an opportunity to explore velocities of 
interplanetary dust in the inner solar system.  Using the Wisconsin 
H-Alpha Mapper (WHAM) spectrometer, Reynolds et al. (2004) were the 
first to obtain accurate measurements of the centroid velocities and line 
profiles of the scattered solar Mg I $\lambda$5184 absorption line in the 
zodiacal light, both along the ecliptic equator and at high ecliptic 
longitudes.

The models for migration of dust particles are discussed in Section 2.  
These results are based on integrations of the orbital evolution of dust 
particles developed by Ipatov et al. (2004) and Ipatov \& Mather 
(2006a-b) (all our recent papers can be found at astro-ph and at 
\url{http://www.astro.umd.edu/$\sim$ipatov}). 
In Sections 3 and 4, based on the results of our runs, we 
consider our model calculations of the radial velocity profile of the 
scattered Mg~I line.  We compare plots of centroid velocities and line 
widths of the Mg~I line versus solar elongation with WHAM observations 
obtained by Reynolds et al. (2004).  Studies of typical eccentricities and 
inclinations of zodiacal dust particles that fit the WHAM observations are 
presented in Section 5. In previous papers, observations have been 
compared with analytical models, but this is the first comparison with 
numerical integrations of the particles.

\section{Models for Migration of Dust Particles}

Our studies of the Mg~I line shifts use the results of following the 
orbital evolution of about 15,000 asteroidal, cometary, and 
trans-Neptunian dust particles under the gravitational influence of 
planets (excluding Pluto for asteroidal and cometary particles), the 
Poynting-Robertson drag, radiation pressure, and solar wind drag.  
Results of some of these integrations were presented by Ipatov et al. 
(2004) and Ipatov \& Mather (2006a-b). As in Liou et al. (1999) and 
Moro-Martin \& Malhotra (2002), we assume the ratio of solar wind drag to 
Poynting--Robertson drag to be 0.35. In this section we describe
the model used for our studies of the migration of dust particles.

In our runs for asteroidal and cometary particles, the values of $\beta$, 
the ratio of the sun's radiation pressure force to gravitational force, 
varied from $\le$0.0004 to 0.4.  Burns et al. (1979) obtained 
$\beta$=$0.573 Q_{pr}/(\rho s)$, where $\rho$ is the particle's density in 
grams per cubic centimeter, $s$ is its radius in micrometers, and $Q_{pr}$ 
is the radiation pressure coefficient, which is close to unity for 
particles larger than 1 $\mu$m. For silicates at density of 2.5 g/cm$^3$, 
the $\beta$ values 0.004, 0.01, 0.05, 0.1, and 0.4 correspond to particle 
diameters $d$ of about 120, 47, 9.4, 4.7, and 1 microns, respectively. For 
water ice, $d$ is greater by a factor of 2.5 than that for silicate 
particles. The orbital evolution of dust particles was studied for a wider 
range of masses (including particles up to several milimeters) of 
asteroidal and cometary particles than in the papers by other authors 
(e.g., Dermott et al. 2001, 2002; Gorkavyi et al. 2000;  Grogan et al. 
2001; Kortenkamp \& Dermott 1998; Liou et al. 1995, 1996, 1999; Liou \& 
Zook 1999; Moro-Martin \& Malhotra 2002, 2003;  Ozernoy 2001;  Reach et 
al. 1997). Smaller particles have a larger surface per unit of mass of 
particles. It has been proposed (Gr\"un et al. 1985) that the main 
contribution to the zodiacal light is from particles that range from 20 to 
200 $\mu$m in diameter (for silicate particles, this range corresponds to 
$\beta$ about 0.002-0.02).  According to Gr\"un et al. (2000), the Lorentz 
force is comparable to solar gravitational interaction for particles of 
$d$$\sim$0.1 $\mu$m at 1 AU and of $d$$\sim$1 $\mu$m at 50 AU. 
Interstellar particles dominate among such small particles, and are not 
believed to be significant contributers to the zodiacal light. We 
considered mainly larger interplanetary particles, so we did not include 
the Lorentz force in our runs.

Migration of dust particles was integrated using the Bulirsh-Stoer method 
(BULSTO) with the relative error per integration step less than $10^{-8}$. 
The SWIFT integration package by Levison \& Duncan (1994) was modified to 
include the additional forces of radiation pressure, Poynting-Robertson 
drag, and solar wind drag. The integration continued until all of the 
particles either collided with the sun or reached 2000 AU from the sun. 
For small $\beta$, lifetimes exceeded 50-80 Myr (240 Myr for 
trans-Neptunian particles). In each run (with a fixed source of particles 
and $\beta$=const) we took $N$$\le$250 particles, because for $N$$\ge$500 
the computer time per calculation for one particle was several times 
greater than for $N$=250. The total number of particles in several tens of 
runs (with different $\beta$ and different sources of particles) was about 
15,000. In our runs, orbital elements were stored with a step of $d_t$ of 
20 yr for asteroidal and cometary particles and 100 yr for trans-Neptunian 
particles during all considered time intervals.  The stored orbital 
elements of all particles during their dynamical lifetimes were then used 
in our studies of the velocities of particles presented in the next 
section.

The planets were assumed to be material points.  However, using orbital 
elements obtained with a step $d_t$, Ipatov and Mather (2006a) calculated 
the mean probability of a collision of a particle with the planet during 
the particle lifetime. The total probability of collisions of all 
particles with planets in one run is small. Therefore, in most runs the 
results obtained will be exactly the same whether or not we consider 
collisions of particles with planets during integration. Even if there 
were a particle that actually collides with a planet, the sum of dynamical 
lifetimes of all particles and the distribution of particles over their 
orbital elements during their dynamical lifetimes will be practically the 
same as in the model for which planets are considered as material points. 
Destruction of particles due to their collisions with other particles is 
also not considered in the present paper. The destruction affects mainly 
lifetimes of particles and their size distributions at different distances 
from the sun, but not the distribution of the particles’ orbital elements 
during their migration in the zodiacal cloud. In future models, which will 
consider the size distributions of particles, we plan to take into account 
the destruction of particles.

The initial positions and velocities of asteroidal particles used in our 
models were the same as those of the first $N$ numbered main-belt 
asteroids (JDT 2452500.5), i.e., dust particles are assumed to leave the 
asteroids with zero relative velocity. The initial positions and 
velocities of the trans-Neptunian particles were the same as those of the 
first trans-Neptunian objects (JDT 2452600.5). The initial positions and 
velocities of cometary particles were the same as those of Comet 2P/Encke 
($a_o$$\approx$2.2 AU, $e_o$$\approx$0.85, $i_o$$\approx$12$^\circ$), or 
Comet 10P/Tempel 2 ($a_o$$\approx$3.1 AU, $e_o$$\approx$0.526, 
$i_o$$\approx$12$^\circ$), or Comet 39P/Oterma ($a_o$$\approx$7.25 AU, 
$e_o$$\approx$0.246, $i_o$$\approx$2$^\circ$), or long-period comets 
($e_o$=0.995, $q_o$=$a_o$$\cdot$(1-$e_o$)=0.9 AU or $q_o$=0.1 AU, $i_o$ varied 
from 0 to 180$^\circ$ in each run, particles started at perihelion; 
these runs are denoted as ‘lp’ runs). We 
considered Encke particles starting near perihelion (runs denoted as 2P), 
near aphelion (runs denoted as `2P 0.5t'),  
and when the comet had orbited for $P_a/4$ after perihelion passage, where 
$P_a$ is the period of the comet (runs denoted as `2P 0.25t').
Note that semimajor axes and eccentricities of dust particles differed 
from those of parent bodies, but inclinations were the same.

For cometary particles, the initial value of time $\tau$ when perihelion 
was passed was varied for different particles with a step $d \tau$=1 day 
or $d \tau$=0.1 day near the actual value of $\tau$ for the comet. Comet 
10P is an example of a typical Jupiter-family comet moving inside 
Jupiter's orbit; Comet 39P moves outside Jupiter's orbit. Comet 2P is the 
only known high-eccentricity comet with aphelion distance $Q$$<$4.2 AU. It 
comes close to the sun and produces a lot of dust. Our initial data for 
dust particles are different from those in previous papers, and the parent 
comets producing dust are different (exclusive of Comet 2P) from those 
considered earlier.

\section{Model Used for Calculation of the Scattered Line Profile}

We calculated how the solar spectrum was changed after the light was 
scattered by the dust particles and observed at earth. This was carried 
out by first considering all orbital elements of dust particles during a 
single run, which were stored in computer memory with a step $d\tau$ (see 
the previous section). Based on these stored orbital elements, we 
calculated velocities and positions of particles and the earth during the 
dynamical lifetimes of the particles. For each pair of positions of a 
particle and the earth, we then calculated many ($\sim$10$^2$-10$^4$, 
depending on a run) different positions of the particle and the earth 
during the period $P_{rev}$ of revolution of the particle around the sun, 
assuming that orbital elements do not vary during $P_{rev}$. This number 
is smaller for runs with larger maximum dynamical lifetimes of particles. 

The model, which is based on all positions and velocities of dust 
particles during their dynamical lifetimes, represents the zodiacal cloud 
for the case when small bodies continuously produce dust at a constant 
rate along their orbits; seasonal effects and jumps in production of dust 
are not considered in this model. The model allows one to study the 
principal differences between spectra corresponding to particles produced 
by asteroids, comets, and trans-Neptunian objects. More complicated models 
to study the fractions of particles of different origin can be considered 
in the future.

The choice of a scattering function was based on analyses of dependences 
of scattering functions on angles $\theta$ and $\epsilon$ (see below) and wavelength presented 
in several papers (e.g., Giese 1963; Giese \& Dziembowski 1969; Leinert 
1975; Leinbert et al. 1976; Weiss-Wrana 1983; Hong 1985; Lamy \& Perrin 
1986), which mainly followed the Mie theory for scattering. The scattering 
function depends on the composition of particles, their sizes, and other 
factors.  However, we considered three simple scattering functions: (1) $g 
\propto 1/\theta$ for $\theta < c$, and $g \propto 1+$($\theta -c$)$^2$ 
for $\theta > c$, where $\theta$ is the angle between the earth and the 
sun, as viewed from the particle in radians and $c = 2\pi$/3 radian; (2) 
$g$ having the same functional form as above, but with $\theta$ replaced 
by elongation $\epsilon$, the angle between the particle and the sun, as viewed from 
the earth (eastward from the sun); (3) isotropic scattering.  For all 
three functions, the intensity $I$ of light that reaches the earth was 
considered to be proportional to $\lambda^2$$\cdot$$(R*r)^{-2}$, where $r$ 
is the distance between the particle and the earth, $R$ is the distance 
between the particle and the sun, and $\lambda$ is the wavelength of 
light. Since we considered the scattering near a single spectral line, 
wavelength $\lambda$ in our runs was essentially a constant.  Except for lines of 
sight close to the sun, these three scattering functions give virtually 
the same results (see below).  Since the differences between the 
functions that we considered were much greater than the differences 
between scattering functions presented in the publications cited above, we 
conclude that we need not worry about the precise form of the scattering.

For each particle position, we calculated the velocity relative to the 
sun, the velocity relative to the earth, and its respective distance to 
the earth and to the sun, $r$ and $R$. 
These parameters and the scattering 
function were then used to construct the solar spectrum received at the 
earth after light had been scattered by all the particles located within 
some limited solid angle (beam) along a line of sight from the earth. The 
direction of the beam is characterized by its eastern elongation from the 
sun $\epsilon$, and by its inclination above the ecliptic plane. 
Particles in beams of diameter 2$^\circ$ (Fig. 1) and 2.5$^\circ$ (other 
figures) were considered. 
Projections of relative velocities of particles on the lines of sight from the 
earth and the sun to particles were used for analysis of the Doppler shift.
In each run, particles all had the same size, 
the same $\beta$, and the same source (e.g., asteroidal).

\section{Variations in Solar Spectrum Caused by Scattering by Dust 
Particles}

\subsection{Centroid velocity shifts}

Figure~1 shows sample spectra of scattered Mg~I $\lambda$5184 obtained 
from these calculations toward sightlines in the antisolar 
direction (Fig.~1a,c) and toward the ecliptic pole (Fig.~1b,d).  The spectra 
consist of intensity vs. wavelength shift $\Delta \lambda$ with respect to 
5183.62 Angstrom.
The thinnest 
line in Figure 1 denotes the initial (unscattered) solar spectrum. For 
asteroidal ('ast') and Comet Encke (`2P') particles, the solar spectrum 
received at the earth is found to be nearly independent of the scattering 
function, while for trans-Neptunian (`tn') particles, the results for 
scattering function (3), denoted as `tn3', differ slightly from (1) and 
(2), denoted as `tn 1' and 'tn 2', respectively. In the legend in the 
figure, the first number (0.2 in Fig.~1a,c) denotes $\beta$, and the next number 
in Figures 1a,c denotes elongation $\epsilon$ (in degrees).
The WHAM observations are also presented for 
comparison. These observations 
and all other plots in Figure~1c,d 
were stretched so that the minimum in the 
line was at approximately the same depth as that for the initial solar 
spectrum. The continuum levels were also set to be the same (equal to 1) 
for all plots.
Similar plots at $\epsilon$=90$^\circ$ and $\epsilon$=270$^\circ$ for zero
inclination above the ecliptic plane were presented by Ipatov et al. (2005).
Unlike results by Clarke et al. (1996) 
who considered spectrum near 4861 Angstrom ($H_{\beta}$ line of hydrogen),
our modeled spectra don't exhibit strong asymmetry. 
We similary found that minima in the plots of dependencies 
of the intensity of light on its wavelength near 5184 Angstrom are not as 
deep as those for the initial solar spectrum. 
Note that the value of $\epsilon$ in the paper by Clarke et al. (1996)
is measured in the opposite clockwise direction than in our present paper, so our 90$^\circ$
corresponds to 270$^\circ$ in their paper.

At the North Ecliptic Pole, the calculated spectrum was 
shifted slightly to the left relative to the solar spectrum for asteroidal 
particles and slightly to the right (to the red) for particles originating 
from Comet 2P (Fig. 1b,d).  These shifts may be due to small asymmetries in 
the model particle distributions with respect to the ecliptic plane.  The 
spectra of 10P and 39P particles and those from long-period comets were 
very similar to each other.  For cometary particles, the line profile has 
a flatter bottom than that for asteroidal particles, but it was not so 
wide as the observed spectrum presented in (Reynolds et al. 2004).
None of our model runs matched the large width of the observation toward 
the ecliptic pole.  This issue will not be addressed in this paper, but 
will be a topic for future investigation.

The calculated model spectra along the ecliptic plane are 
in general agreement with the observations.  
For different values of $\epsilon$, based on the model spectra, 
which use the distribution of velocities and positions of dust particles 
in our run, we determined the Doppler shift 
of the scattered Mg~I line 
with respect to the centroid wavelength of the unscattered solar profile.
The plot of this characteristic 
velocity vs. the solar elongation $\epsilon$ along the ecliptic plane is 
called the `velocity-elongation' plot. For `velocity-elongation' plots 
marked by $c$, we considered the shift of the centroid (the `center of 
mass' of the line), while `velocity-elongation' plots marked by $m$ denote 
the shift of the minimum of the line.  `Velocity-elongation' plots for 
different scattering functions are denoted as $c1$ and $m1$ for the 
scattering function 1, as $c2$ and $m2$ for the function 2, and as $c3$ 
and $m3$ for the function 3. The lines in Figure 1 are nearly symmetric, 
so the results for `c' and `m' in Figure 2 differ only a little.  Brief 
discussions of `velocity-elongation' plots obtained in our runs with 
observations have been made by Ipatov et al. (2005, 2006), Madsen et al. 
(2006), and Ipatov and Mather (2006a).  Below we 
compare the `velocity-elongation' plots in more details.

The details of the model spectra depend on $\epsilon$, $\beta$, $i_o$, and 
the source of particles. Different particles populations produce clearly 
distinct model spectra of the zodiacal light.  The `velocity-elongation' 
plots obtained for different scattering functions were similar at 
30$^\circ$$<$$\epsilon$$<$330$^\circ$ (Fig. 2); though the difference was 
greater for directions close to the sun. Some jumps in the plots are 
caused by small number statistics for these runs. In Figures 3-8 the 
results were obtained using only the second scattering function.

A comparison of the observed `velocity-elongation' plot with those 
obtained from our model for dust particles of different sizes (i.e., 
different values of $\beta$) produced by asteroids, comets (2P/Encke, 
10P/Tempel 2, 39P/Oterma, long-period), and trans-Neptunian objects 
allowed us to draw some conclusions about the sources of zodiacal dust 
particles. Asteroidal, trans-Neptunian, and 2P particles populations 
produce clearly distinct model spectra of the zodiacal light. The 
differences between `velocity-elongation' plots for several sources of 
dust 
reached its maximum 
at $\epsilon$ between 90$^\circ$ and 120$^\circ$ (Figs. 3---6). For future 
observations of velocity shifts in of the zodiacal spectrum, it will be 
important to pay particular attention to these elongations.

We consider the amplitude of `velocity-elongation' curves as 
$v_a$=($v_{\max}$-$v_{\min}$)/2, where $v_{\min}$ and $v_{\max}$ 
are the minimum and maximum values of velocities at 90$^\circ$$<$$\epsilon$$<$270$^\circ$. 
The observational value of $v_a$ is about 12 km/s (if we smooth the plot).
For several dust sources, the characteristic values of $v_a$, $v_{\min}$, and $v_{\max}$ are presented in Table 1.
Mean eccentricities $e_z$ and mean inclinations $i_z$ at distance 
from the sun 1$\le$$R$$\le$3 AU are also included in the table.
For asteroidal dust, the `velocity-elongation' plots had 
lower amplitudes than the observations (Fig. 3a--b).  The plots obtained 
at different $\beta$ differed little from each other, especially at 
$\beta$$<$0.01. For 10P particles, the `velocity-elongation' amplitudes 
were also lower than that of the observations (Fig. 3c--d). The difference 
between the plots obtained for 10P particles at different $\beta$ was 
greater than that for asteroidal particles, but was usually less than that 
for other sources of particles considered. On the other hand, the 
`velocity-elongation' curve corresponding to particles produced by Comet 
2P have slightly larger amplituides than the observed curve (Fig. 4a-b).  
So perhaps a combination of cometary 2P dust particles and asteroidal 
particles could provide a result that is close to the observational 
`velocity-elongation' curve. The velocity amplitudes $v_a$ for 
particles originating from long-period comets are much greater than those for the 
observational curve (Fig. 6b). 

The orbit of Comet 39P is located outside Jupiter's orbit.  Studies of the 
migration of 39P particles thus give some information about the the 
migration of particles originating beyond Neptune's orbit that have 
reached the orbit of Jupiter.  For 39P particles and 
0.01$\le$$\beta$$\le$0.2 at 60$^\circ$$<$$\epsilon$$<$$150^\circ$, 
`velocity-elongation' amplitudes were smaller than those for the observations (Fig. 
5a), while for $\beta$$\le$0.004, the curves more closely match the 
observations (Fig. 5b). For such small $\beta$, only a small number of 
particles entered inside Jupiter's orbit, and statistics were poor. The 
distribution of particles over their orbital elements could be somewhat 
different if we had considered a greater number $N$ of particles in one 
run, but the difference between the curves obtained at different $N$ will 
not be more than the difference between the curves obtained at adjacent 
values of $\beta$ presented in the figure. At $\beta$=0.0001, particles 
spent on average more time in the zodiacal cloud than for other runs, but 
it was due mainly only to one particle, which during about 9 Myr moved 
inside Jupiter's orbit before its collision with the sun.  In reality such 
a particle could sublimate or be destroyed in collisions during its 
dynamical lifetime. Small number statistics may also be responsible for 
the spread of results for the trans-Neptunian dust in Figure 6a.  If there 
had been a greater number of particles in the runs, the differences 
between the corresponding plots would probably have been smaller.
There was a zero vertical shift for the observational plot. Plots for asteroidal, 
10P, and 39P particles are shifted a little down, and that for 2P particles 
was shifted up relative to the observational plot. So a combination of different 
sources of particles could give a zero vertical shift.

For trans-Neptunian dust particles, velocity amplitudes $v_a$ were greater than 
the observational values at $\beta$$\ge$0.05 and were about the same at 
$\beta$=0.01 (Fig. 6a). So together with cometary particles, trans-Neptunian 
particles can compensate small values of $v_a$ for asteroidal particles.
The difference between different plots in Fig. 6a 
may be caused partly by small statistics, as only a few trans-Neptunian particles 
in each run (at fixed $\beta$) entered inside Jupiter's orbit. 
The observational plot was mainly 
inside the region covered by trans-Neptunian plots obtained for different $\beta$, but at 
180$^\circ$$<$$\epsilon$$<$270$^\circ$ it was mainly above
the trans-Neptunian curves. 


In the future we plan to explore the fractions of particles of different 
origin in the overall dust population based on various observations and 
taking into account a model for the size distribution of particles. Here 
we present estimates based on a much simpler, two-component zodiacal dust 
cloud that fits the observations.  For example, with a velocity amplitude 
$v_a$=9 km/s for asteroidal dust and at $v_a$=14 km/s for 2P particles, 
the fraction $f_{ast}$ of asteroidal dust would need to be 40\%. If all of 
the cometary particles in the zodiacal dust were from long-period comets 
($v_a$=33 km/s), then $f_{ast}$ = 88\%. 
The contribution of {\it lp} particles to the zodiacal light can not be
 large also because their inclinations are large and IRAS observations showed
(Liou et al. 1995) that most of the zodiacal light is due to particles with
inclinations $i$$<$30$^\circ$.
At $\beta$$\ge$0.004, {\it lp} particles are quickly ejected from the solar system,
so among zodiacal dust we can find only {\it lp} particles greater than 100 $\mu$m. The 
contribution of {\it lp} particles to the total mass of the zodiacal cloud is greater than 
their contribution to the brightness $I$, as a surface of a particle of radius $r_p$ is proportional to $r_p^2$, 
and its mass $M$ is proportional to $r_p^3$ (i.e., $M/I$$\propto$$r_p$). 
Dynamical lifetimes of {\it lp} particles at 
$\beta$$\le$0.002 (i.e., at diameters greater than 200 $\mu$m) can exceed 
several Myrs (i.e.,  exceed mean lifetimes of asteroidal and 2P particles),
so the fraction of {\it lp} particles in the zodiacal cloud can be greater than
their fraction in the new particles that were produced by small bodies or came 
from other regions of the solar system. 
Dynamical lifetimes of dust particles are greater for smaller $r_p$ 
(Ipatov and Mather 2006a), 
and some particles can be destroyed in collisions with other particles. 
So the mass distributions of particles produced by small bodies are
different from the mass distributions of particles located at different distances from the sun.

A significant fraction of cometary dust in the near-earth space was 
proposed by Liou et al. (1995) and Zook (2001). Liou et al. (1995) 
based his estimates of the fraction of cometary dust (3/4 to 2/3) on 
comparison of the migration 
of 9 $\mu$m diameter dust particles that produced by comet Encke-type parent 
bodies with the IRAS observations of the shape of the zodiacal cloud. 
Based on cratering rates from an ensemble of earth- 
and lunar-orbiting satellites, Zook (2001) estimated that the cometary contribution 
to the near-earth flux of particles is $\sim$75\%.
Ozernoy (2001) compared the computer simulation results of migration of 
1 $\mu$m and 5 $\mu$m dust particles with the COBE/DIRBE data and showed 
that the trans-Neptunian dust contributes as much as $\sim$1/3 of the total 
number density near the earth. 
Grogan et al. (2001) suggested that at least 30\% of zodiacal 
dust comes from the break-up of asteroids in order to explain formation of dust 
bands. The conclusions on a considerable fraction of cometary dust are also in agreement with 
earlier studies of the dynamics of Jupiter-family comets (Ipatov \& Mather 
2003, 2004a-b, 2006a), which showed that some former cometary objects 
could get high eccentric orbits located entirely inside Jupiter's orbit 
and move in such orbits for a long time. Some of these 
objects could disintegrate producing a substantial amount of dust. 
Comparison of the number density 
of migrating dust particles with the observation that number density is 
constant at 3-18 AU from the sun (Ozernoy 2001, Landgraf 2002, Ipatov \& Mather 2005, 2006a) is also 
consistent with that a fraction of cometary dust particles is significant 
(perhaps dominant) inside Saturn's orbit.

\subsection{Line widths}

Observations by Reynolds et al. (2004) also provided the FWHM (full width 
at half maximum) of the Mg~I line in the zodiacal light. In Figure 7, 
their values are marked by crosses. 
Most of the observed values were 
between 70 and 80 km/s, with a mean value of 76.6. In Figure~7, we compare 
these observations with the FWHM values obtained from our model line profiles. 
The values are different for different $\beta$ and different 
sources of the dust.  For particles started from asteroids, comets 2P, 10P 
and 39P, and long-period comets, the mean (at 
30$^\circ$$<$$\epsilon$$<$330$^\circ$) values of FWHM are about 74, 81-88, 76-77, 
76-77, 73-86 km/s, respectively. For 2P particles, the values of FWHM depend on 
$\beta$ and the place of origin from the orbit. The mean values of FWHM 
obtained for particles started from long-period comets can differ 
considerably in different runs with different $\beta$.
Discrete values of the width in Figure 7 are caused by a discrete model of spectra.
As observational values of the width can differ by 10 km/s for close values of $\epsilon$,
there is no need to calculate accurately the width in our models.

At $\beta$$\le$$0.004$ for particles started from Comet 2P, the width is 
greater at $\epsilon$$\approx$220$^\circ$ than at 
$\epsilon$$\approx$45$^\circ$ (Fig. 7b-d). For other sources of dust, the 
width is approximately independent of $\epsilon$ at 
30$^\circ$$<$$\epsilon$$<$$330^\circ$.  The widths all become relatively 
large at $-15^\circ$$<$$\epsilon$$<$15$^\circ$, where there are no 
observations. At $\beta$$\ge$0.1 for particles started from Comet 2P, the 
widths are generally narrower than that at smaller $\beta$. For asteroidal 
particles, on the contrary, the width is maximum at $\beta$=0.4.  For 10P 
and 39P particles, the mean width is slightly greater for smaller $\beta$ 
at 0.0001$\le$$\beta$$\le$0.1 (Fig. 8). For particles started from 
long-period comets (at $e_z$=0.995, $q$=$a(1-e)$=0.9 AU, and orbital 
inclinations distributed between 0 and 180$^\circ$) the mean width usually 
is even less than that for 2P particles at the same $\beta$.

As it is summarized in Figure 8, the FWHM for asteroidal dust is 
significantly less than the 77~km/s FWHM of the observations.  To fit the 
observations, we need to consider a significant fraction of cometary 
particles.
2P particles and particles started from long-period comets provided large
values of the width in most runs, but the width varied 
considerably in such runs depending on $\beta$ and the place of start 
(i.e., the value of true anomaly) from the comet.

\section{Orbital Elements of Zodiacal Particles}                         

In order to understand the variations in the model line profiles with the 
source and size of particles, we examined the values of mean 
eccentricities and mean orbital inclinations of particles at different 
distances $R$ from the sun (Figs. 9-12). These data show that, in general, 
the velocity amplitudes of the Mg~I line are greater for greater mean 
eccentricities and inclinations, but they depend also on distributions of 
particles over their orbital elements.

The main contribution to the brightness of a dust cloud observed at the 
earth is from particles located at $R$$<$3 AU.  Therefore, because only 
positions of particles at $R$$\ge$1 AU are used for calculation of the 
brightness of particles at 90$^\circ$$\le$$\epsilon$$\le$$270^\circ$, if it is not 
mentioned specially, the mean eccentricities $e_z$ and orbital inclinations $i_z$
refer to 1$\le$$R$$\le$3 AU.  At $\epsilon$=270$^\circ$ prograde particles 
move in the same direction as the earth, but usually slower because they 
are located father from the sun than the earth. Therefore, velocities of 
Mg~I line are usually positive at $\epsilon$=90$^\circ$ and negative at 
$\epsilon$=270$^\circ$.

The values of $e_z$, $i_z$, $v_a$, $v_{\min}$, and $v_{\max}$ for particles from different sources 
are presented in Table 1 for several values of $\beta$. This table and 
Figures~3-6, 9-12 show that at $e_z$$<$0.5 (e.g., for particles started 
from asteroids and Comet 10P) the velocity amplitudes $v_a$ are 
usually smaller than the observed amplitude (12 km/s), while for some runs 
at $e_z$$>$0.5 (e.g., for runs 2P and {\it lp}), $v_a$ is greater than 
the observed amplitude. Thus the WHAM observations correspond to a mean 
orbital eccentricity $e_z$ of about 0.5.  However, 
the velocity amplitudes of the line depend not only on $e_z$, but also on the 
distribution of all orbital elements of dust particles.  For example, for 
trans-Neptunian particles at 0.05$\le$$\beta$$\le$0.4, we have relatively 
large amplitudes ($v_a$$\approx$16 km/s), even though $e_z$$<$0.5; the values 
of $i_z$ in this case were a little greater than for other sources of 
particles at $e_z$$<$0.5. Also, for trans-Neptunian particles, the values 
of $v_a$ were smaller at $\beta$=0.01 than at 0.05$\le$$\beta$$\le$0.4, 
though $e_z$ was greater at $\beta$=0.01. 

The velocity amplitudes also depend on inclinations, because high 
inclination orbits 
have smaller projections of their orbital velocities on 
the lines of sight from the earth and the sun
to particles near the ecliptic plane. 
For particles originating from 
long-period comets ({\it lp} particles), the values of $v_a$ are 
significantly greater than those for dust particles from other sources, 
and the values of $i_z$ ($>$105$^\circ$) were much greater than for other 
runs.  For {\it lp} runs at $\beta$=0.002 and $q$=0.9 AU, $e_z$ was smaller ($<$0.25) 
than in most other runs, but $v_a$ was large (41 km/s), because $i_z$ was 
large (154-160$^\circ$). Note that mean initial inclinations for {\it lp} 
runs were about 90$^\circ$ (initial orbital inclinations were distributed 
uniformly between 0 and 180$^\circ$), but the mean inclinations of 
migrating particles were greater than 90$^\circ$ (see Fig. 12d).

The velocities of Mg~I line depend not only on mean eccentricities and 
inclinations of particles, but also on the distributions of the orbital elements.  The 
distribution of orbital parameters and the resulting scattered line 
profile is dependent upon $\beta$, because $\beta$ influences the lifetime 
of the particle.  Dynamical lifetimes of particles are greater for smaller 
$\beta$. In the case of particles started from Comet 2P, 
for runs presented in Figures 13a,c,d, the maximum lifetimes were 
less than 5 Kyr, and for runs presented in Fig. 13b,f they were about 
50-60 Kyr. Particles that migrate more slowly into the sun (smaller 
$\beta$) interact with planets for a longer time, and therefore they 
exhibit a wider range of parameter values, even though they have the same 
origin. This explains the difference between the plots in 
Figure~13 obtained at different $\beta$ 
(wider range of $e$ was obtained at smaller $\beta$). 
As the motion is stochastic and 
the number of particles in our runs is not large, there may be no strict 
dependence on $\beta$.
The plots of the velocities of MgI line for `2P 0.25t' and `2P 0.5t' runs 
presented in Fig. 4c-d  
are higher than the observational line at large $\beta$. In this case, 
all eccentricities are large at $a$$>$1 AU (Fig. 13a,c,d), 
lifetimes of all particles are close to each other and are very short ($<$5 Kyr), 
and all particles moved practically in the same way. 
For $\beta$=0.05, the maximum lifetime of 2P-particles 
started at aphelion (`2P 0.5t' run) was greater by a factor of 4 than that for 
`2P 0.25t' run. Therefore for $\beta$=0.05, variations in $e$ at the same $a$
 were greater for `2P 0.5t' run than for `2P 0.25t' run.
For many particles other than 2P particles, lifetimes are several tens or several 
hundreds of thousands of years 
and can reach tens of millions of years (Ipatov and Mather 2006a).
Besides, more particles start from Comet 2P in its perihelion
(in this case, the plots do not differ much from the observational plot
even at large $\beta$) than in aphelion.
Therefore the zodiacal light contribution is very small for particles with 
$\beta$$\ge$0.05 started from Comet 2P at aphelion or in the middle of the orbit.

Finally, we examine the number density $n(R)$ of particles with distance $R$ 
from the sun.  In Figures 14a-c we present the values of $\alpha$ in $n(R) 
\propto R^{-\alpha}$ for $R$ equal to 0.3 and 1 AU (a), at $R$=0.8 and 
$R$=1.2 AU (b), and at $R$ equal to 1 and 3 AU (c). These results can then 
be compared to values of $\alpha$ deduced from observations of the actual 
zodiacal cloud. The 
micrometeoroid flux ($10^{-12}$ g - $10^{-9}$ g, i.e., at $d$$<$5 $\mu$m 
for $\rho$=2.5 g/cm$^3$, or at $\beta$$\ge$0.1) measured on board Helios 1 
during 1975 is compatible with a number density $n(R) \propto R^{-1.3}$ at distance 
$R$ from the sun between 0.3 and 1 AU
(Gr\"un et al. 1977, Leinert et al. 1981). Pioneer 10 observations between the 
earth's orbit and the asteroid belt yielded 
$n(R) \propto R^{-1.5}$ for particles of mass $\sim$$10^{-9}$ g (Reach 1992).
    
Note that in our models, at 0.3$\le$$R$$\le$1 AU and 0.001$\le$$\beta$$\le$0.2 all values of 
$\alpha$ exceed 1.9 for 2P-particles and are smaller than 1.1 for 
asteroidal particles (Fig. 14a). At $\beta$$\ge$0.005, the values of $\alpha$ for particles started 
from other considered comets were less than 1.5, but were mainly greater than those 
for asteroidal particles. 
We can get $\alpha$=1.3 for two component dust cloud if we consider 86\% of particles
with $\alpha$=1.1 and 14\% of particles with $\alpha$=2. 
So the fraction of 2P particles needed to fit the Helios observations
can be less than 15\%. 
At $\beta$$\ge$0.1 and 0.8$\le$$R$$\le$1.2 AU, 
the mean value of $\alpha$ for all points in Figure 14b was a little 
smaller 
than 1.5. For cometary dust, $\alpha$ was mainly greater than for 
asteroidal dust; this difference was greater at $\beta$$\le$0.05 than for 
$\beta$$\ge$0.1. At $\beta$$\le$0.2, the values of $\alpha$ for 2P-particles 
were greater than for other sources of dust. At 1$\le$$R$$\le$3 AU
for most of the dust sources,  
the values of $\alpha$  were mainly greater than 1.5 (Fig. 14c). For 0.1$\le$$\beta$$\le$0.2, the 
values of $\alpha$ for particles started from trans-Neptunian objects and Comet 39P 
better fit the observational value of 1.5 than those for particles from other 
sources (including asteroidal dust).
So our studies of the distributions of particles over $R$ and 
of the velocities of Mg~I line testify that a fraction of trans-Neptunian
particles among zodiacal dust particles can be considerable.

\section{Conclusions}

Our study of velocities and widths of the scattered Mg~I line in the 
zodiacal light is based on the distributions of positions and velocities 
of migrating dust particles originating from various solar system sources. 
These distributions were obtained from our integrations of the orbital 
evolution of particles started from asteroids, comets, and trans-Neptunian 
objects. A comparison of our models with the observations of velocities 
and widths of the zodiacal Mg~I line made by Reynolds et al. (2004) shows 
that asteroidal dust particles alone cannot explain these observations, 
and that particles produced by comets, including high-eccentricity comets 
(such as Comet 2P/Encke and long-period comets), are needed.  
A considerable fraction of trans-Neptunian dust particles among zodiacal dust
particles also fits different observations.
The mean eccentricity of zodiacal 
dust particles that best fit the WHAM observations is estimated to be 
about 0.5.


\section*{Acknowledgements}

     This work was supported by NASA (NAG5-12265) and by the National 
Science Foundation through AST-0204973.

\begin{figure}   

\includegraphics[width=81mm]{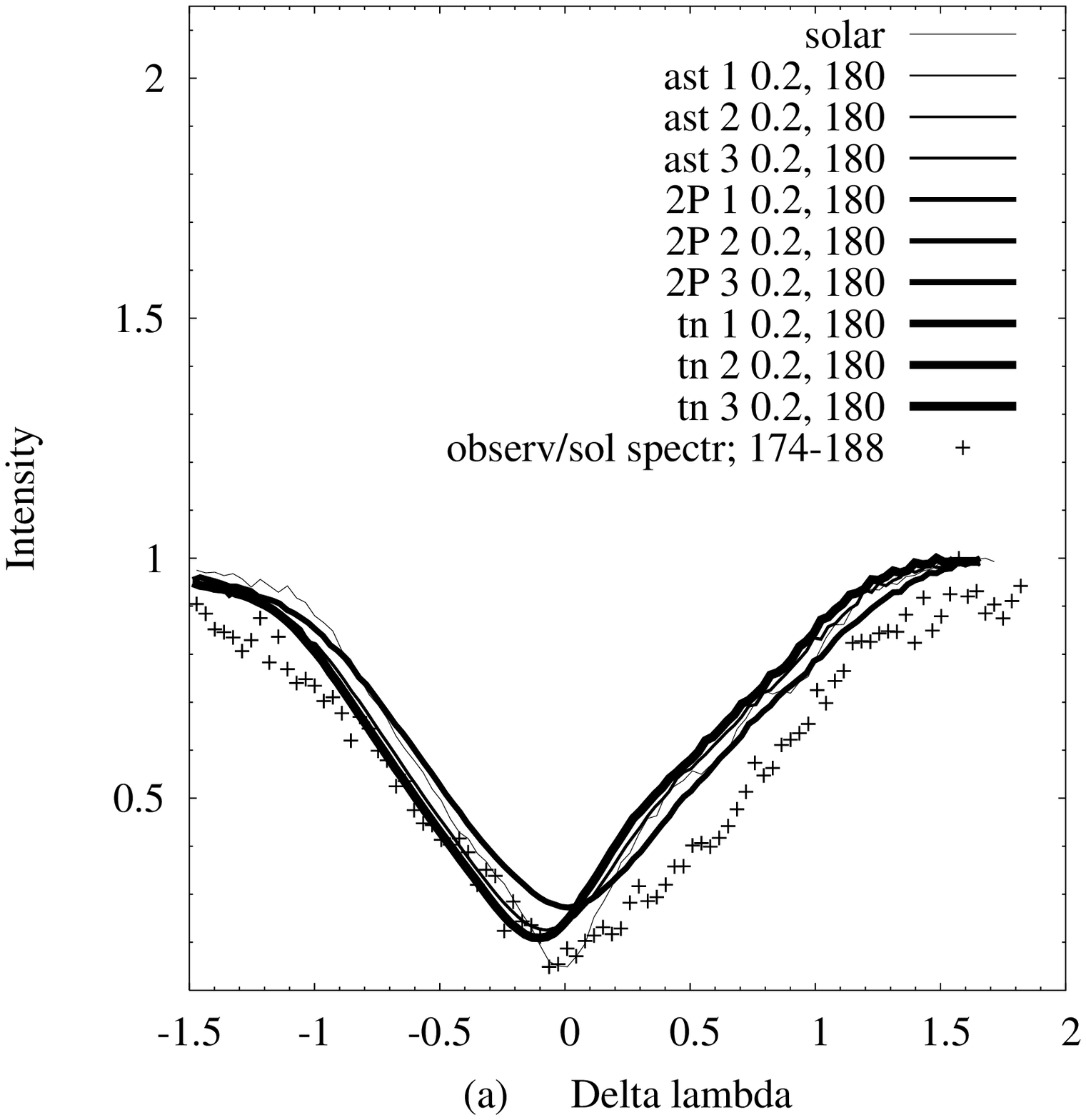} 
\includegraphics[width=81mm]{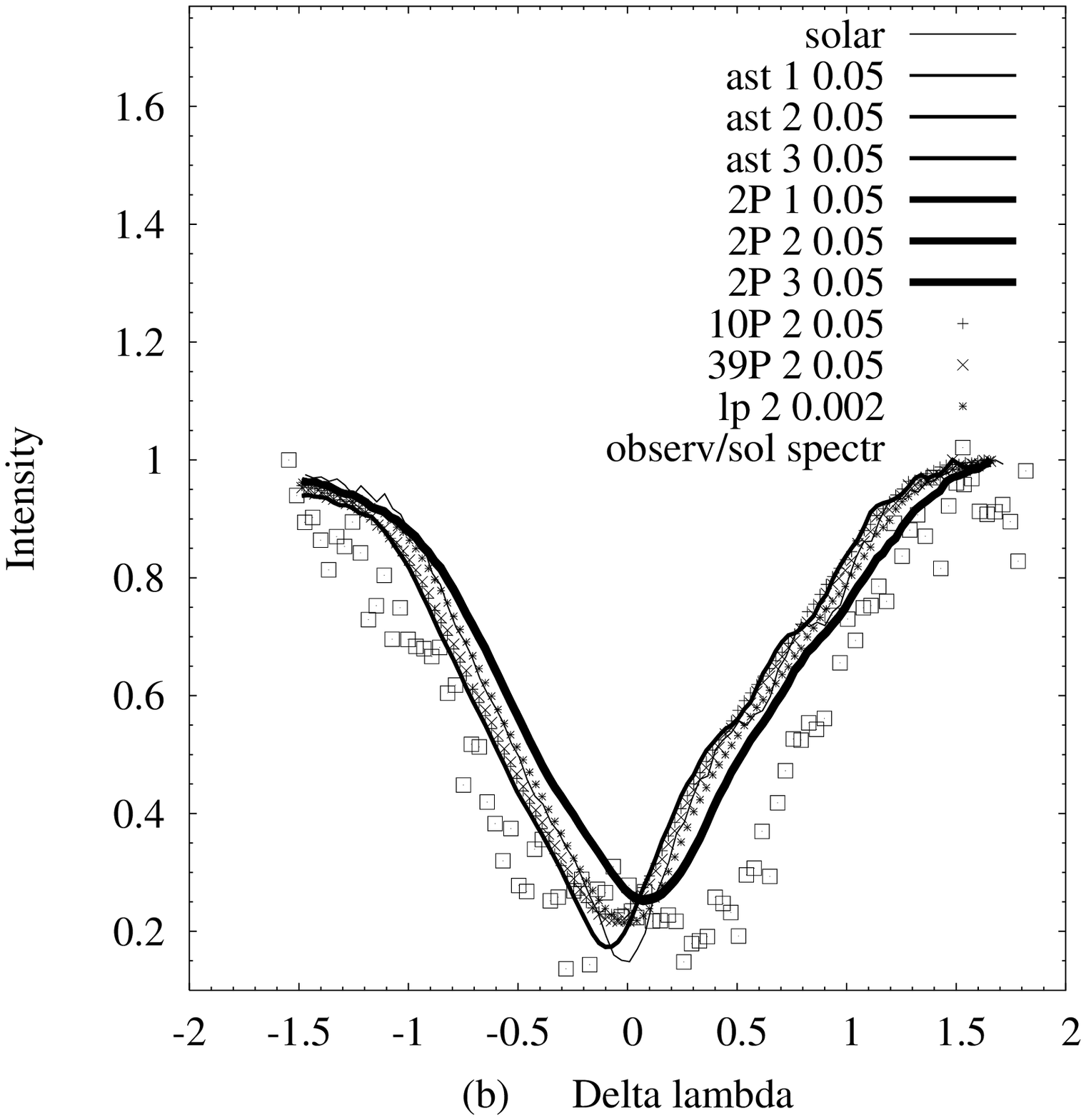} 
\includegraphics[width=81mm]{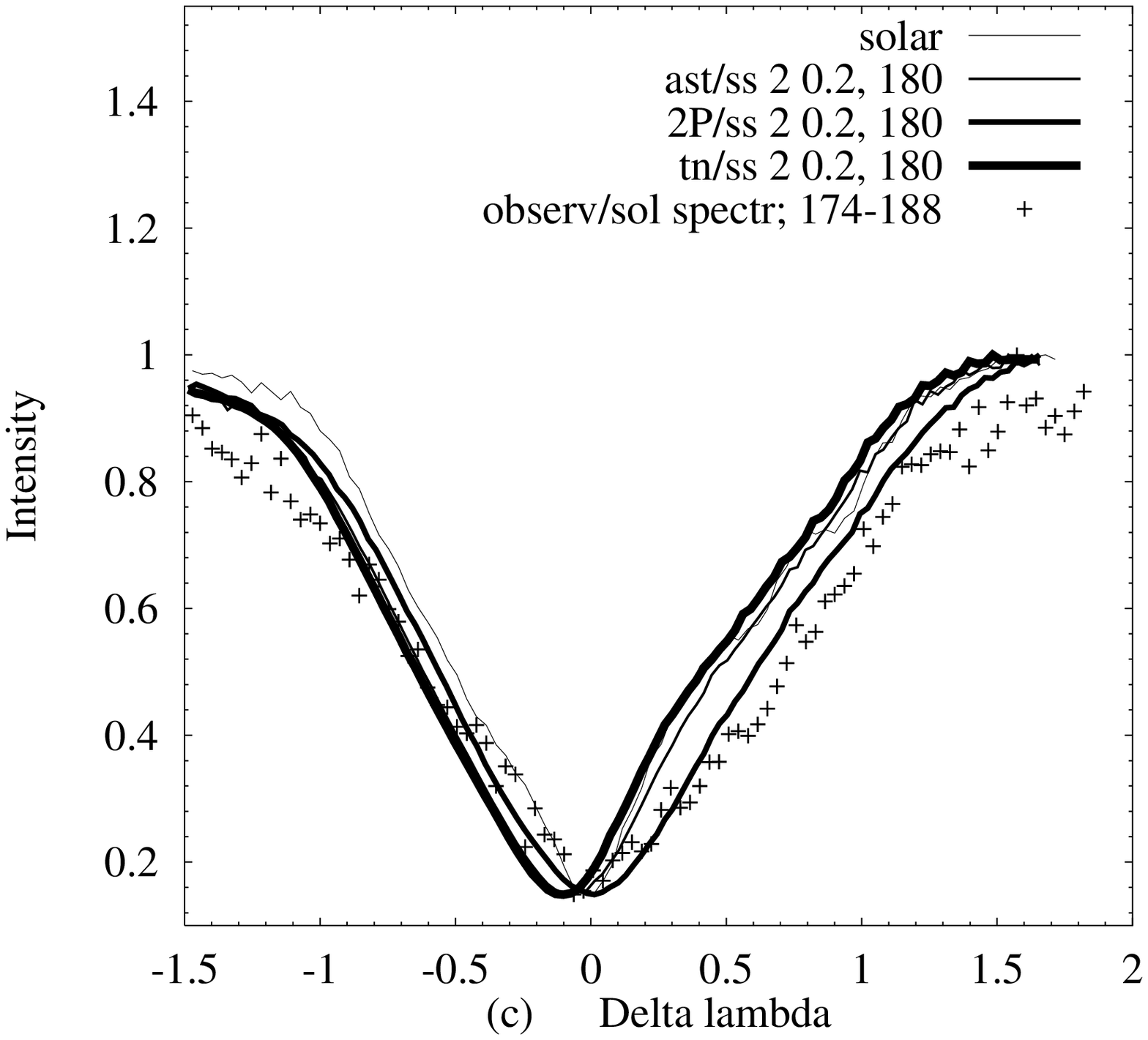} %
\includegraphics[width=81mm]{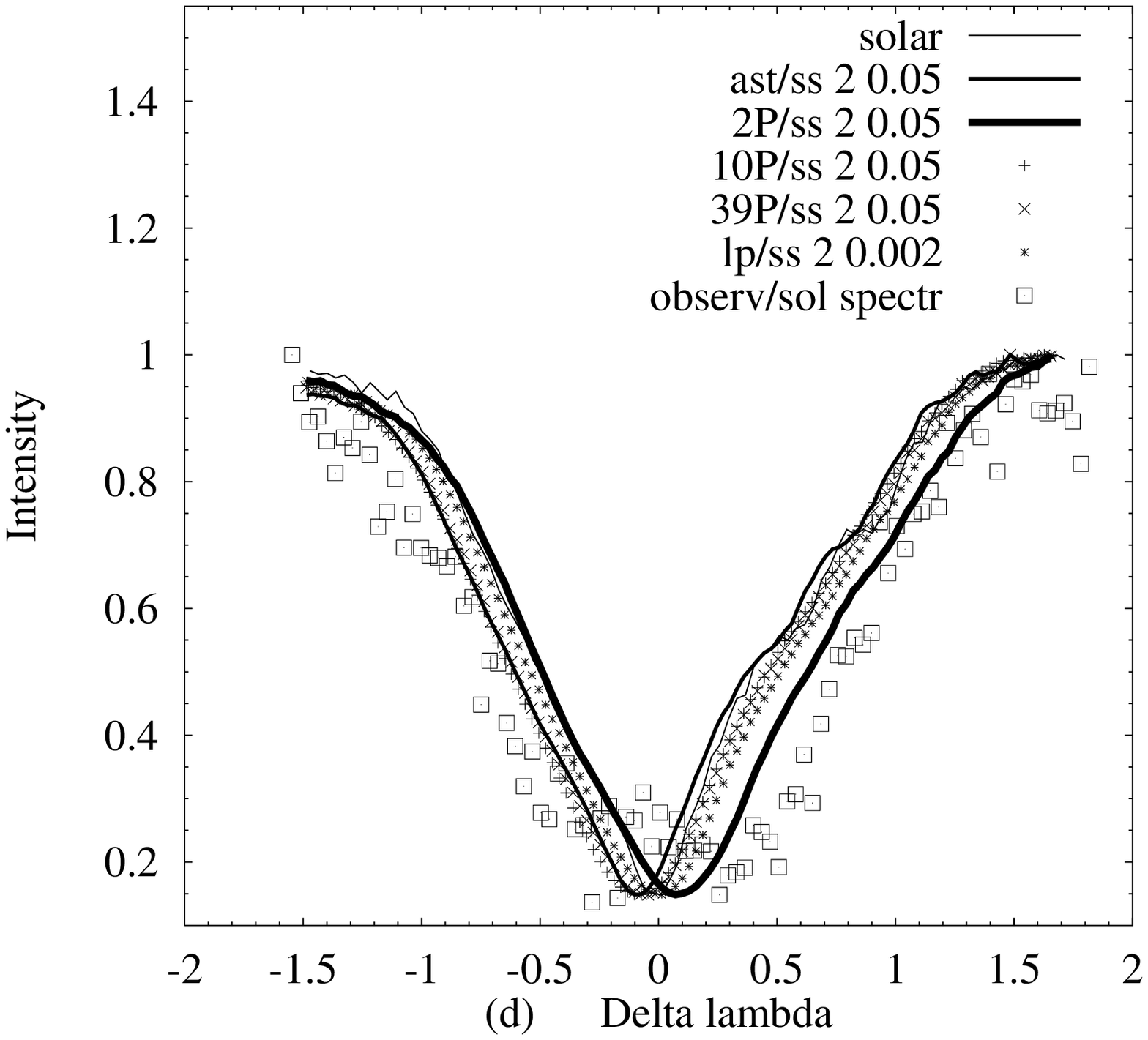} 

\caption{\small Dependence of the intensity of light vs. its wavelength $\lambda$ 
(in Angstrom) at $\beta$=0.2, $\epsilon$=180$^\circ$, in the ecliptic 
plane (a,c) and at 
$\beta$=0.05 (exclusive for lp particles considered at $\beta$=0.002) toward the north ecliptic pole (b,d).
Zero of $\Delta \lambda$=$\lambda$-$\lambda_\circ$ corresponds to 
$\lambda$=$\lambda_\circ$=5183.62 Angstrom.
The plots for dust particles started from asteroids, trans-Neptunian
objects, Comet 2P at perihelion, comets 10P and 39P are denoted by `ast', `tn', 
`2P', `10P', and `39P', respectively. 
Data for particles started from 
long-period comets (lp) at $e_o$=0.995, 
$q_o$=0.9 AU, and $i_o$ distributed between 0 and 180$^\circ$, are marked 
as 
`lp'.
The curves marked by 
`ast 1', `tn 1', and `2P 1' are for scattering function 1, `ast 2', `tn 2',
and `2P 2' are for function 2, and `ast3', `tn3', and `2P 3' are for 
function 3. 
The curves are practically the same for all three functions.
Marks in Fig. (a,c) are for average intensity for observations at 174$^\circ$$\le$$\epsilon$$\le$188$^\circ$),
and marks in Fig. (b,d) are for average observations toward the north ecliptic pole. 
Coordinates of marks for `observ/sol spectr' were obtained from observational data
by making it have the same minimum value as the solar spectrum.
In Fig. (c,d) the minima of plots obtained in all runs were made the same as that for the solar spectrum.
}
\end{figure}%

\begin{figure}   

\includegraphics[width=81mm]{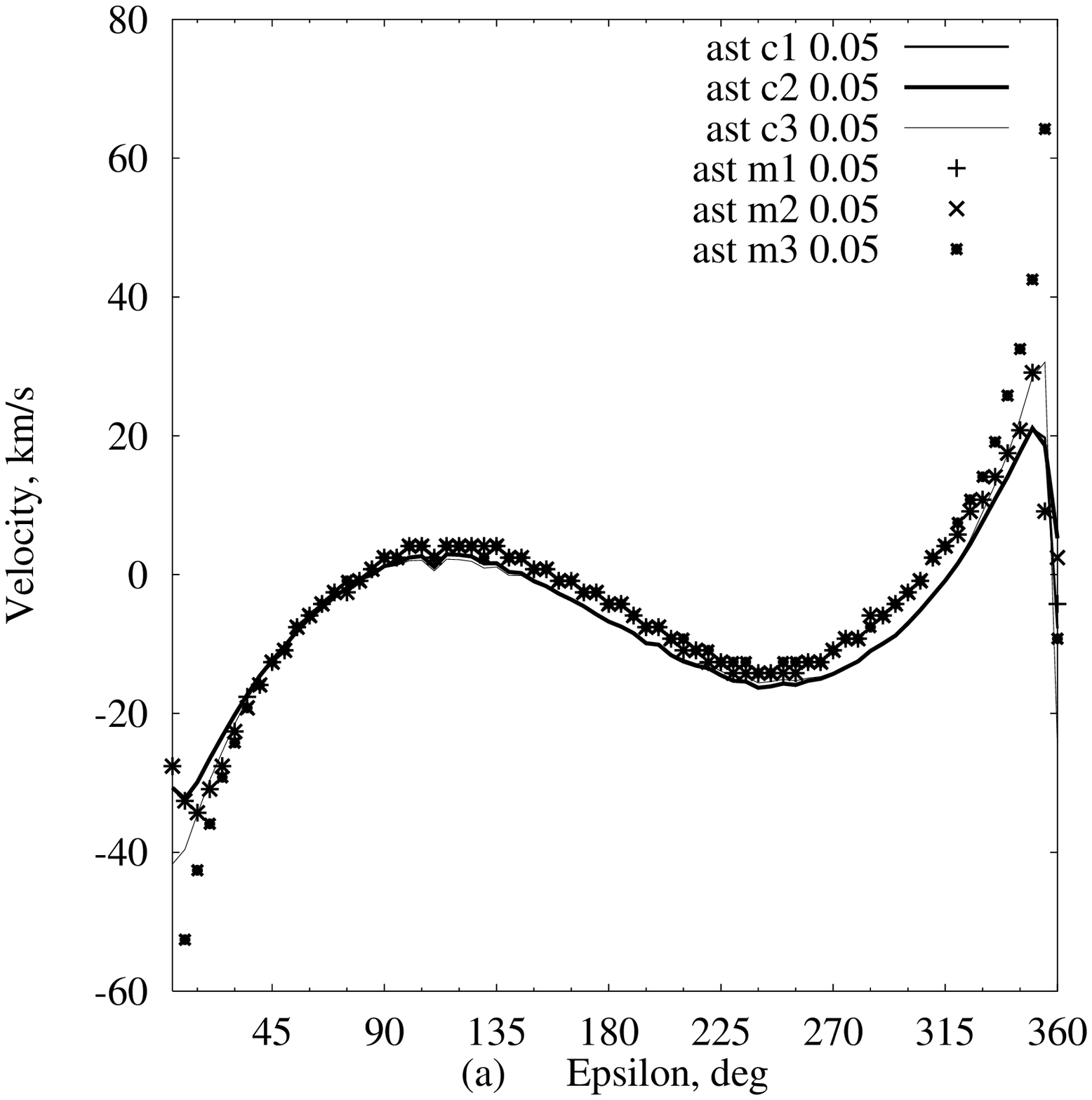} 
\includegraphics[width=81mm]{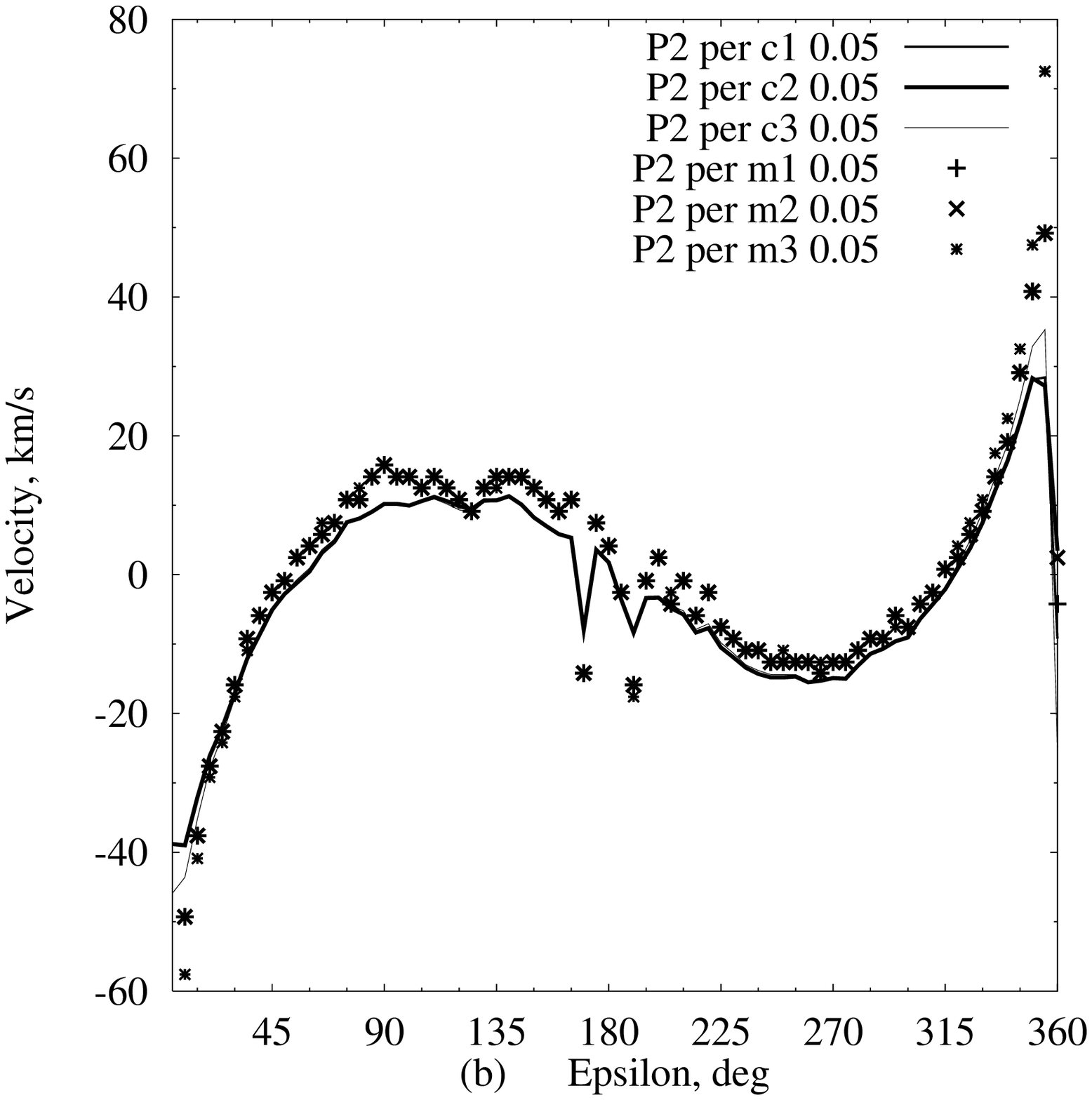} 
\includegraphics[width=81mm]{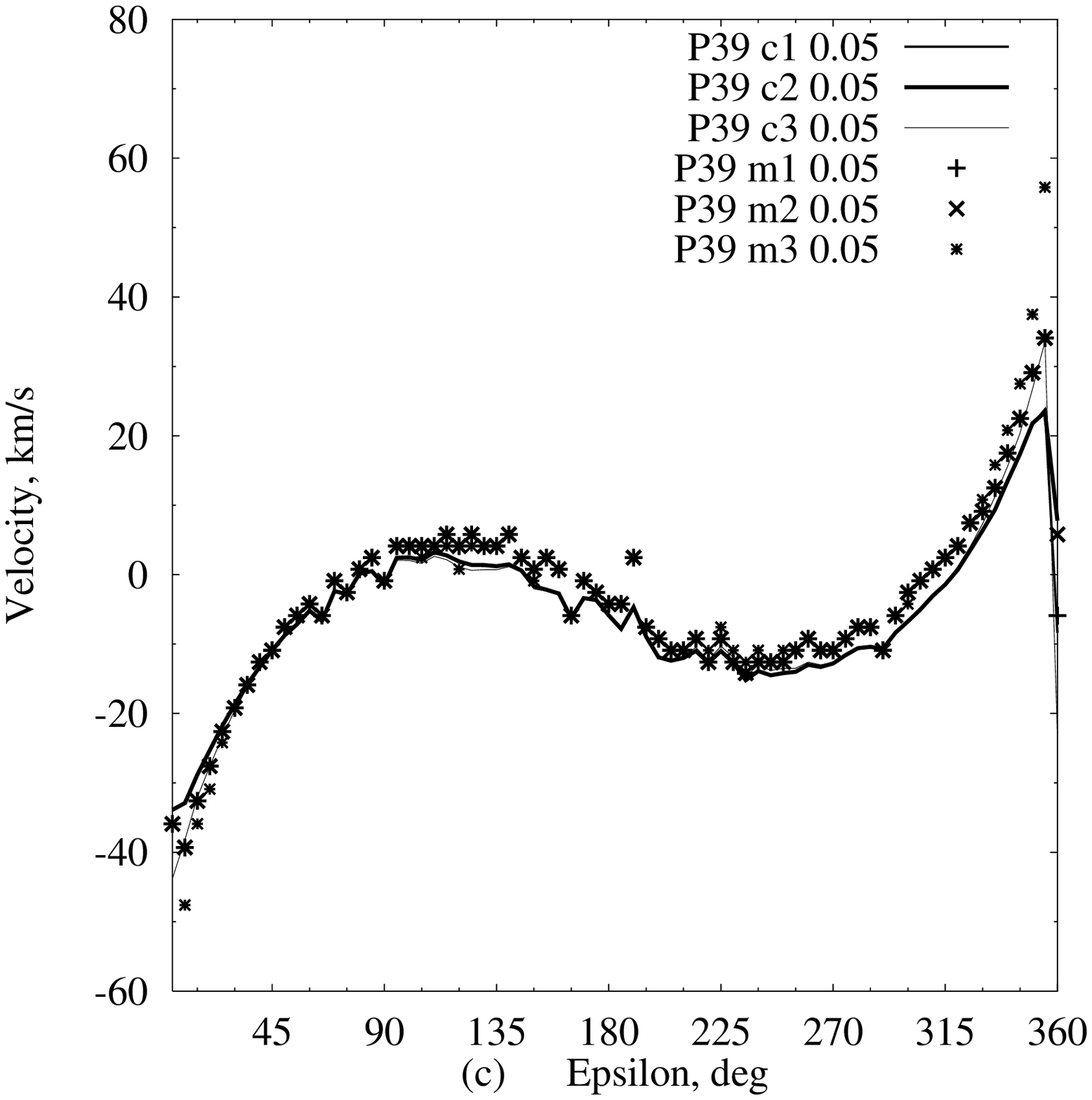} 
\includegraphics[width=81mm]{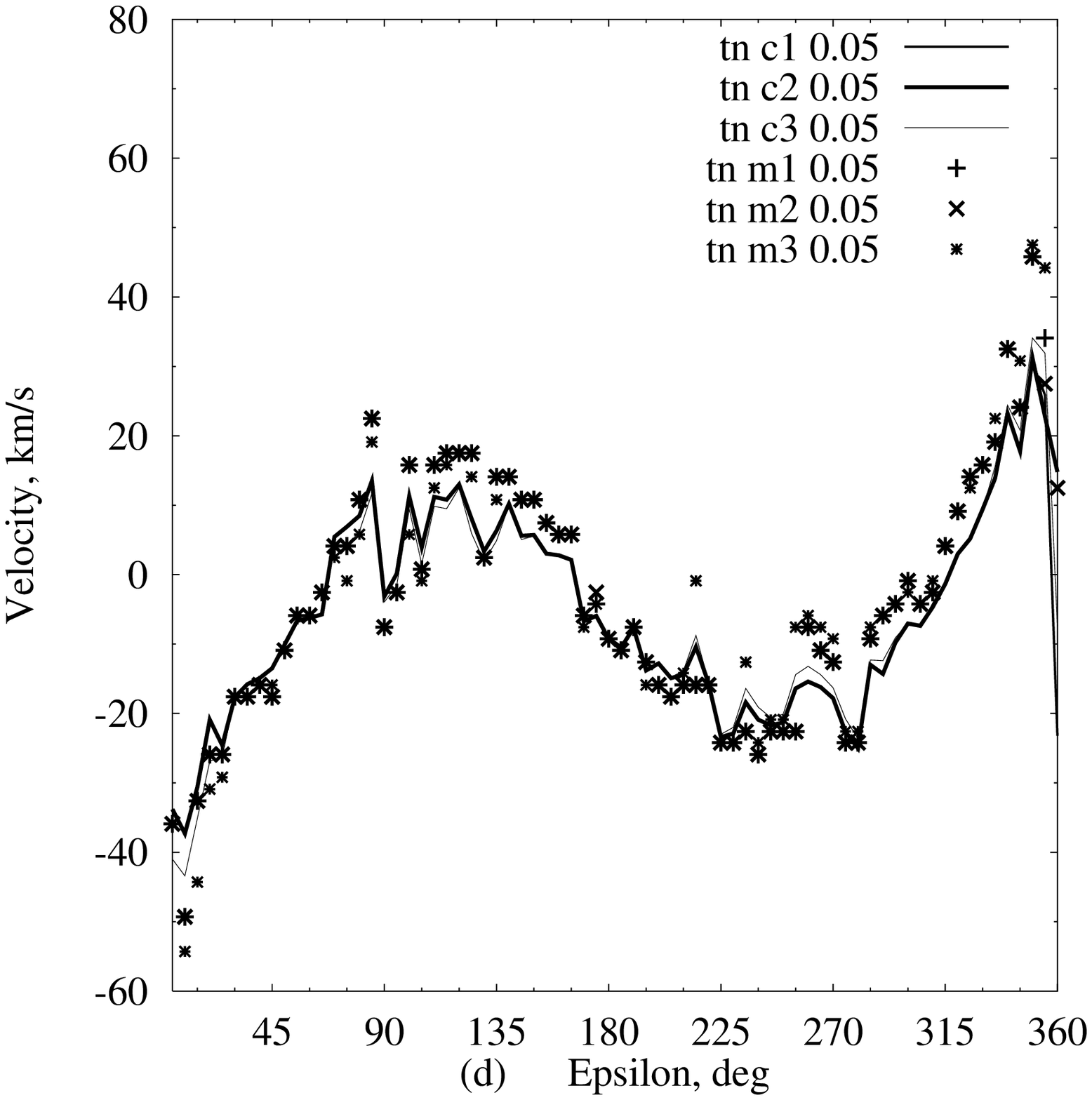} 

\caption{Velocities of Mg I line (at zero inclination) versus elongation 
$\epsilon$ (measured eastward from the sun) at $\beta$=0.05 
for dust particles started from asteroids (a), Comet 2P at perihelion (b),
Comet 39P (c), and trans-Neptunian particles (d). 
Letter `c' denotes the model for which the shift of the plot of 
$I$ vs. $\epsilon$ 
is calculated as a shift of centroid, and letter `m' denotes the model 
for which the shift of the plot is calculated as a shift of the minimum
of the plot. The number after `m' or `c' characterizes the number of a scattering function used.}


\end{figure}%

\begin{figure}

\includegraphics[width=81mm]{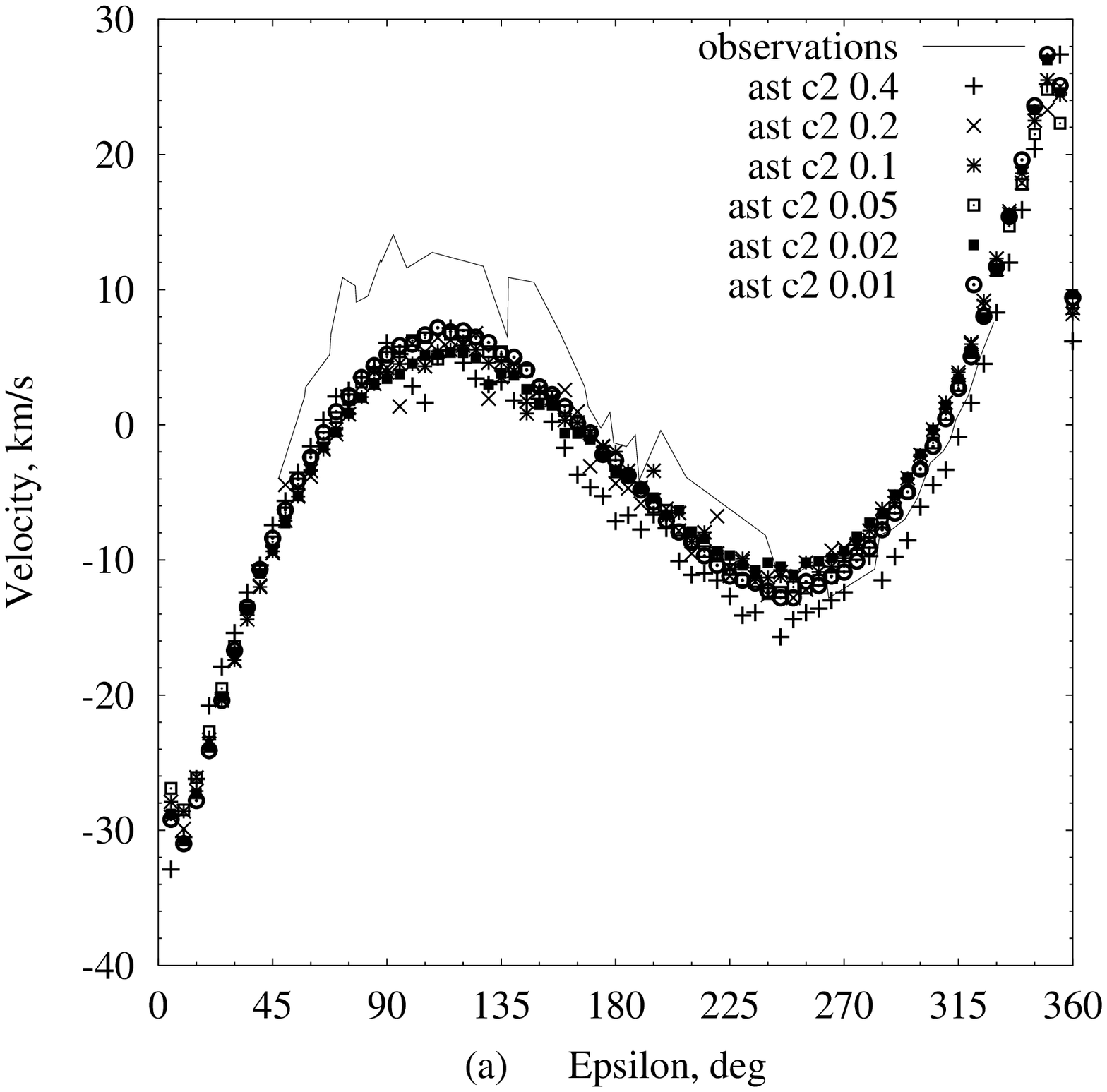} 
\includegraphics[width=81mm]{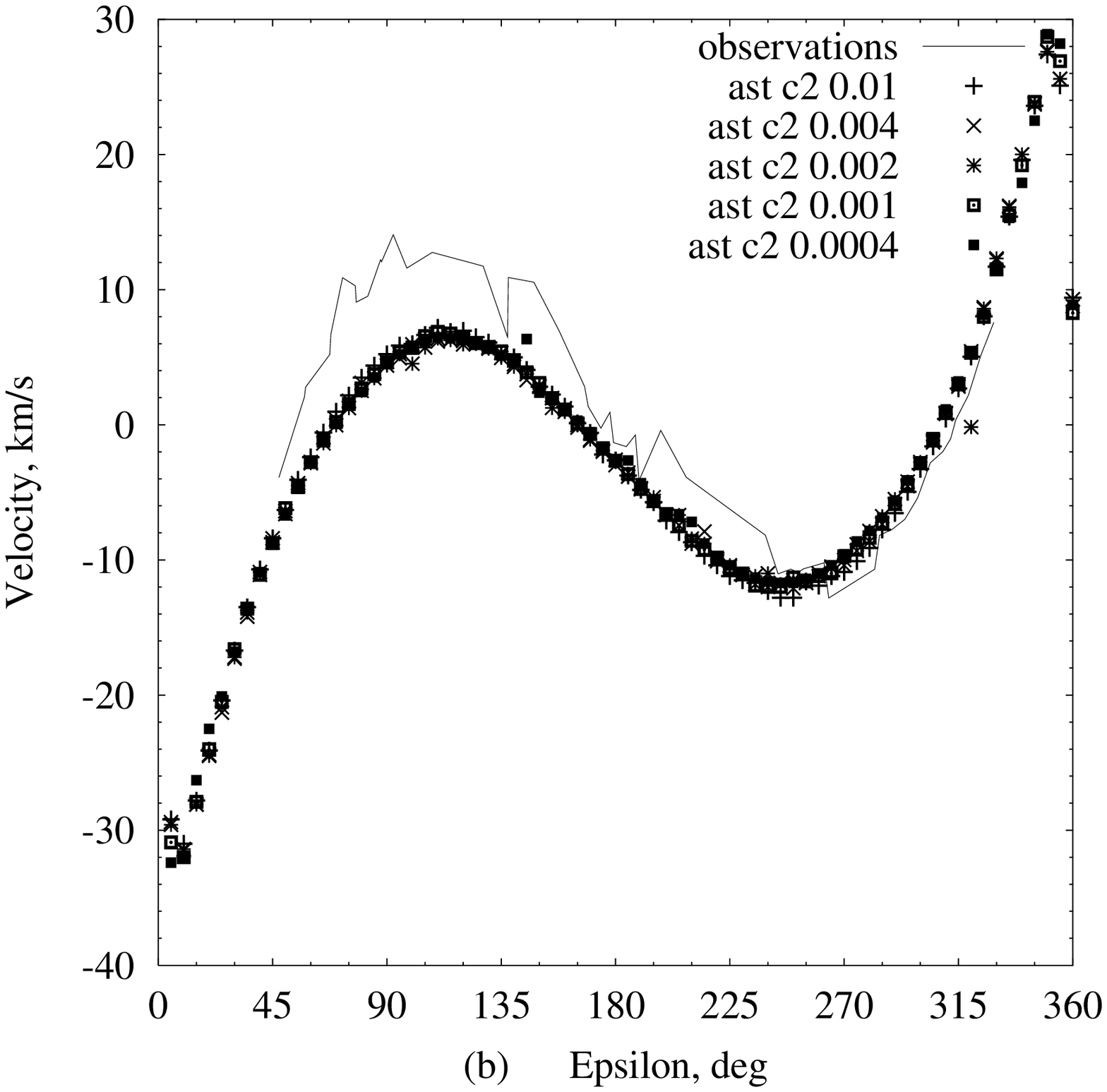} 
\includegraphics[width=81mm]{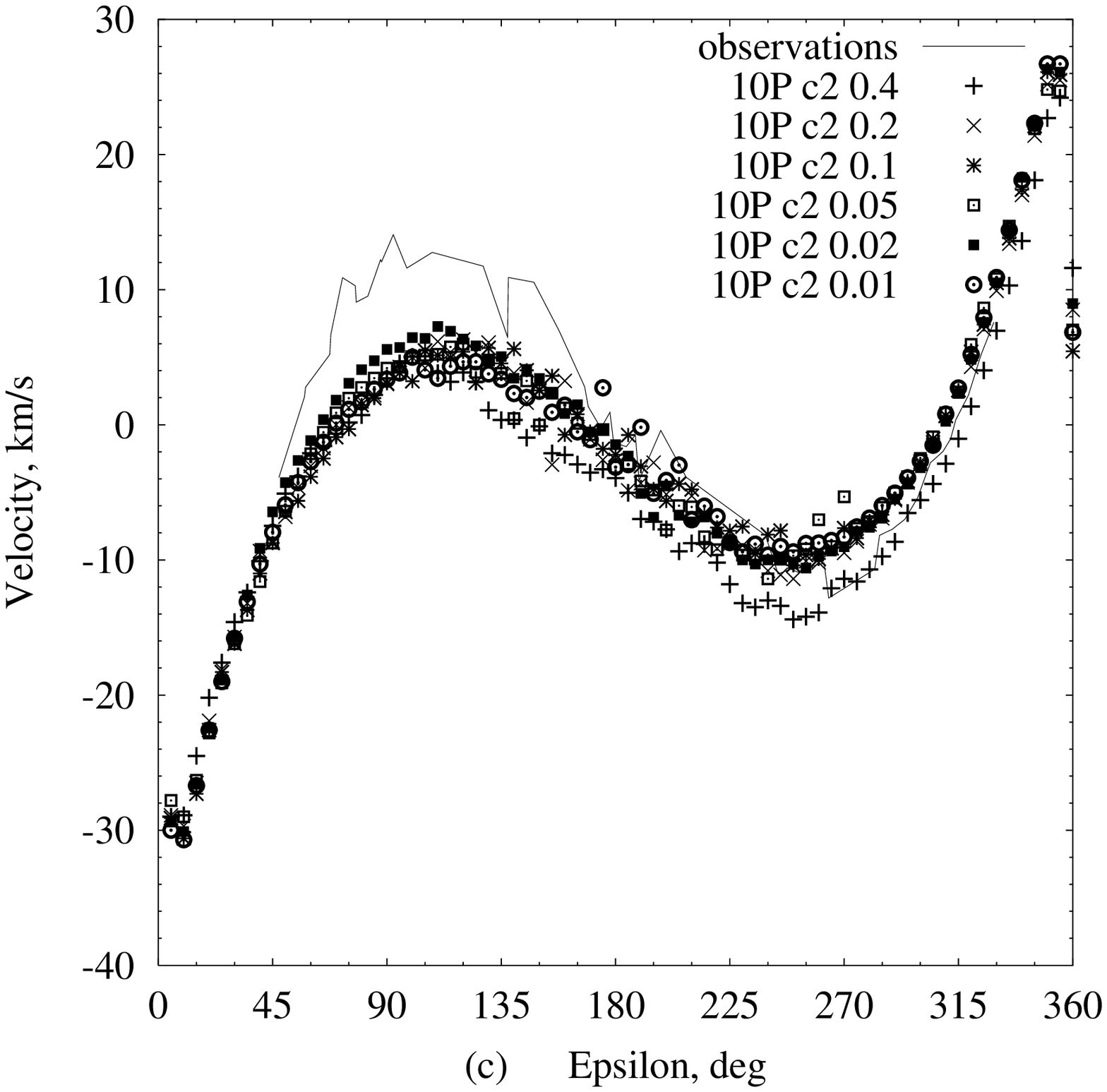} 
\includegraphics[width=81mm]{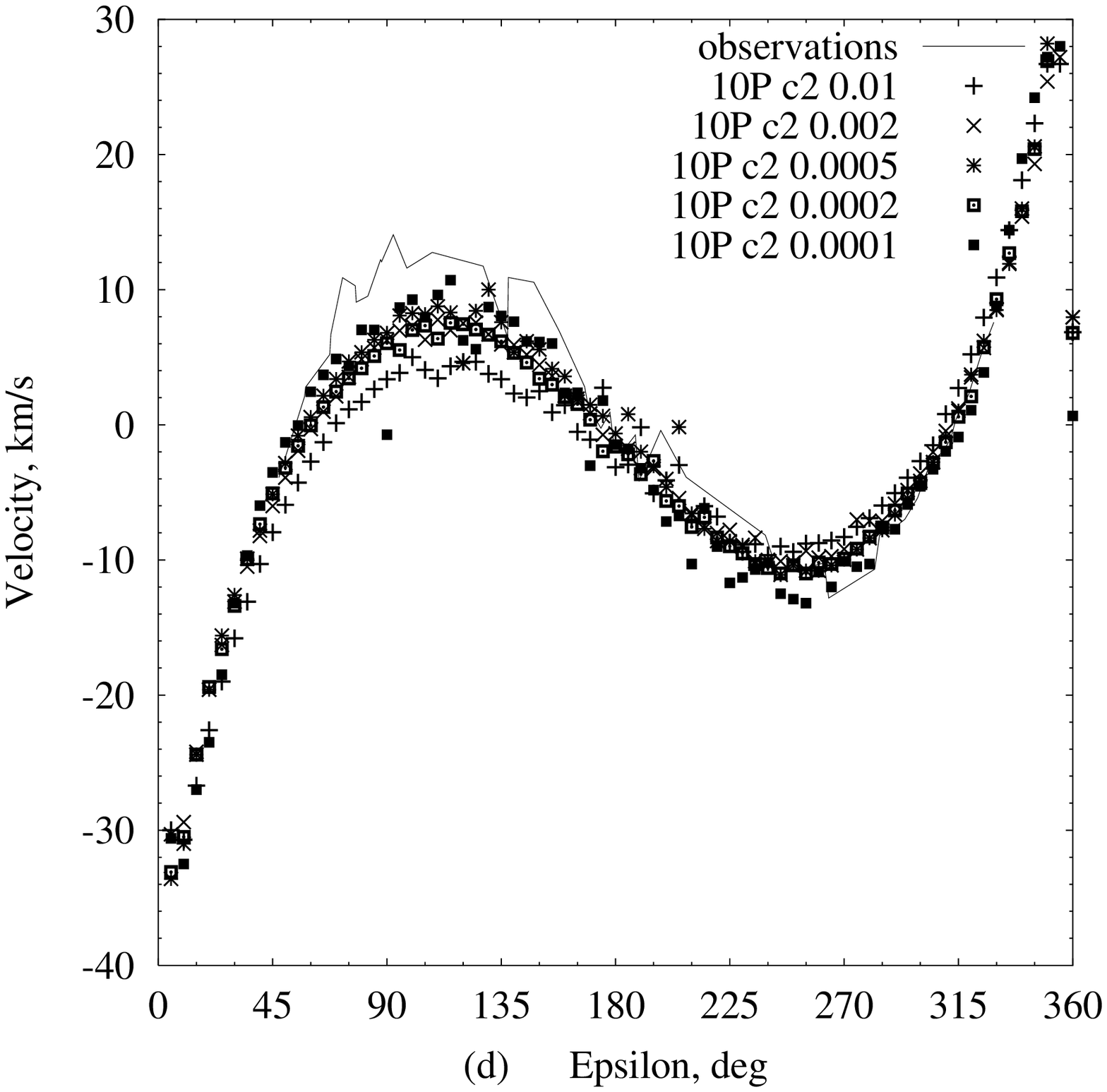} 

\caption{Velocities of Mg I line (at zero inclination) versus elongation 
$\epsilon$ (measured eastward from the sun) at several values of $\beta$
(see the last number in the legend)
for particles started from asteroids (a-b) and Comet 10P (c-d). 
The line corresponds to the observations made by Reynolds et al. (2004).
}

\end{figure}%

\begin{figure}

\includegraphics[width=81mm]{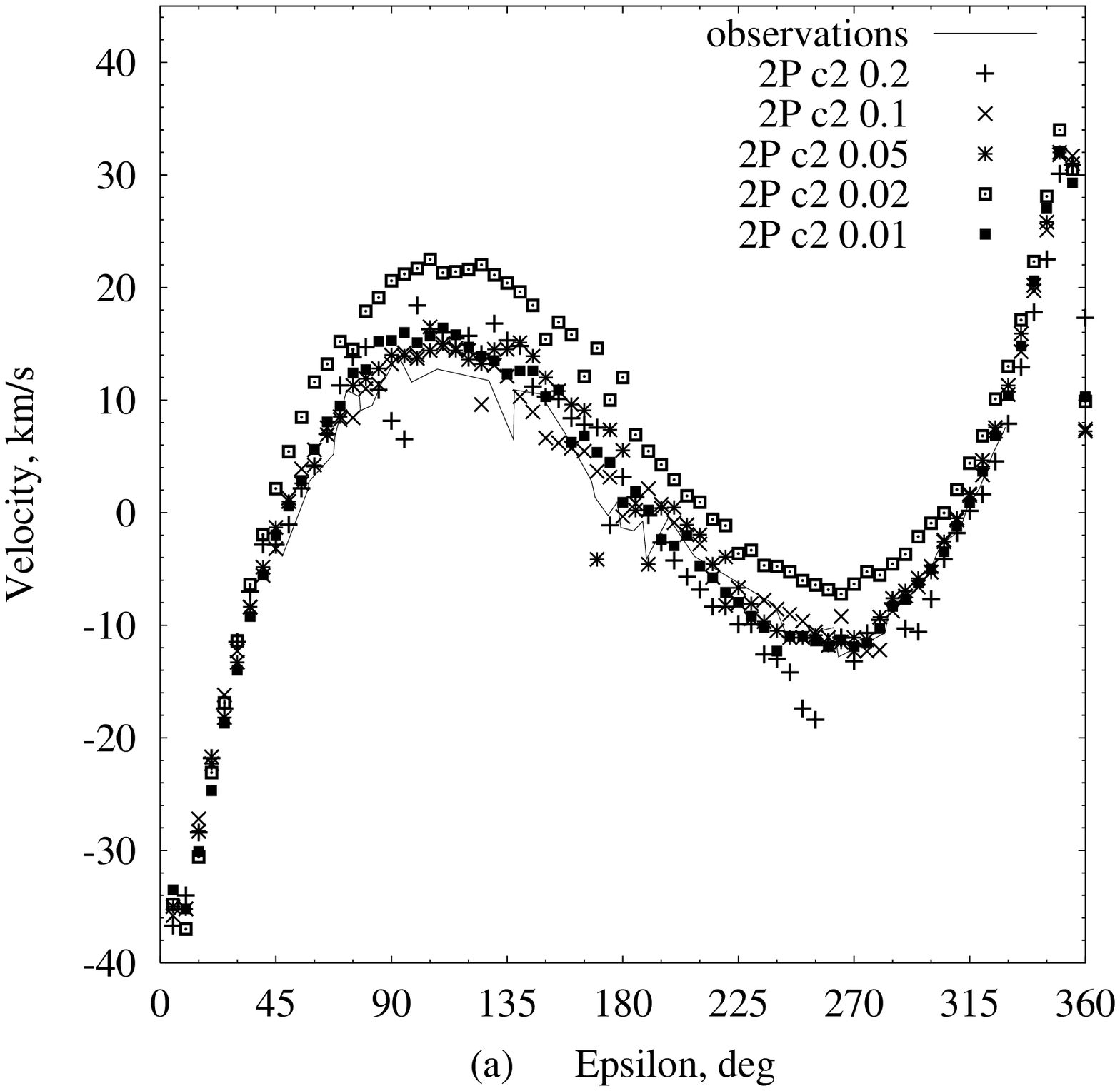} 
\includegraphics[width=81mm]{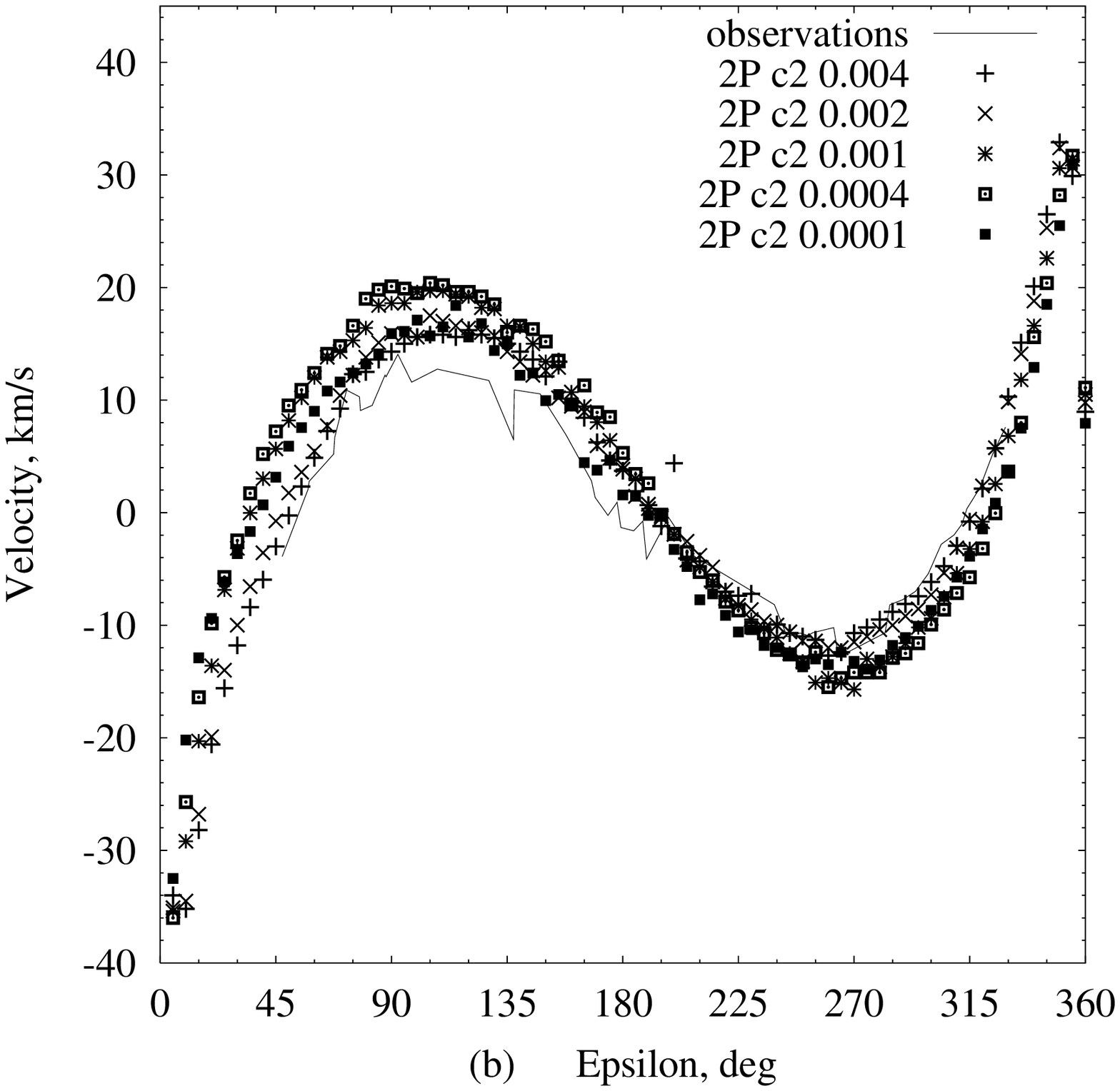} 
\includegraphics[width=81mm]{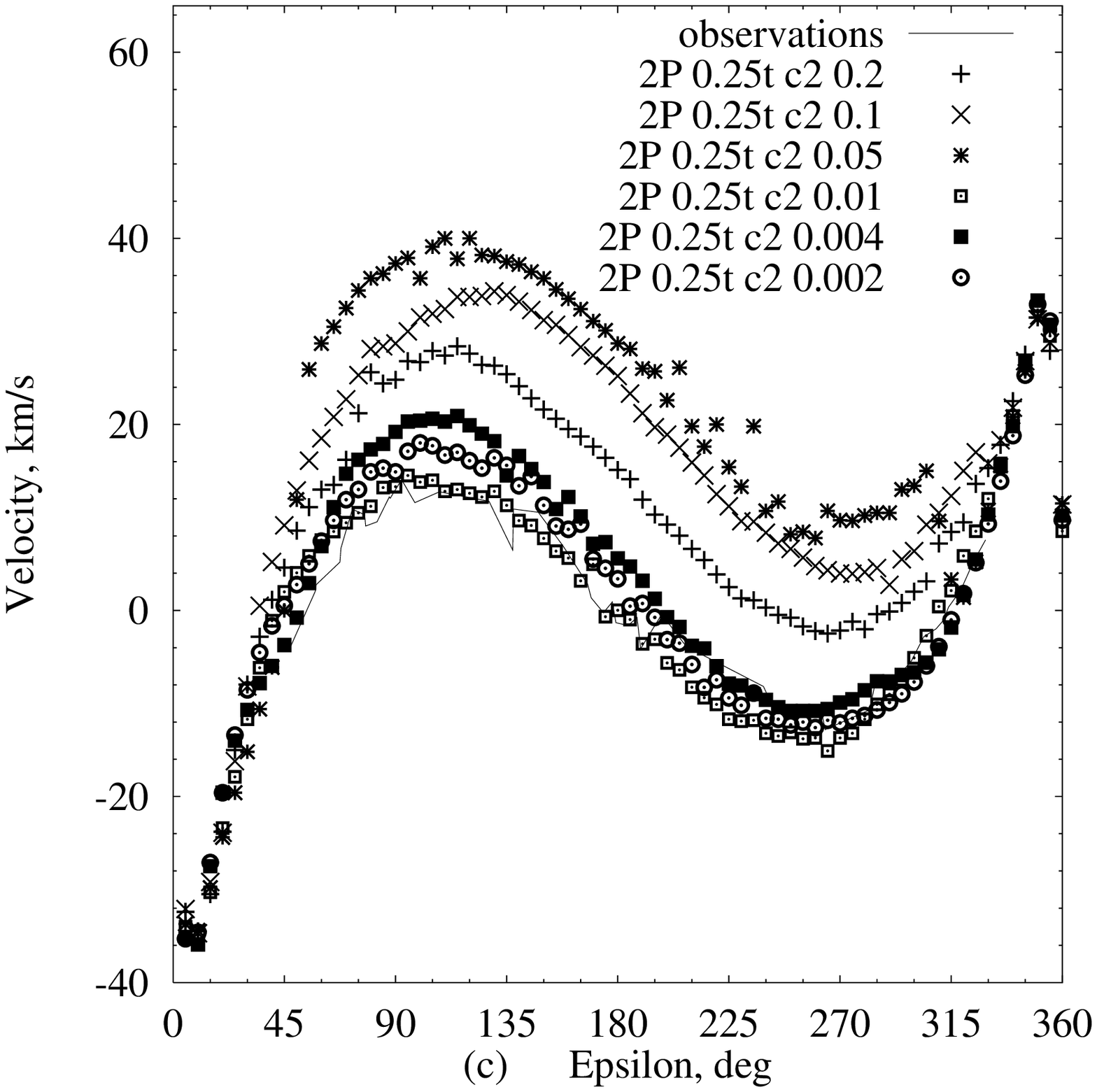} 
\includegraphics[width=81mm]{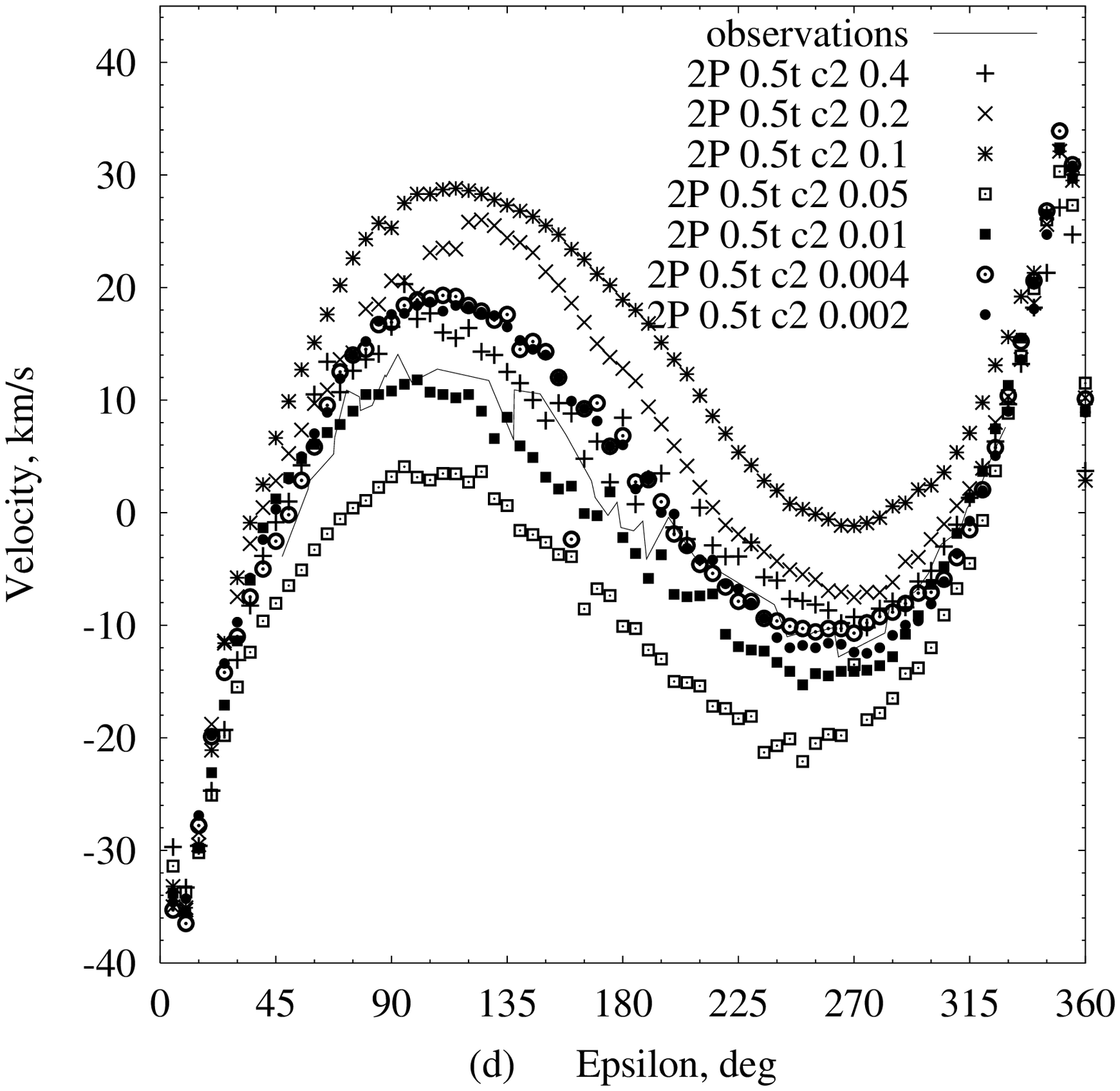} 

\caption{Velocities of Mg I line (at zero inclination) versus elongation 
$\epsilon$ (measured eastward from the sun) at several values of $\beta$
for particles started from Comet 2P at perihelion (a-b), at the middle of the orbit (c),
and at aphelion (d).
The line corresponds to the observations made by Reynolds et al. (2004).
}

\end{figure}%

\begin{figure}   

\plottwo{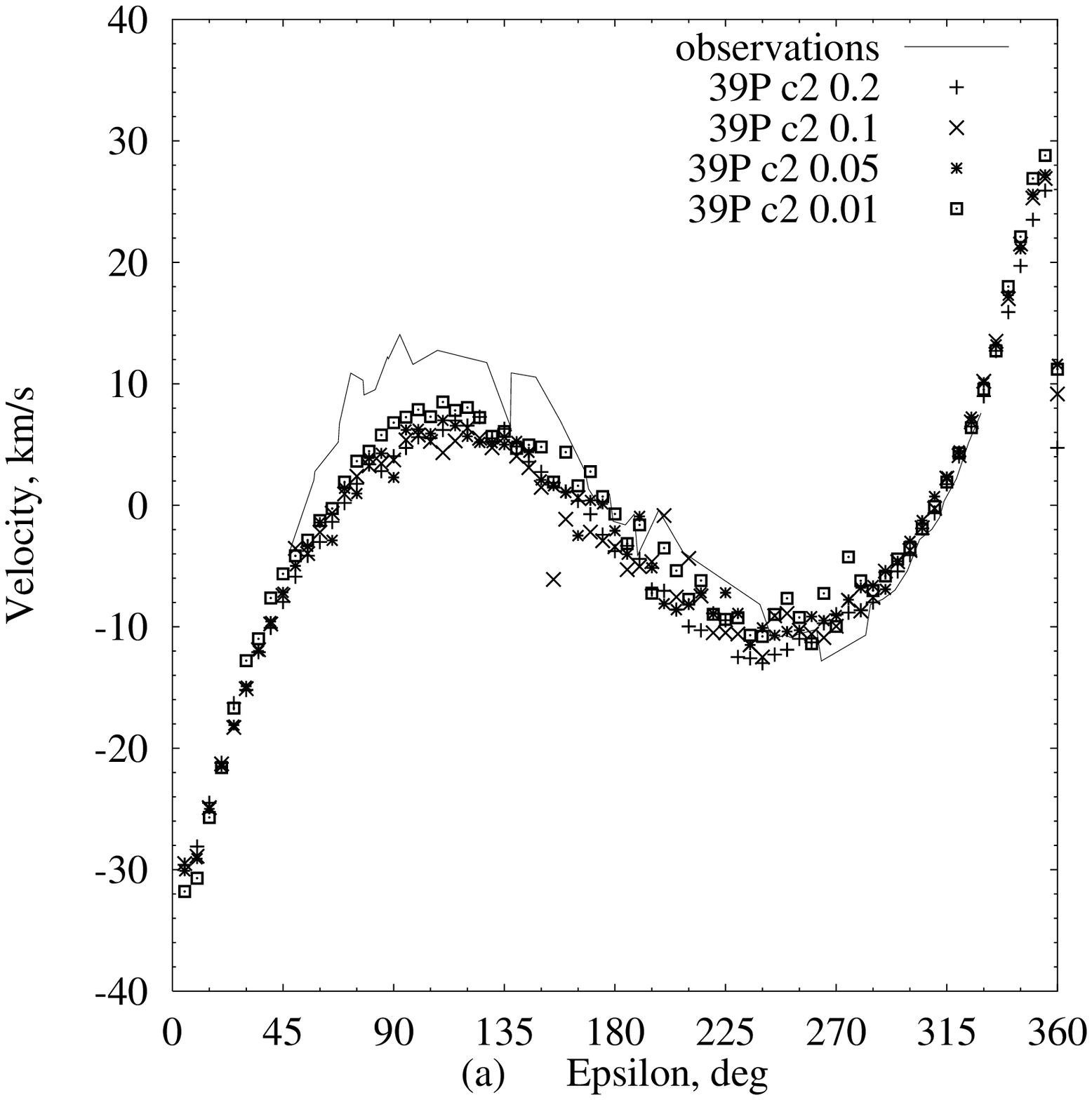}{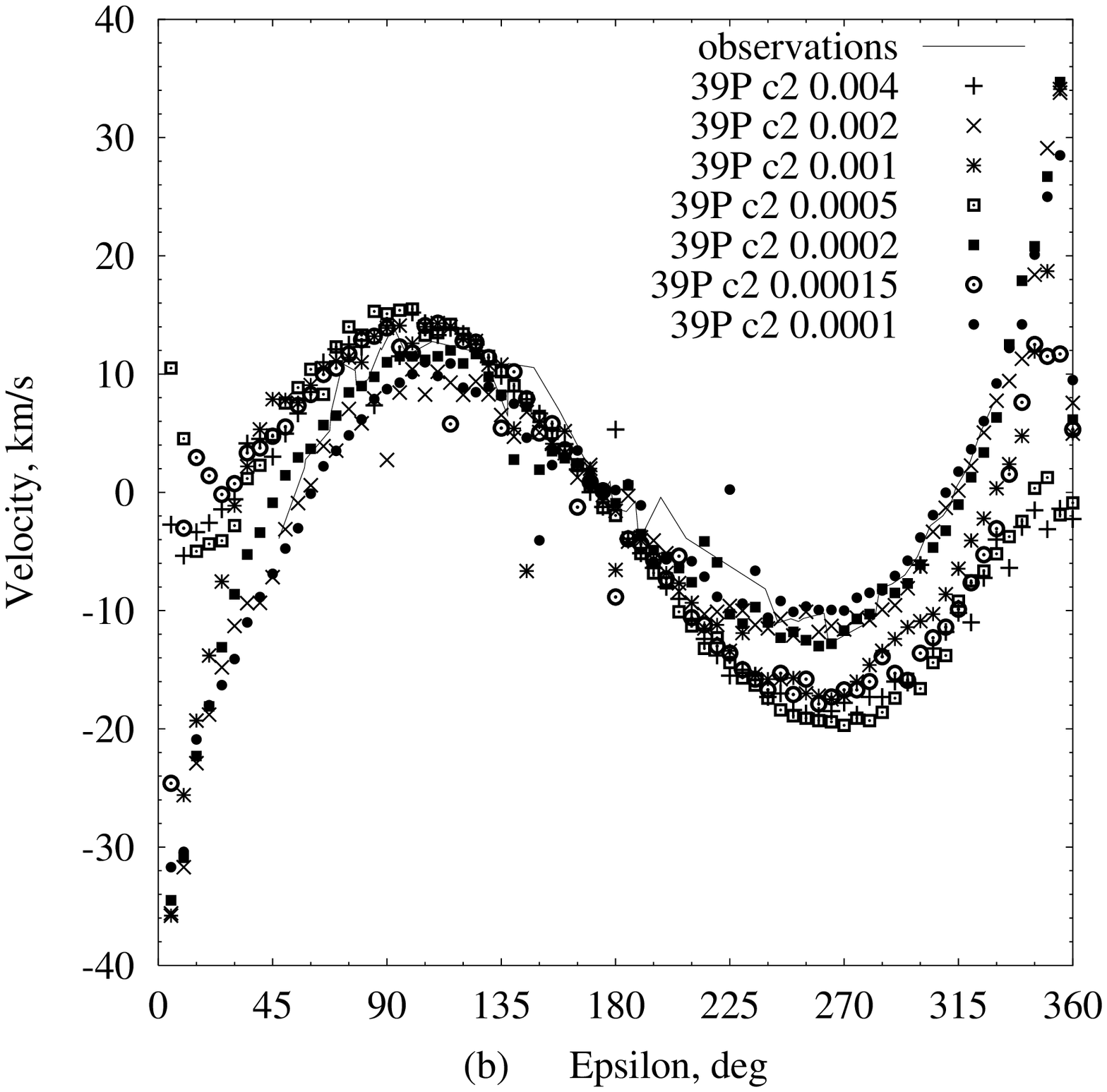} 

\caption{Same as Figs. 3-4, but 
for particles started from Comet 39P.}

\end{figure}%

\begin{figure}

\plottwo{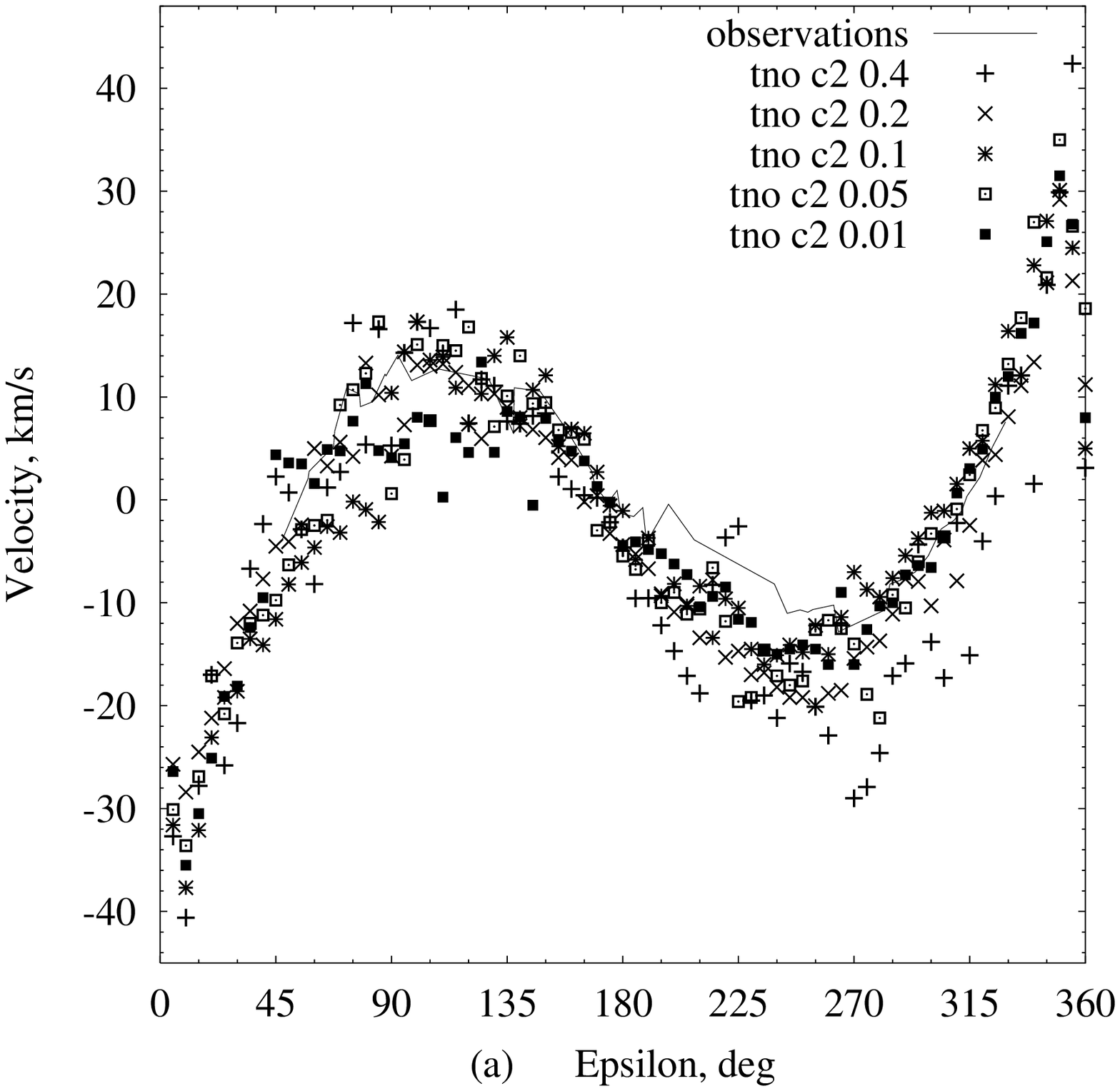}{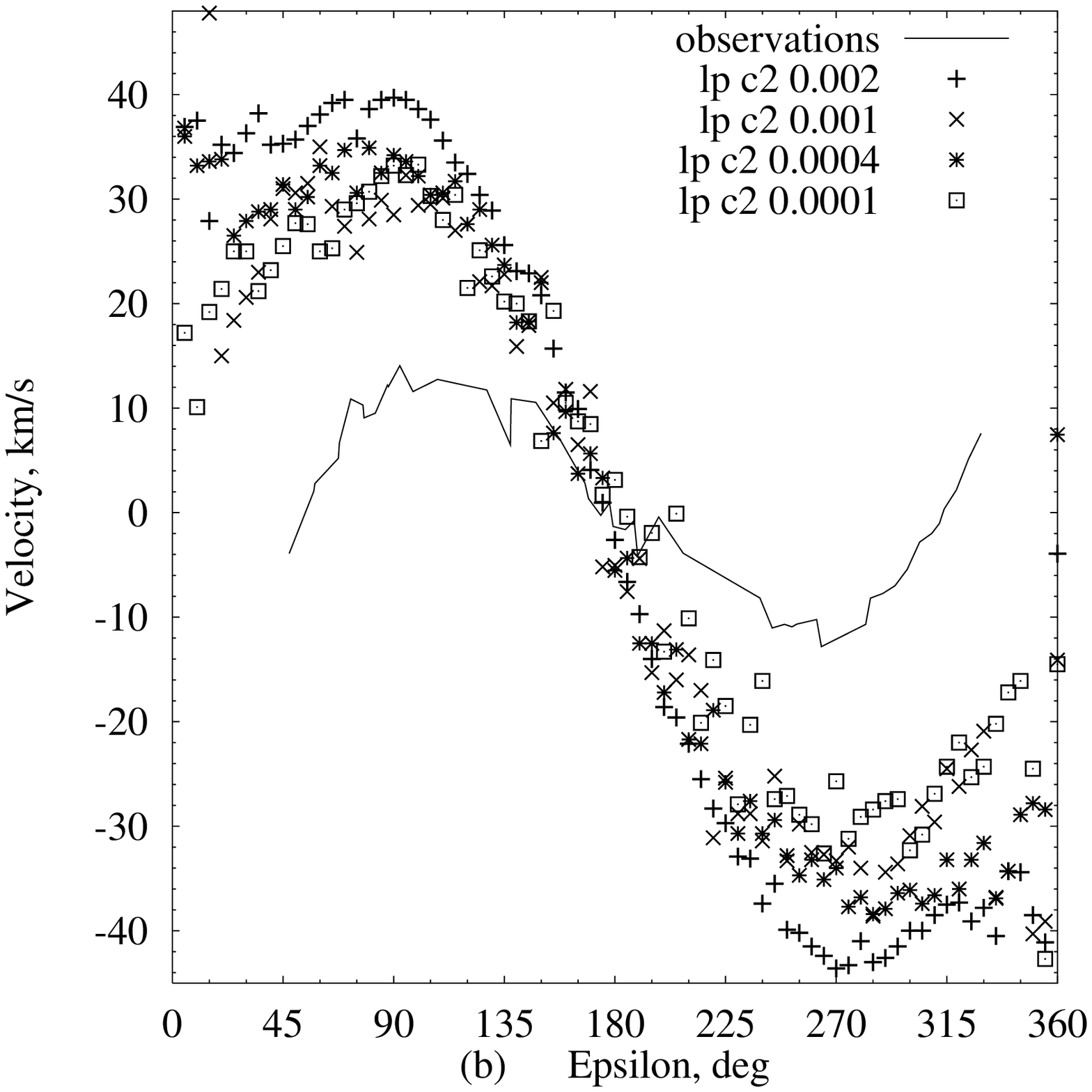} 

\caption{Same as Figs. 3-4, but 
for particles started from trans-Neptunian objects (a)
and long-period comets at $e_o$=0.995, $q_o$=0.9 AU, and $i_o$ distributed 
between 0 and 180$^\circ$ (b).}

\end{figure}%

\begin{figure}   

\includegraphics[width=81mm]{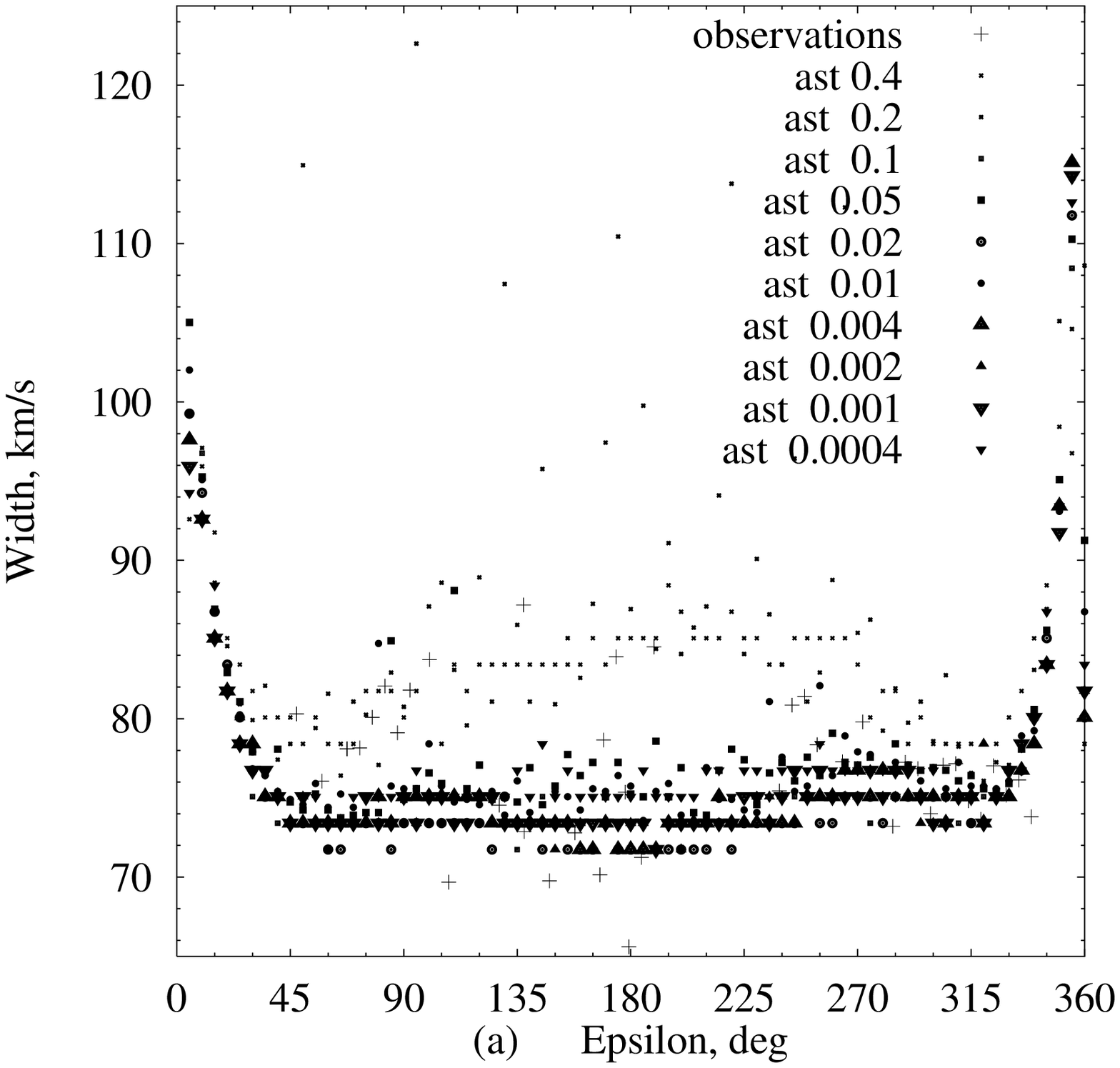} 
\includegraphics[width=81mm]{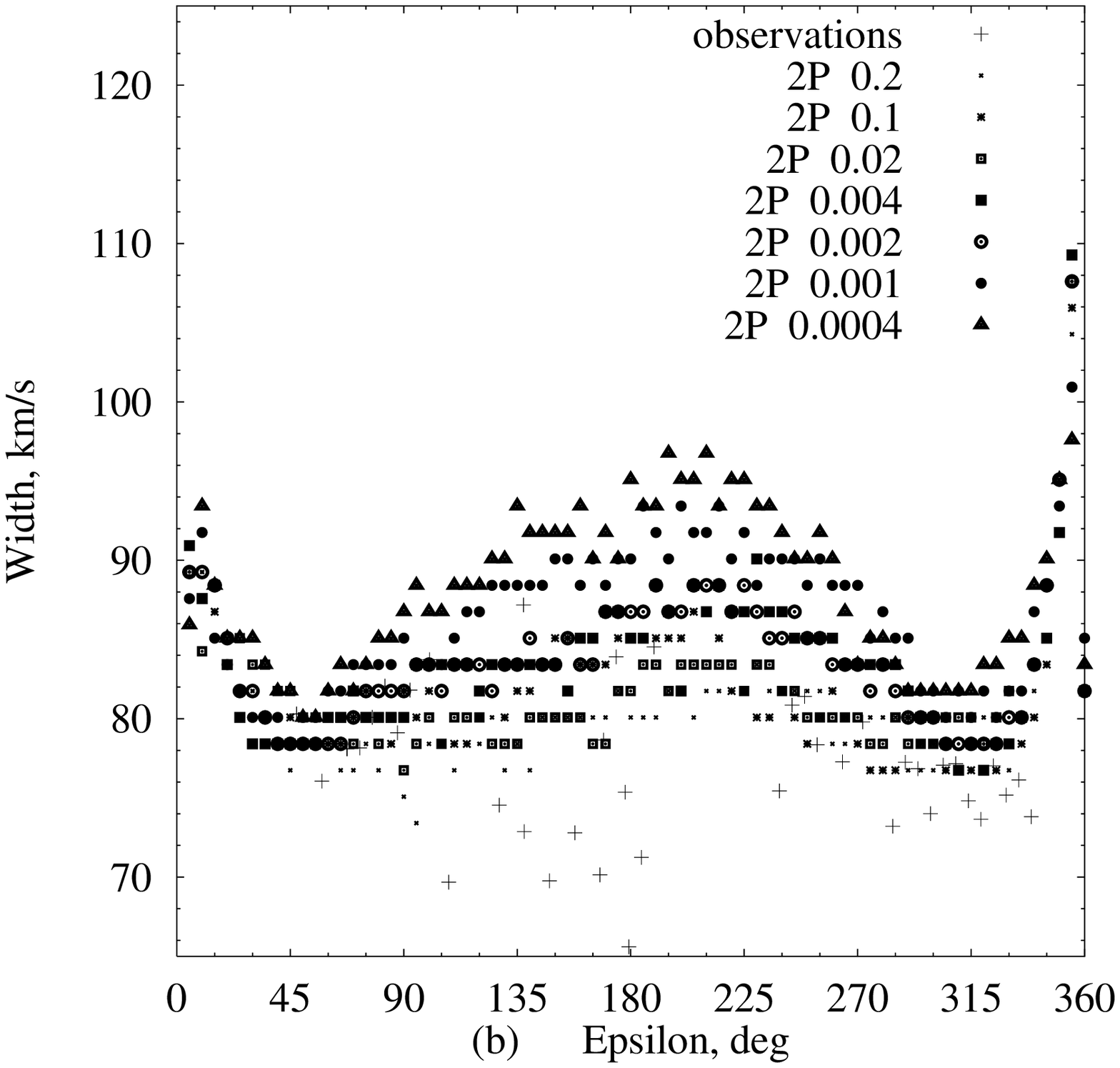} 
\includegraphics[width=81mm]{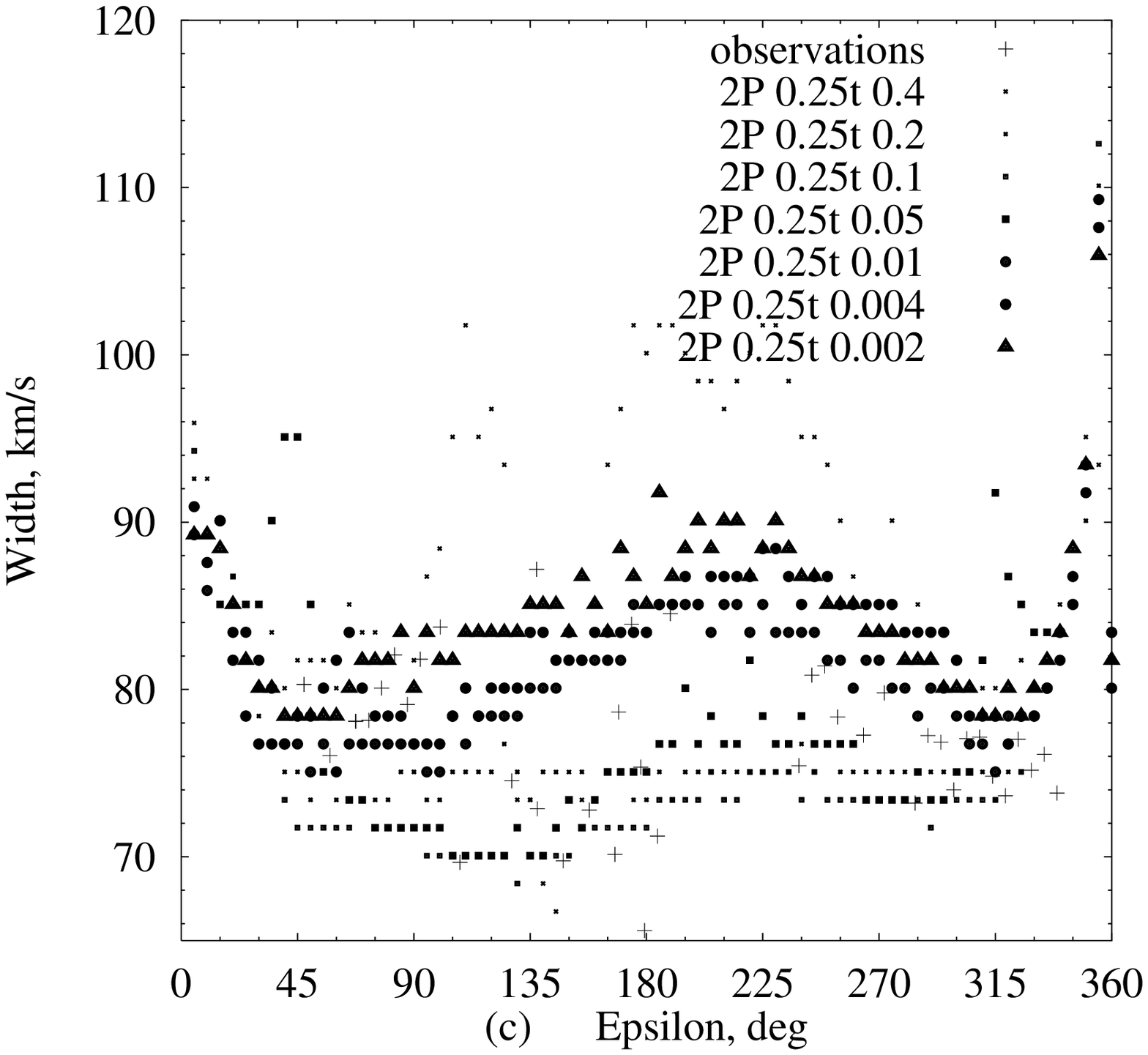} 
\includegraphics[width=81mm]{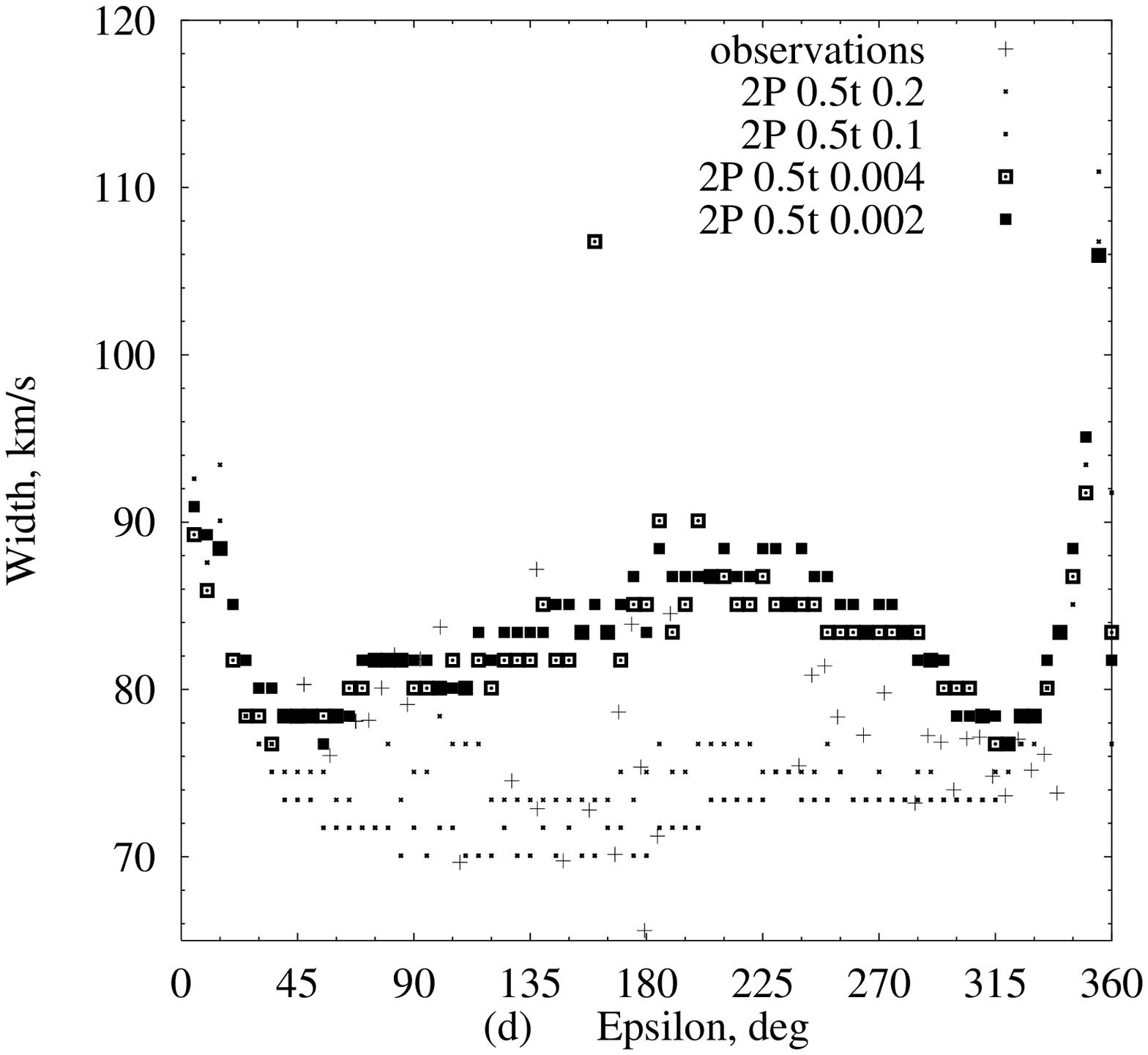} 
\includegraphics[width=81mm]{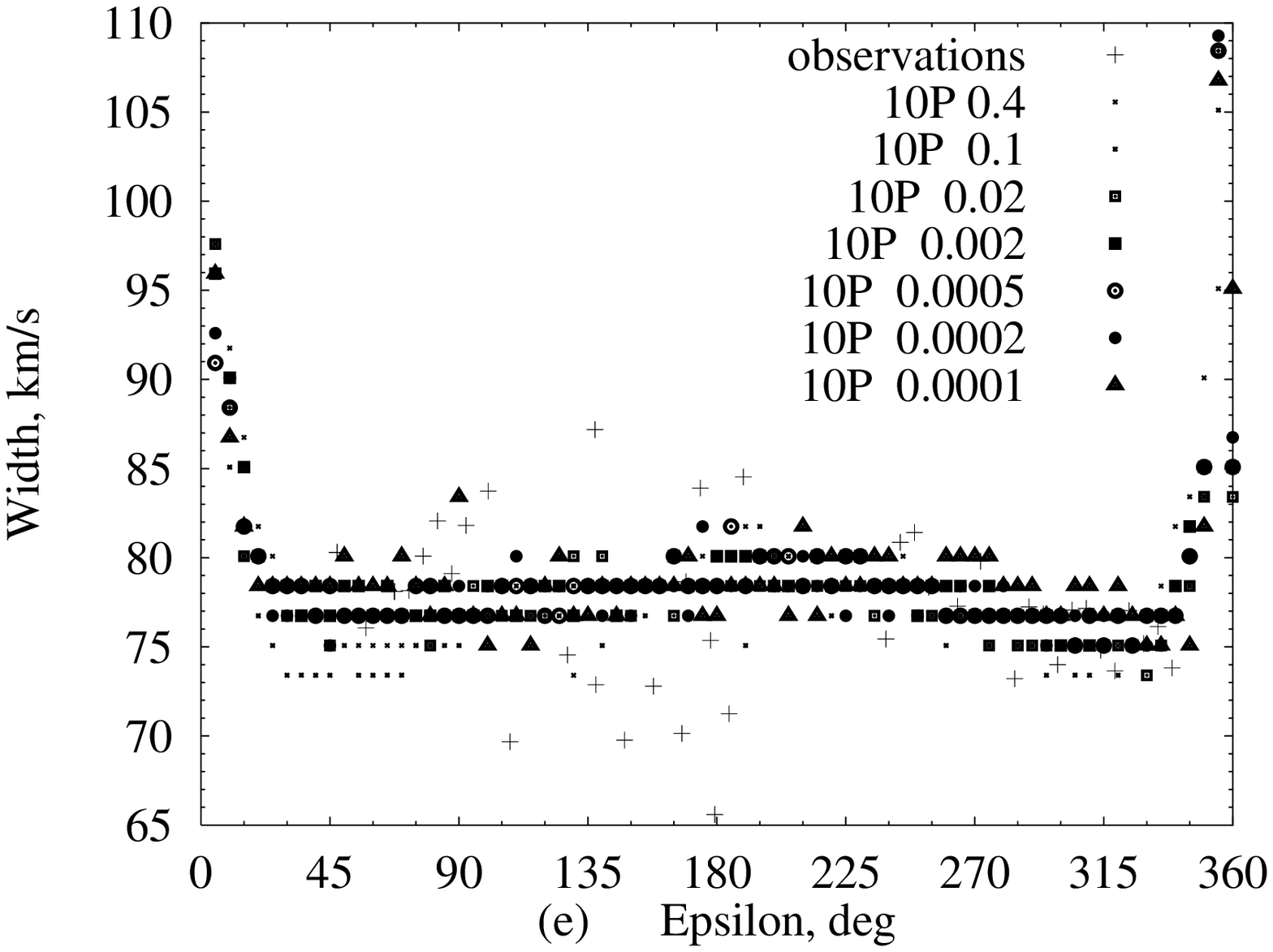} 
\includegraphics[width=81mm]{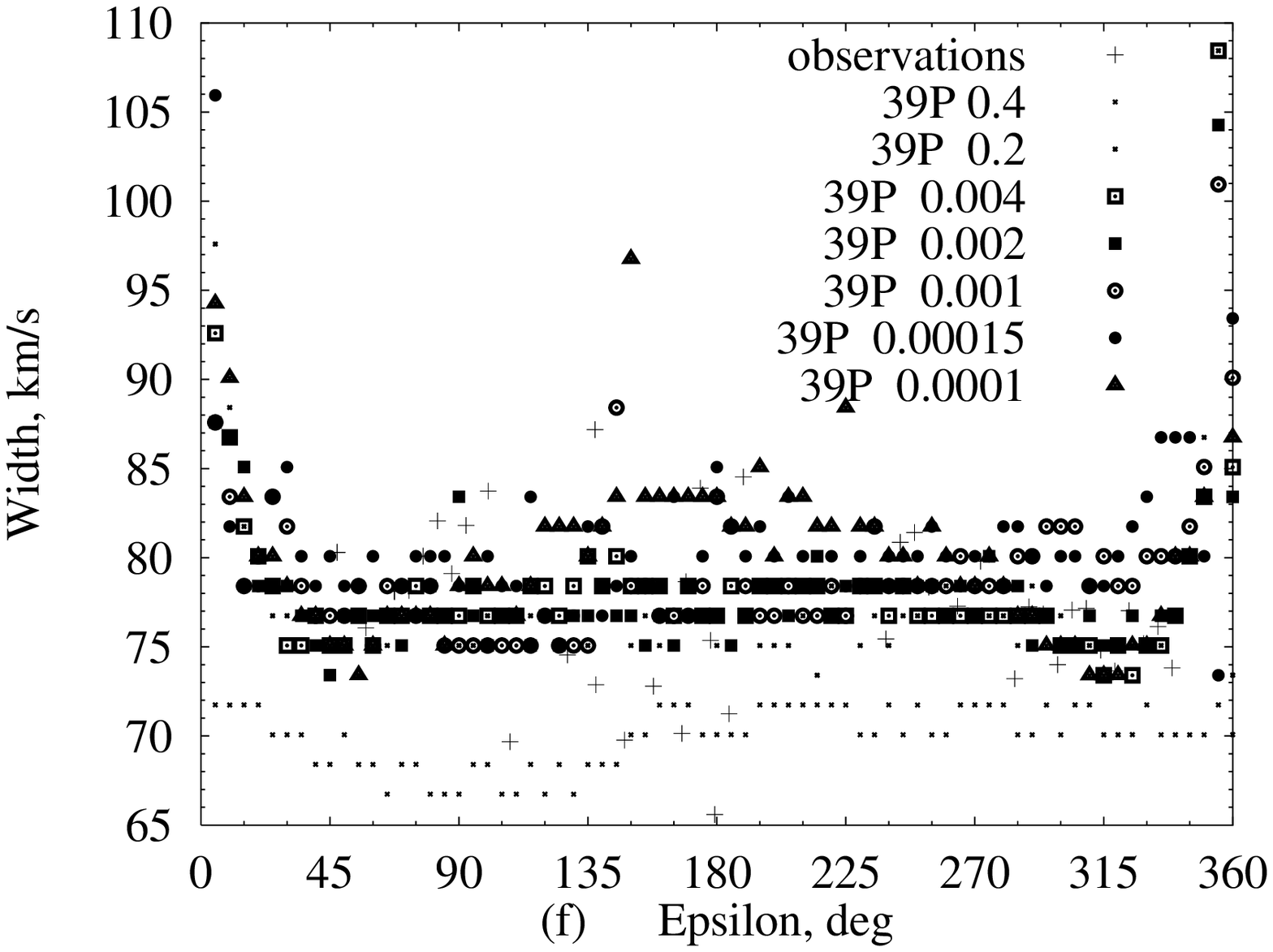} 

\caption{\small Width of Mg I line (in the ecliptic plane) versus elongation 
$\epsilon$ 
at several values of $\beta$
for particles started from asteroids (ast), from Comet 2P at perihelion (2P), 
at the middle of the orbit (2P 0.25t), and at aphelion (2P 0.5t), from Comet 10P (10P), 
and Comet 39P (39P).
The first designation (+) corresponds to the observations 
made by Reynolds et al. (2004).
}

\end{figure}%

\begin{figure}    

\plotone{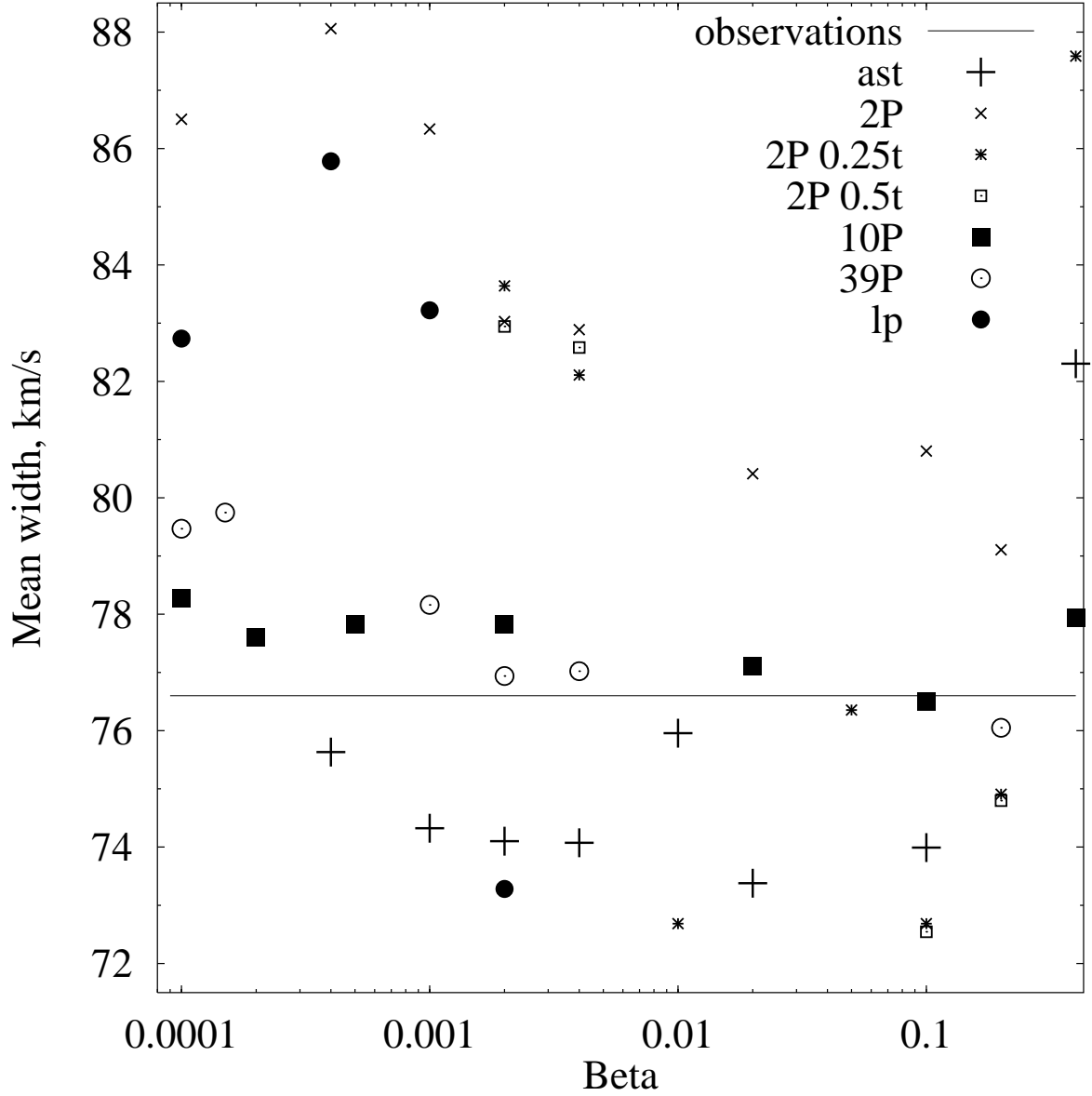}

\caption{Mean value of the width of Mg I line (in the ecliptic plane) for 
elongation 
$\epsilon$ between 30$^\circ$ and 330$^\circ$ at several values of $\beta$
for particles started from asteroids (ast), from Comet 2P at perihelion (2P), 
at the middle of the orbit (2P 0.25t), and at aphelion (2P 0.5t), 
from Comets 10P and 39P (10P and 39P),
and from long-period comets (lp) at $e_o$=0.995, 
$q_o$=0.9 AU, and $i_o$ distributed between 0 and 180$^\circ$.
Solid line corresponds to the observational value.
}

\end{figure}%

\begin{figure}

\includegraphics[width=81mm]{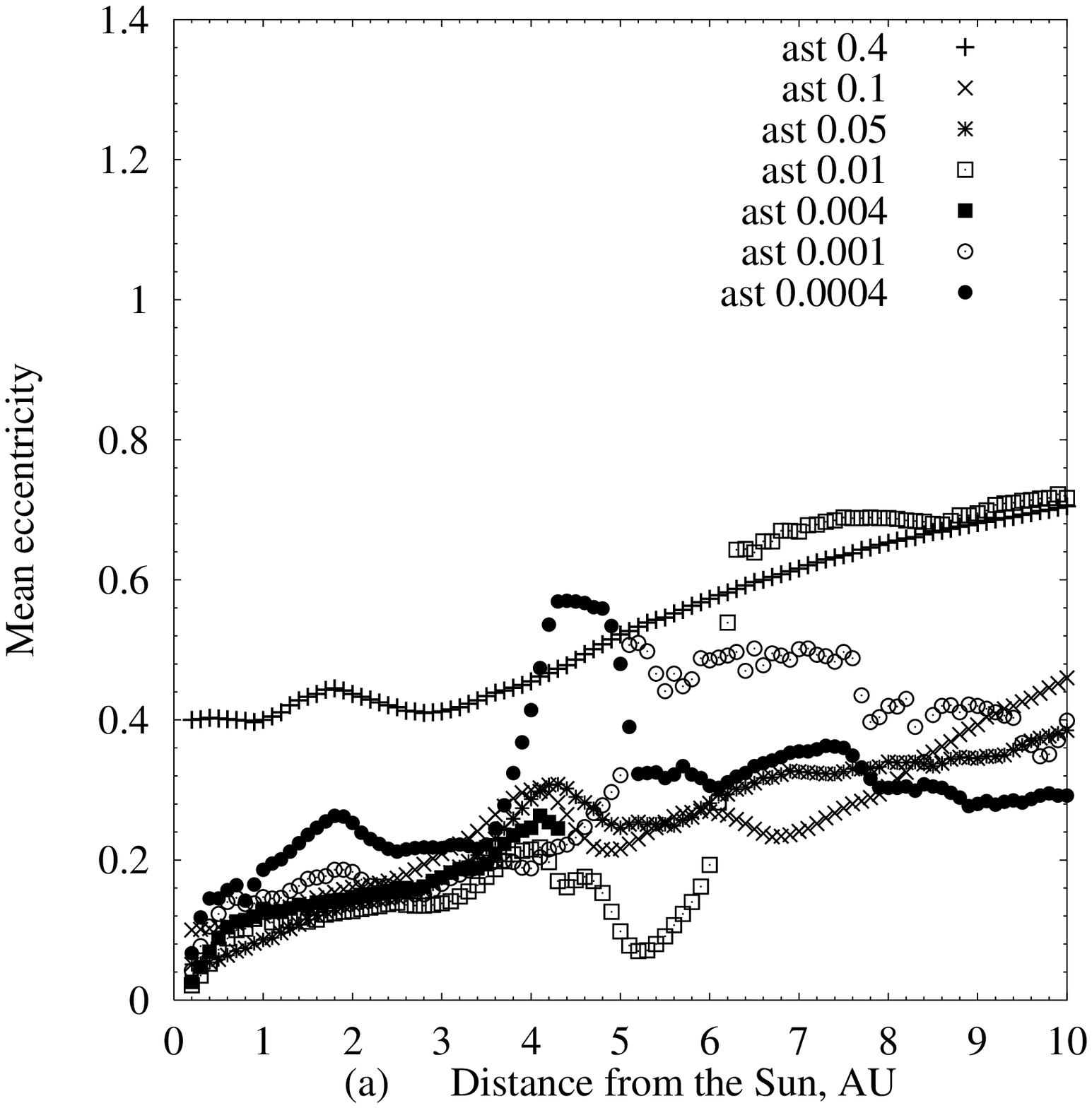} 
\includegraphics[width=81mm]{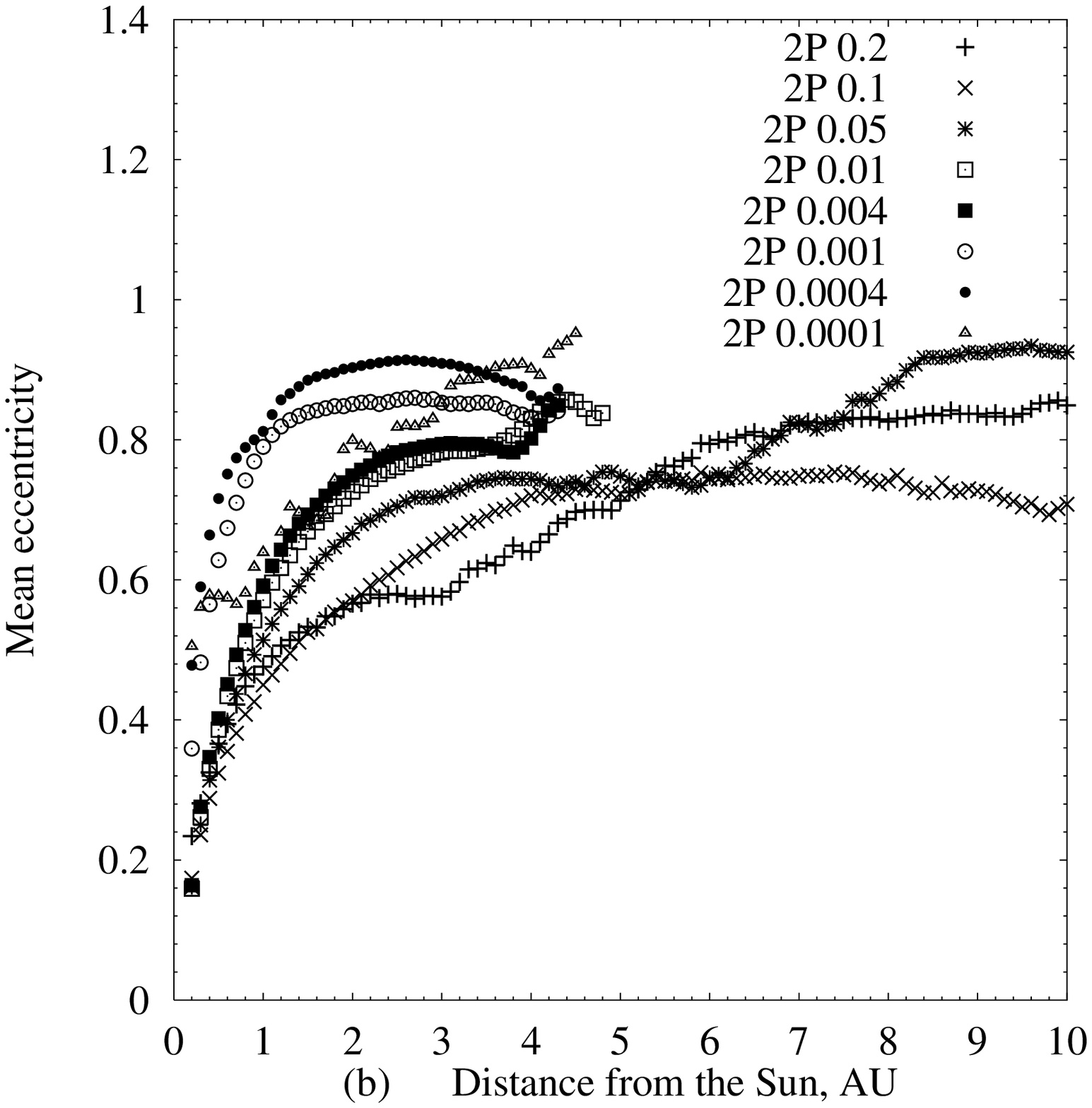} 
\includegraphics[width=81mm]{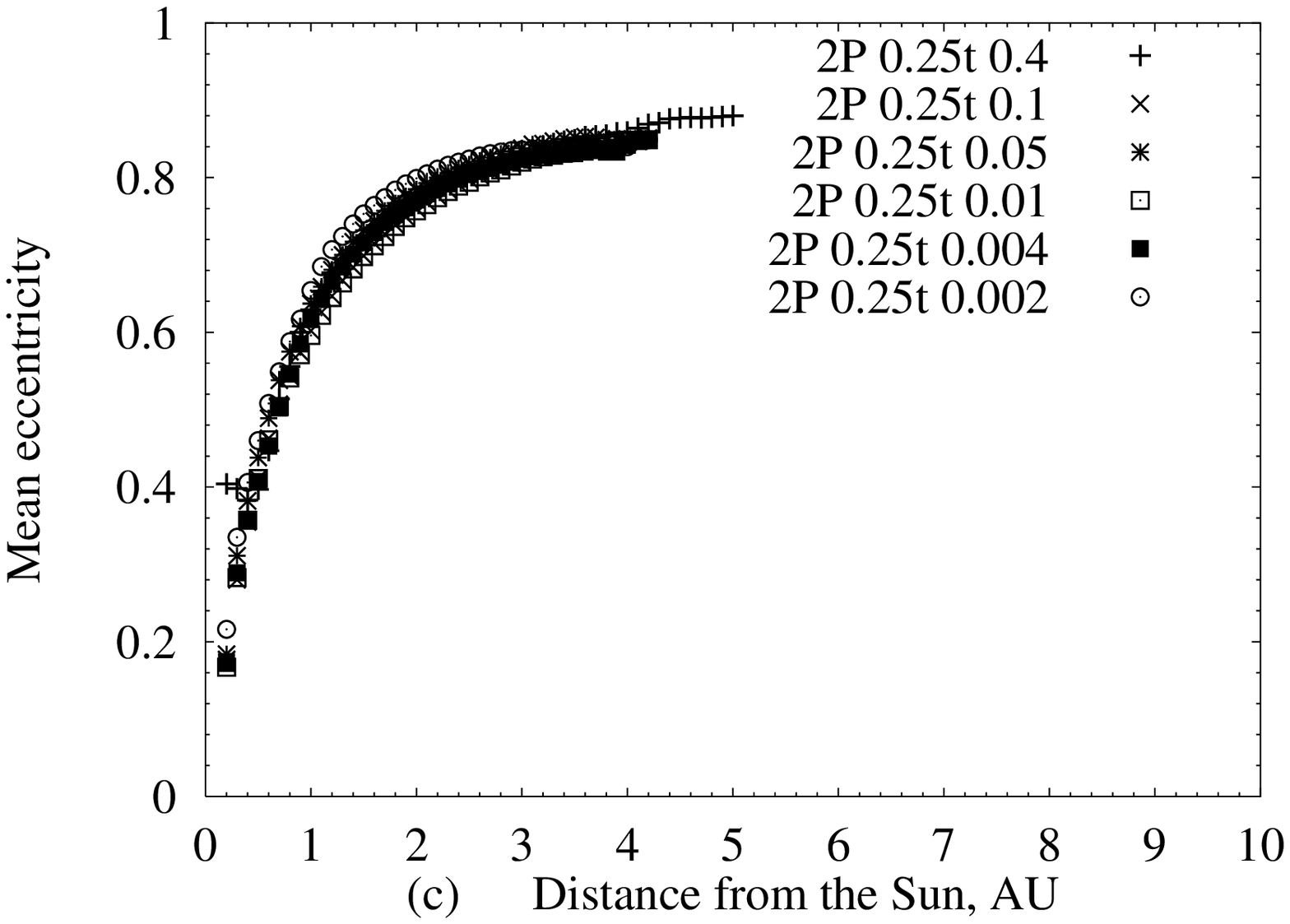} 
\includegraphics[width=81mm]{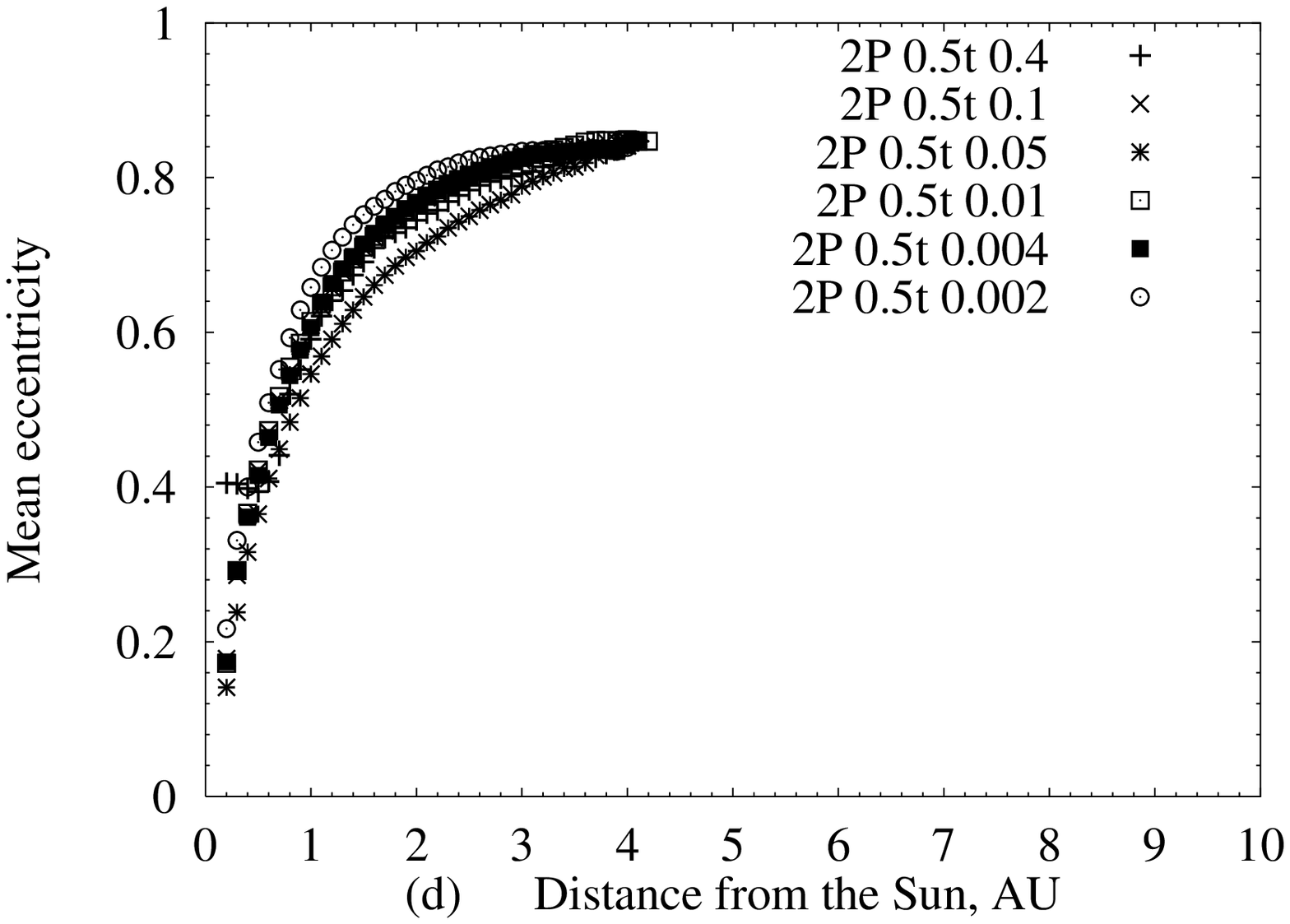} 

\caption{Mean eccentricity of particles at different distances from the sun
at several values of $\beta$ (see the last number in the legend)
for particles started from asteroids (ast), from Comet 2P at perihelion (2P), 
at the middle of the orbit (2P 0.25t), and at aphelion (2P 0.5t).
}
\end{figure}%

\begin{figure}

\includegraphics[width=81mm]{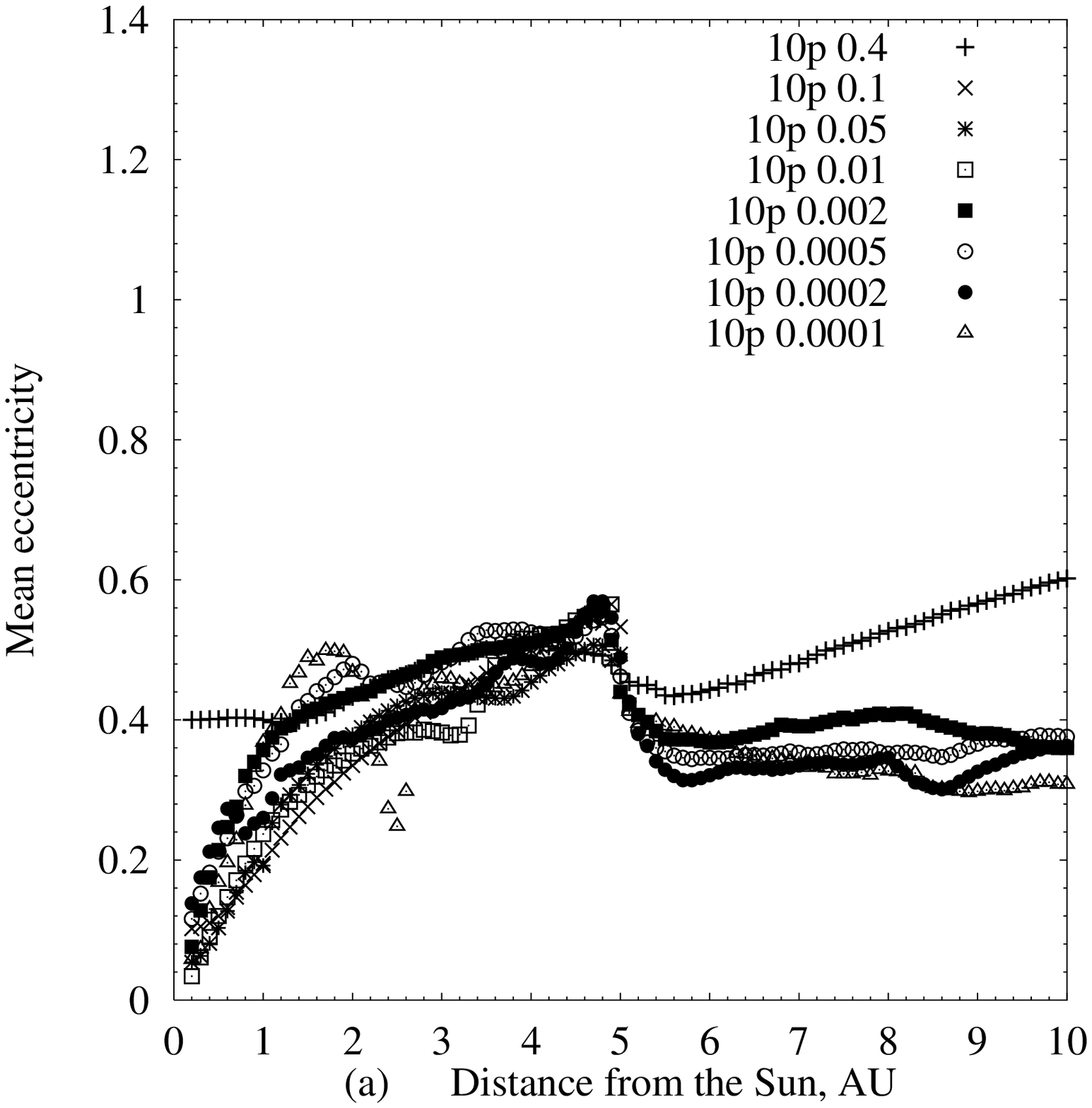} 
\includegraphics[width=81mm]{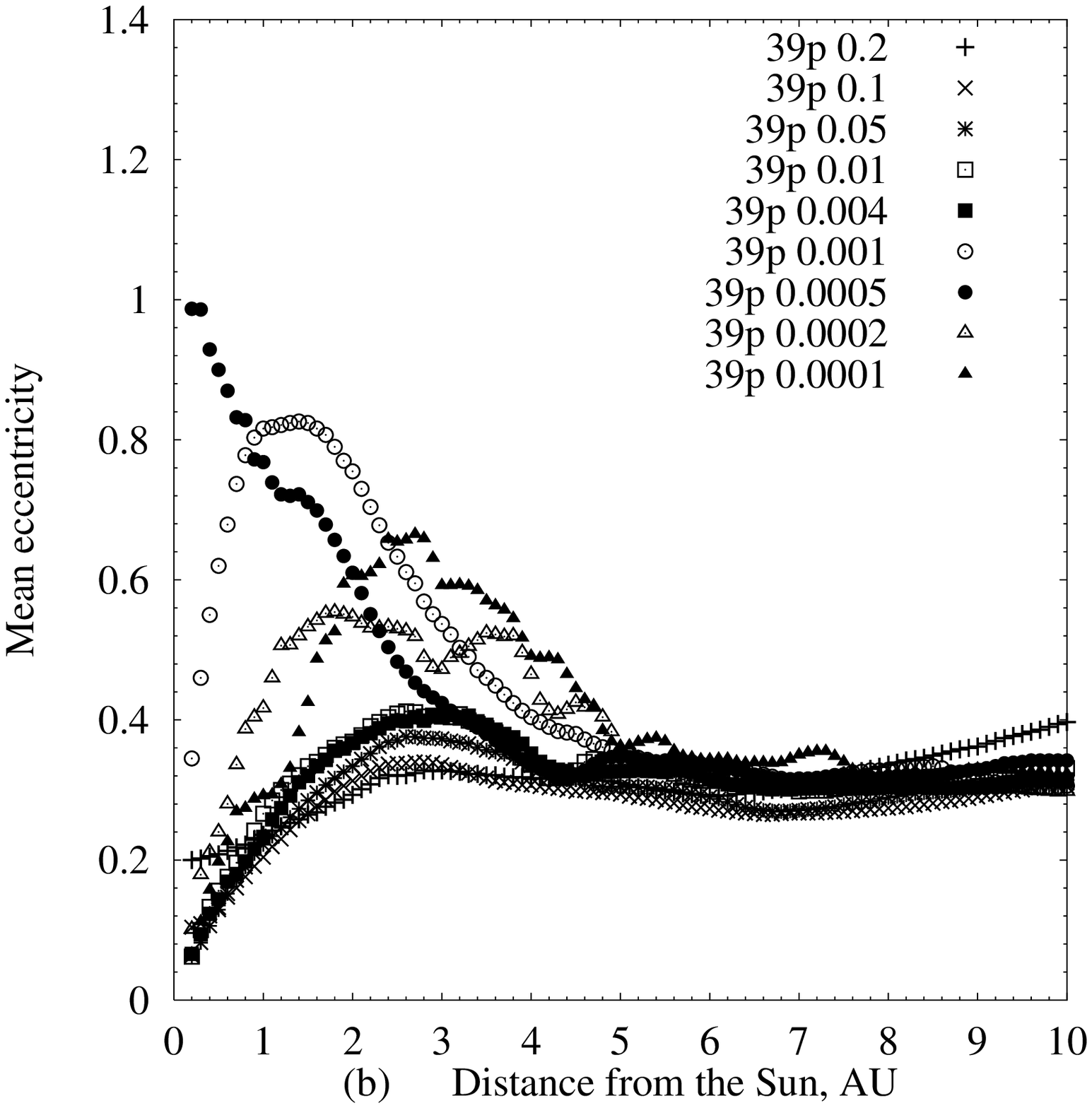} 
\includegraphics[width=81mm]{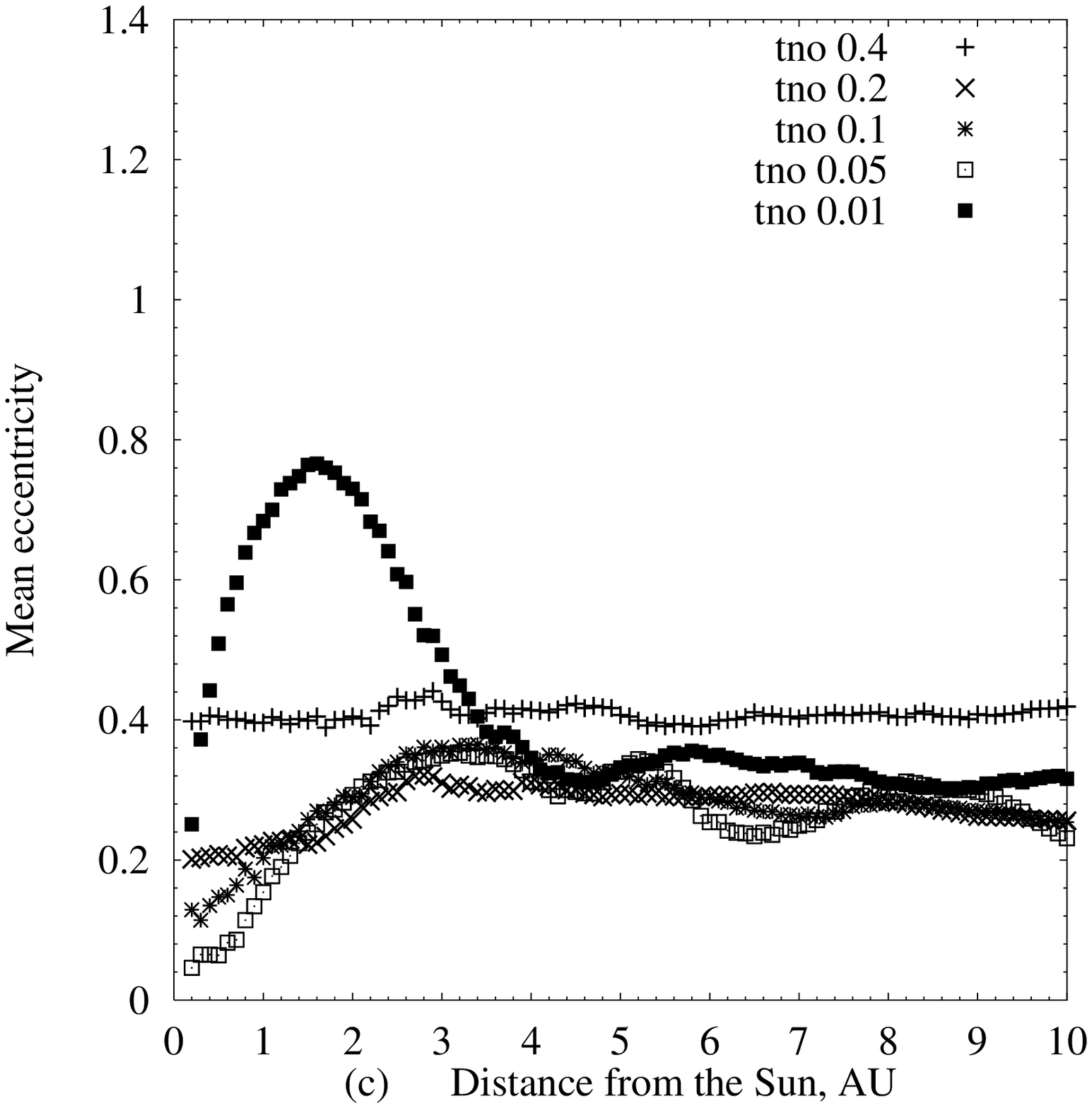} 
\includegraphics[width=81mm]{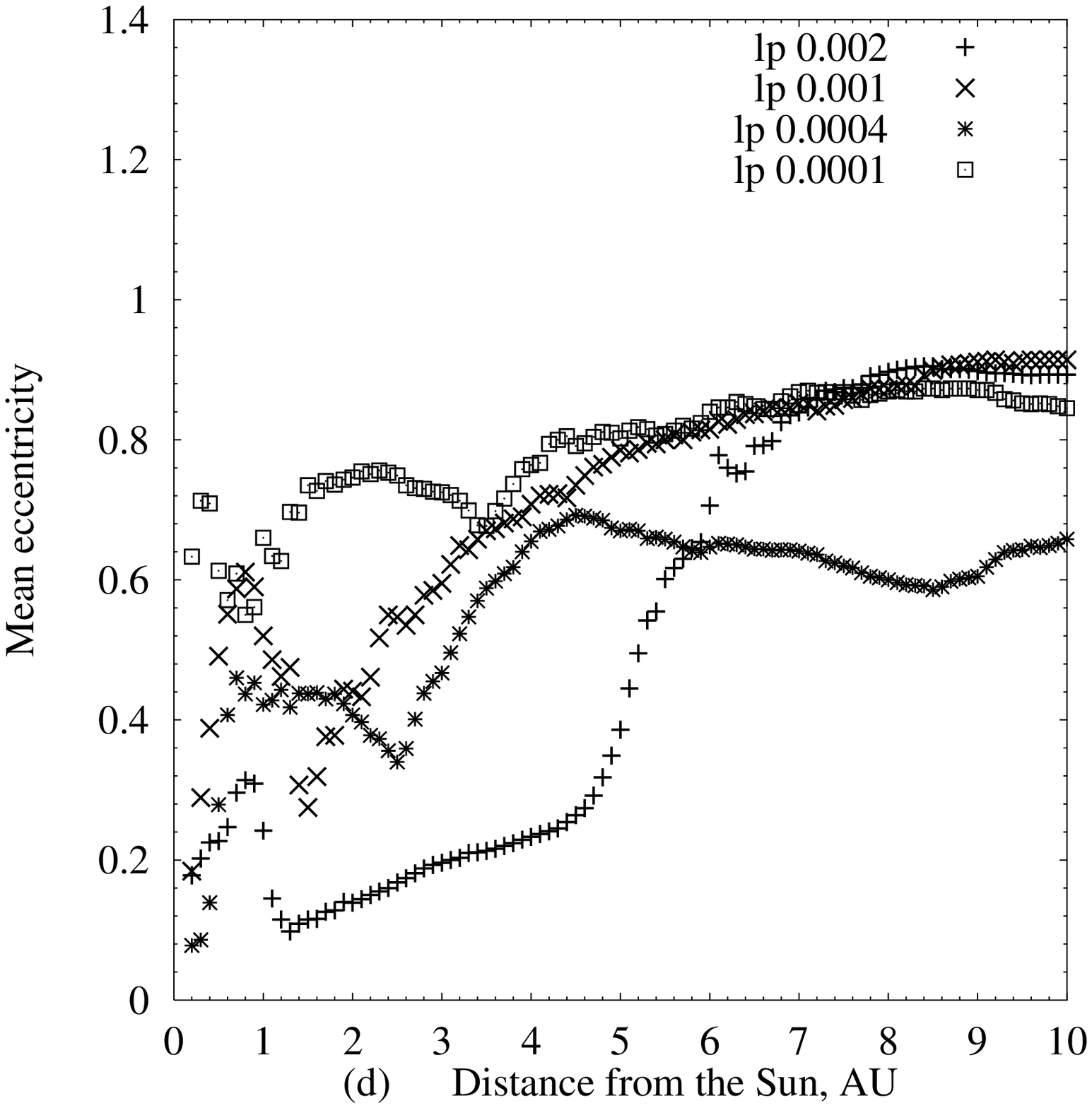} 

\caption{Mean eccentricity of particles at different distances from the sun
at several values of $\beta$ (see the last number in the legend)
for particles started from Comets 10P and 39P (10P and 39P),
from trans-Neptunian objects (tno), and from long-period comets (lp) at 
$e_o$=0.995, 
$q_o$=0.9 AU, and $i_o$ distributed between 0 and 180$^\circ$.
}
\end{figure}%

\begin{figure}

\includegraphics[width=81mm]{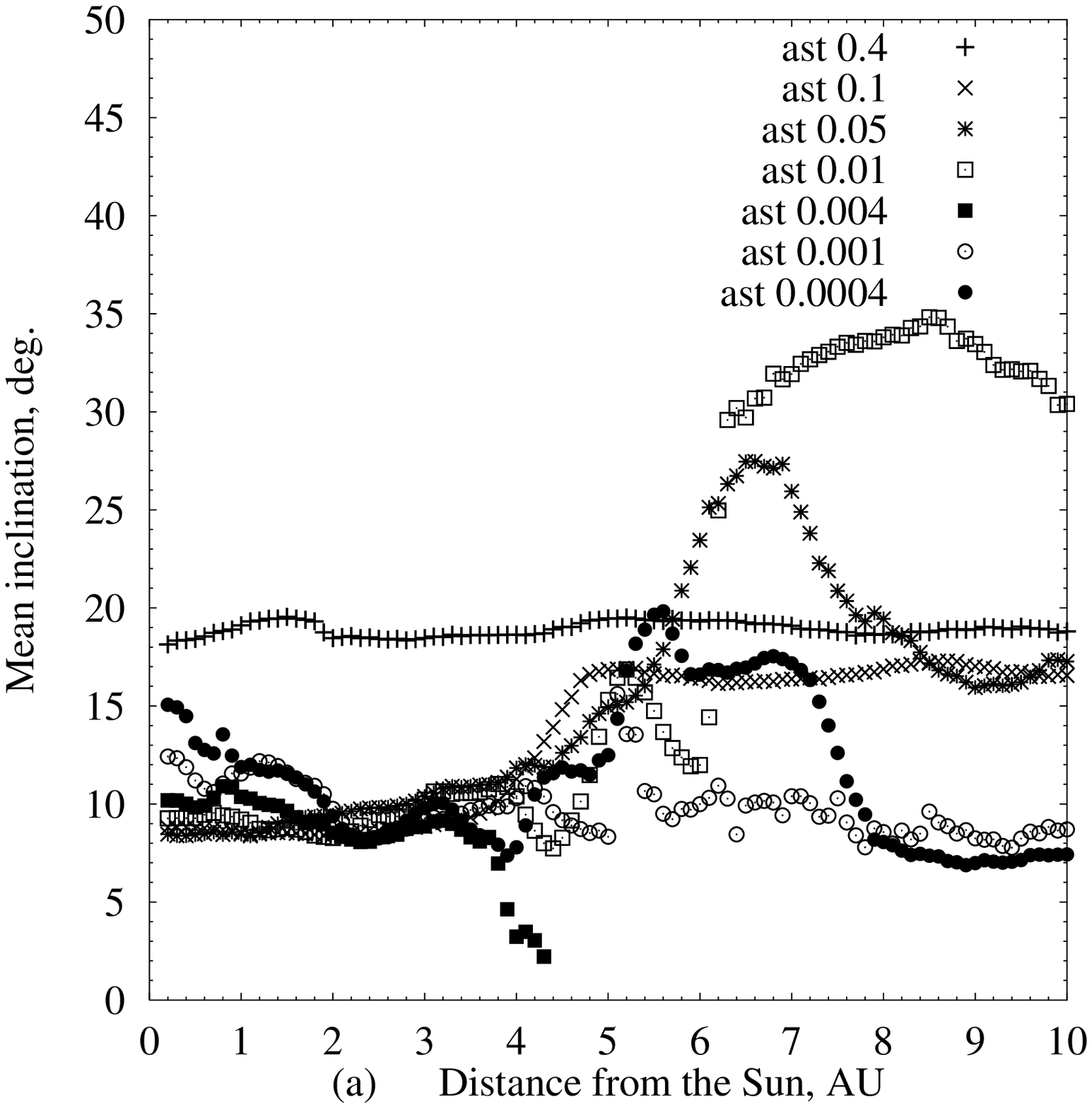} 
\includegraphics[width=81mm]{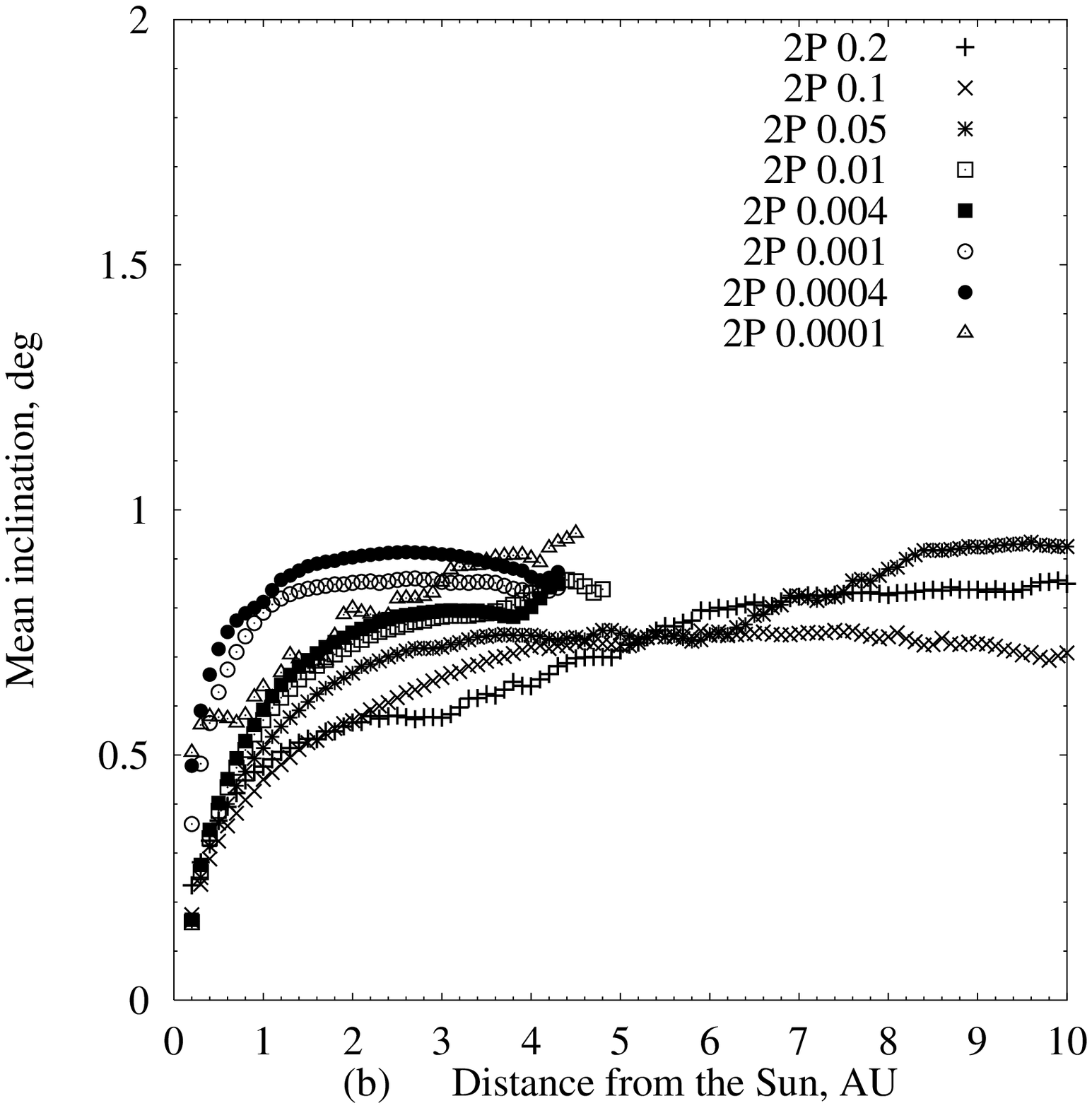} 
\includegraphics[width=81mm]{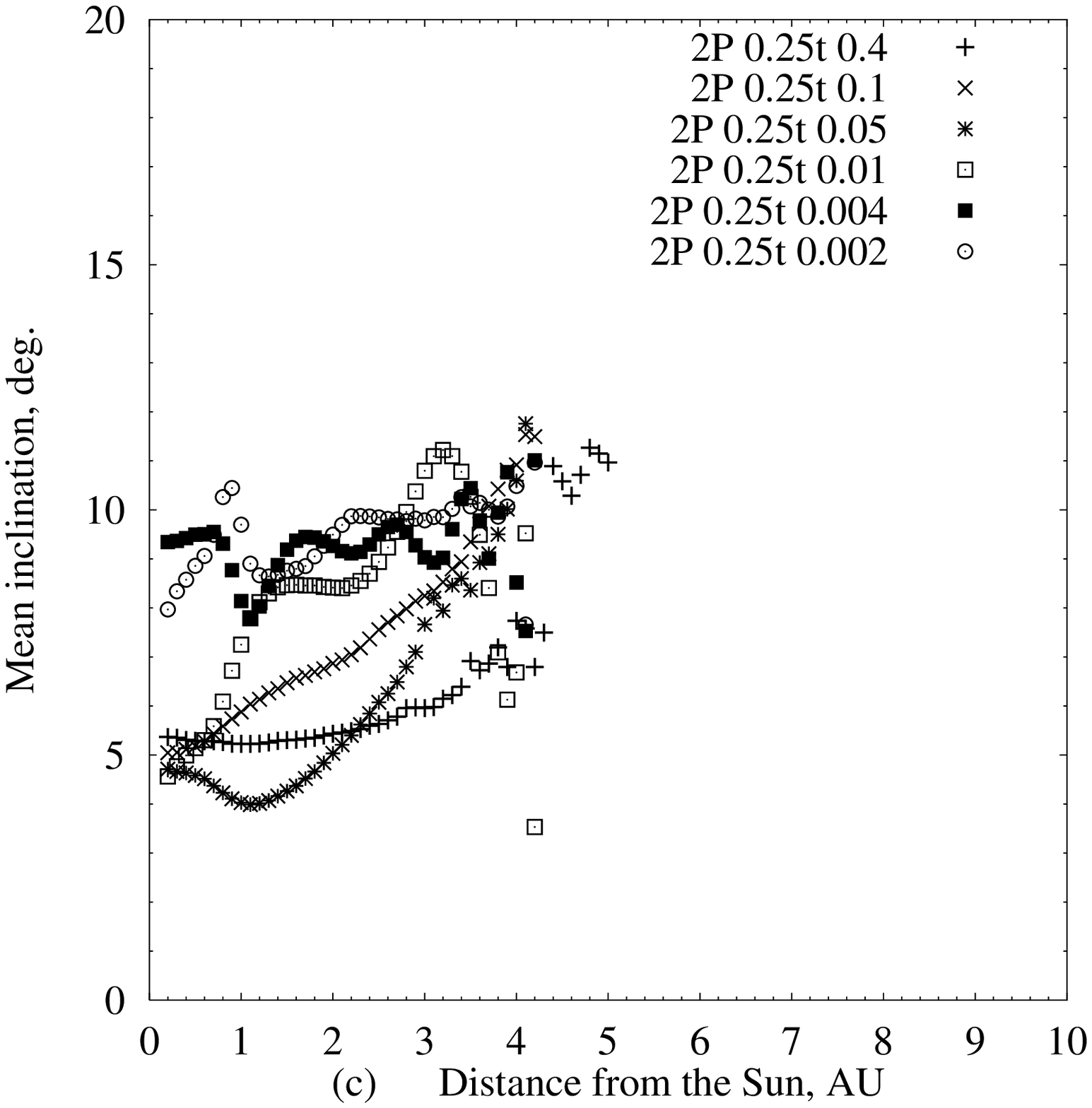} 
\includegraphics[width=81mm]{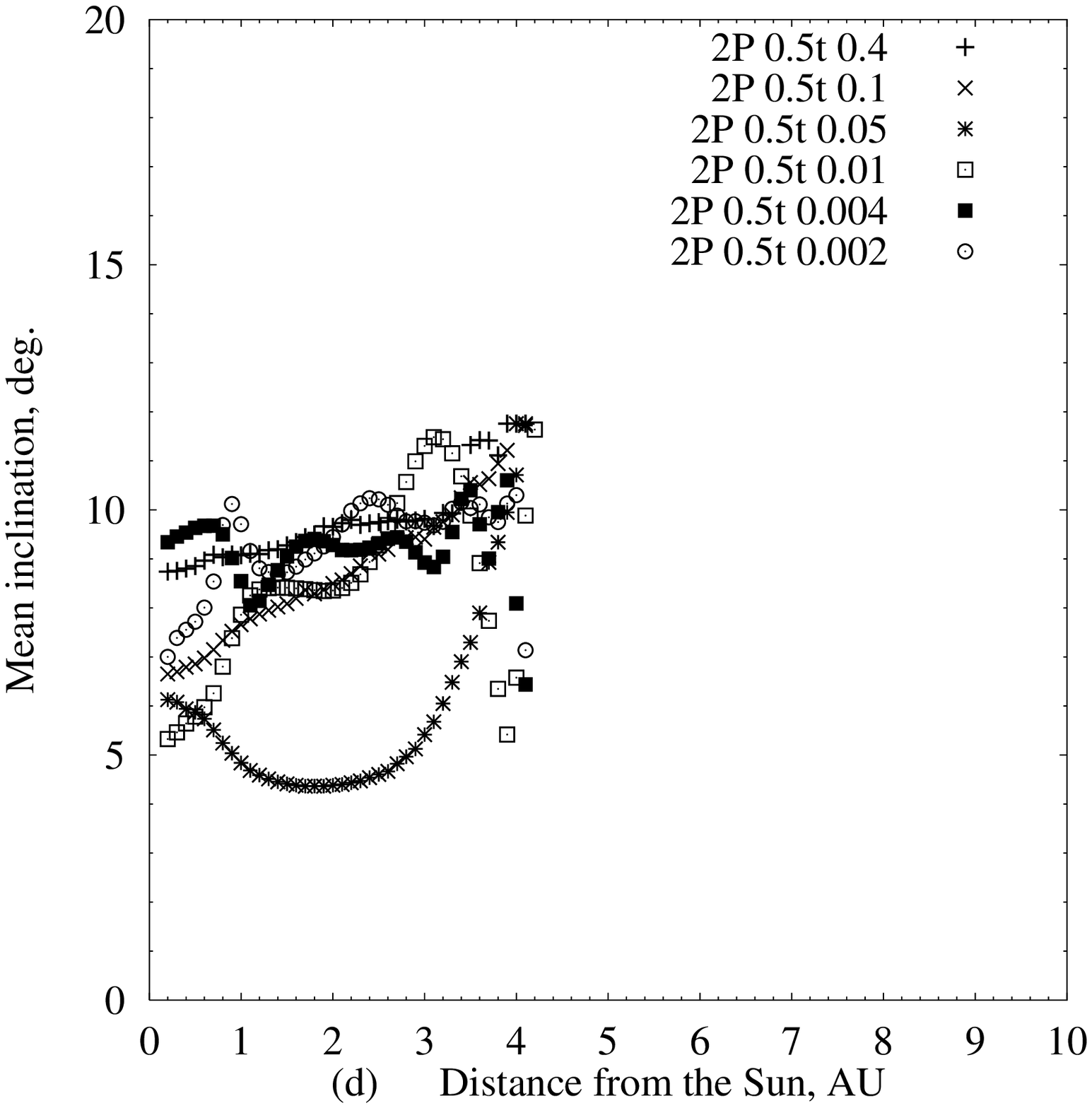} 

\caption{Mean orbital inclination (in degrees) of particles at different 
distances from the sun
at several values of $\beta$ (see the last number in the legend)
for particles started from asteroids (ast), from Comet 2P at perihelion (2P), 
at the middle of the orbit (2P 0.25t), and at aphelion (2P 0.5t).
}
\end{figure}%

\begin{figure}

\includegraphics[width=81mm]{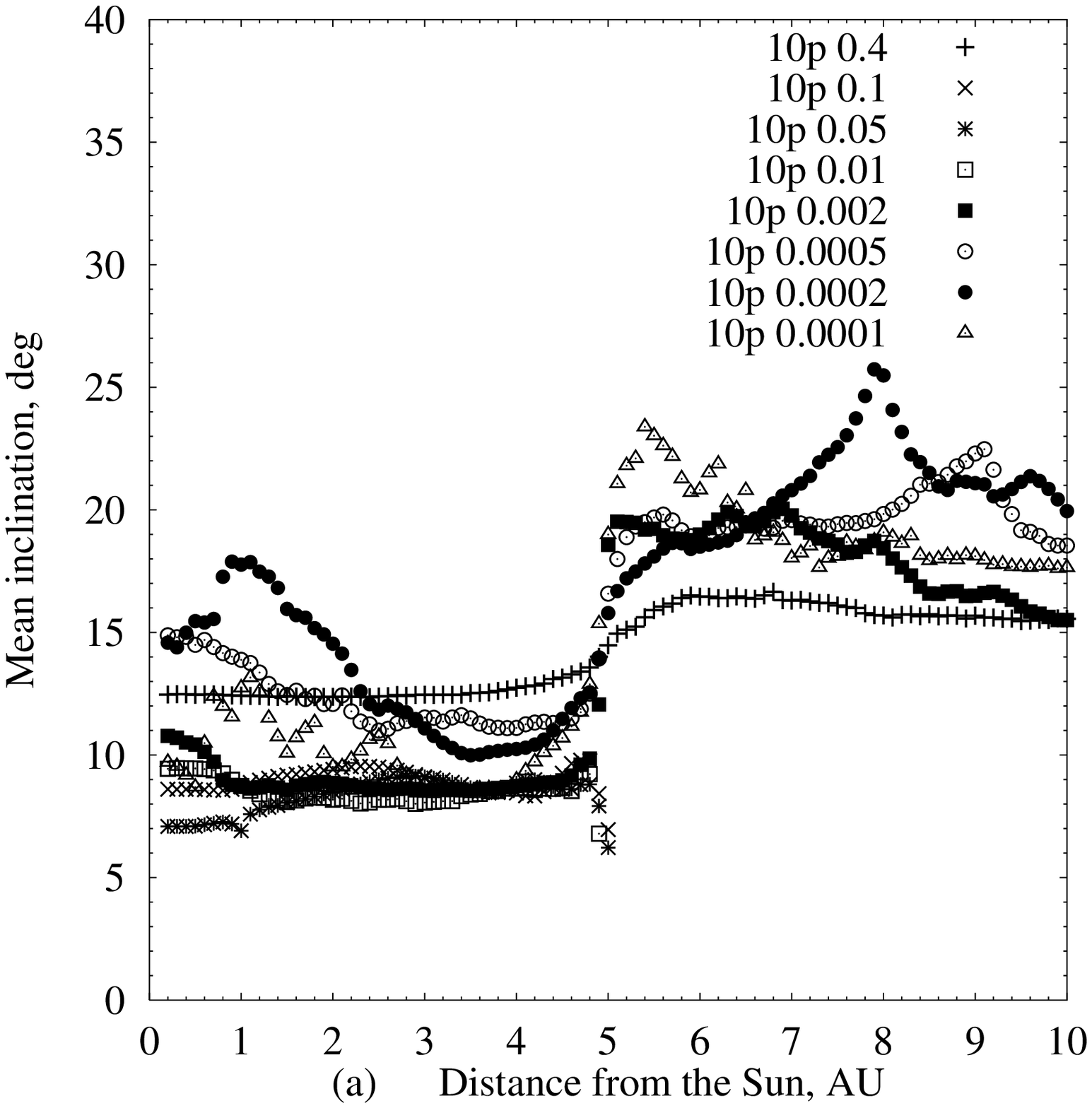} 
\includegraphics[width=81mm]{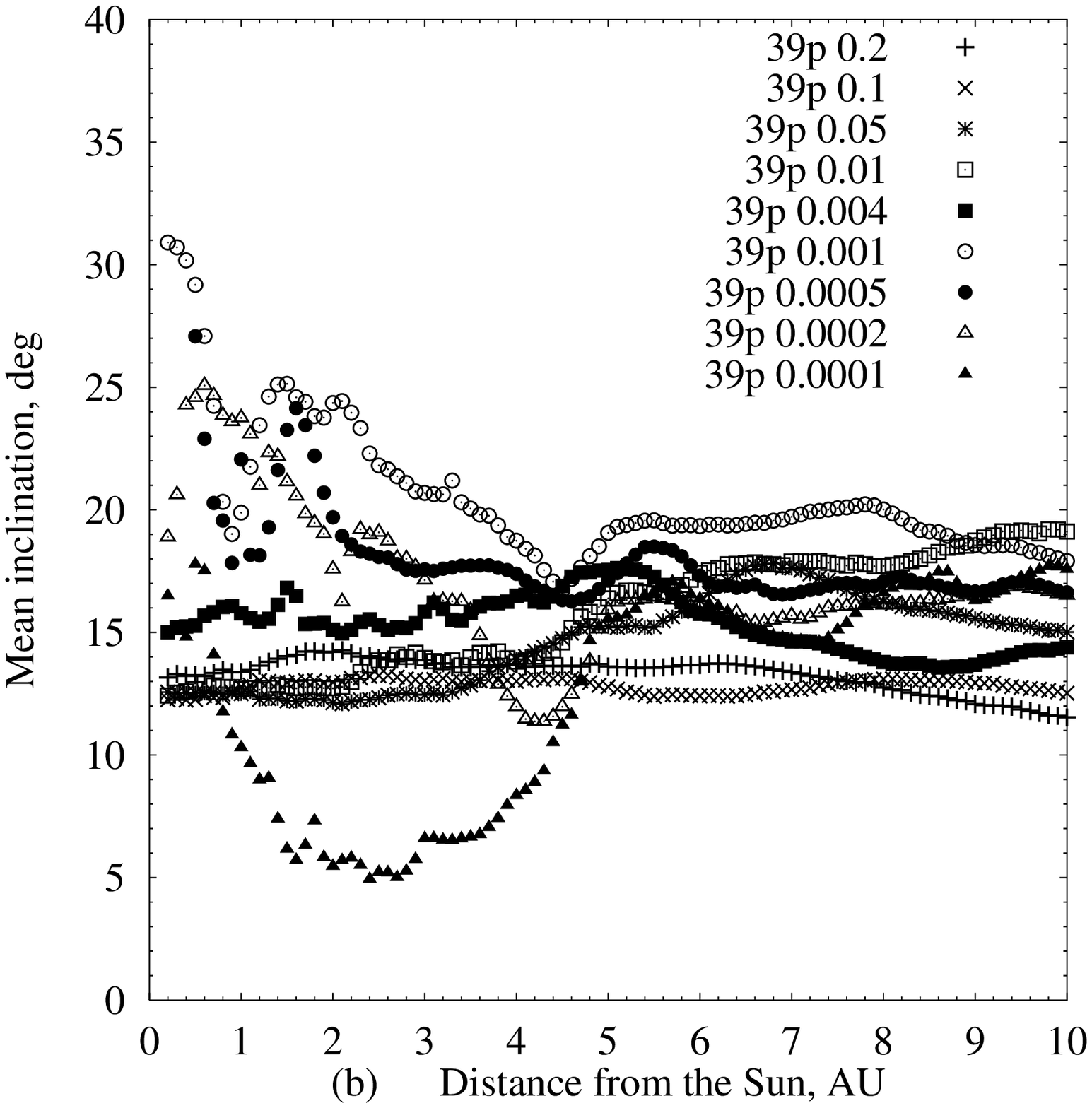} 
\includegraphics[width=81mm]{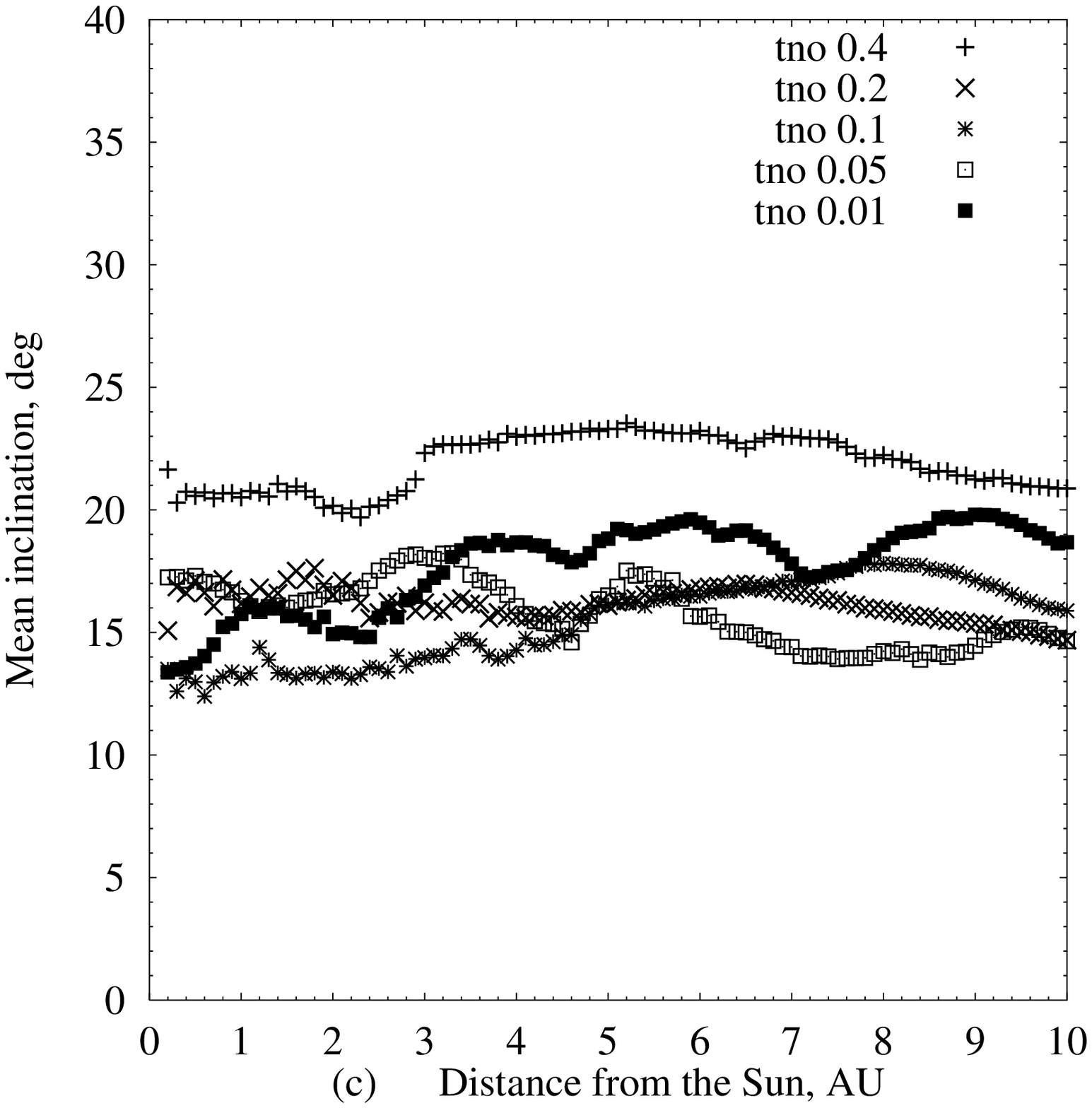} 
\includegraphics[width=81mm]{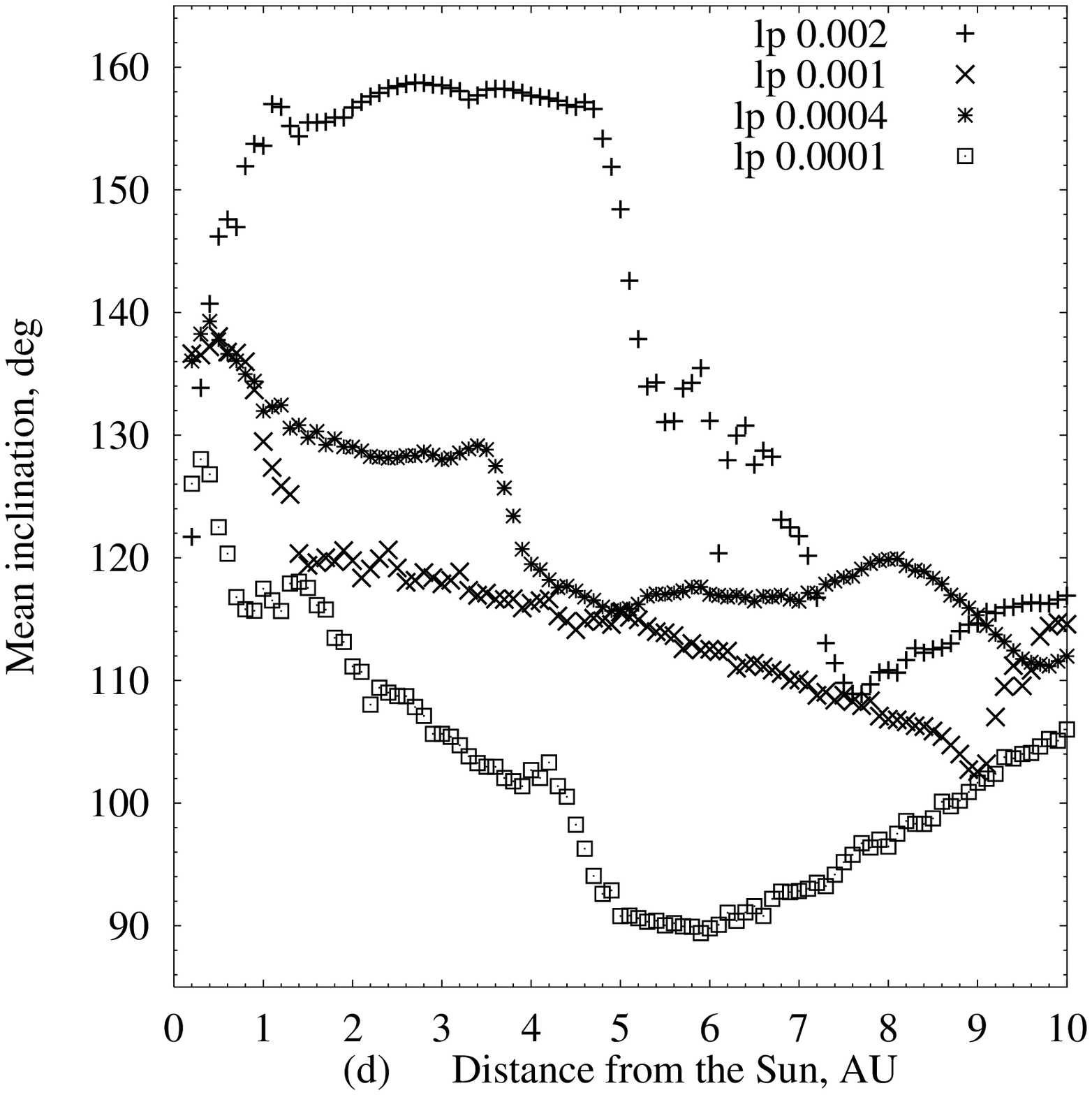} 

\caption{Mean orbital inclination (in degrees) of particles at different 
distances from the sun
at several values of $\beta$ (see the last number in the legend)
for particles started from Comets 10P and 39P (10P and 39P),
from trans-Neptunian objects (tno),
and from long-period comets (lp) at $e_o$=0.995, 
$q_o$=0.9 AU, and $i_o$ distributed between 0 and 180$^\circ$.
}
\end{figure}%

\begin{figure}

\includegraphics[width=81mm]{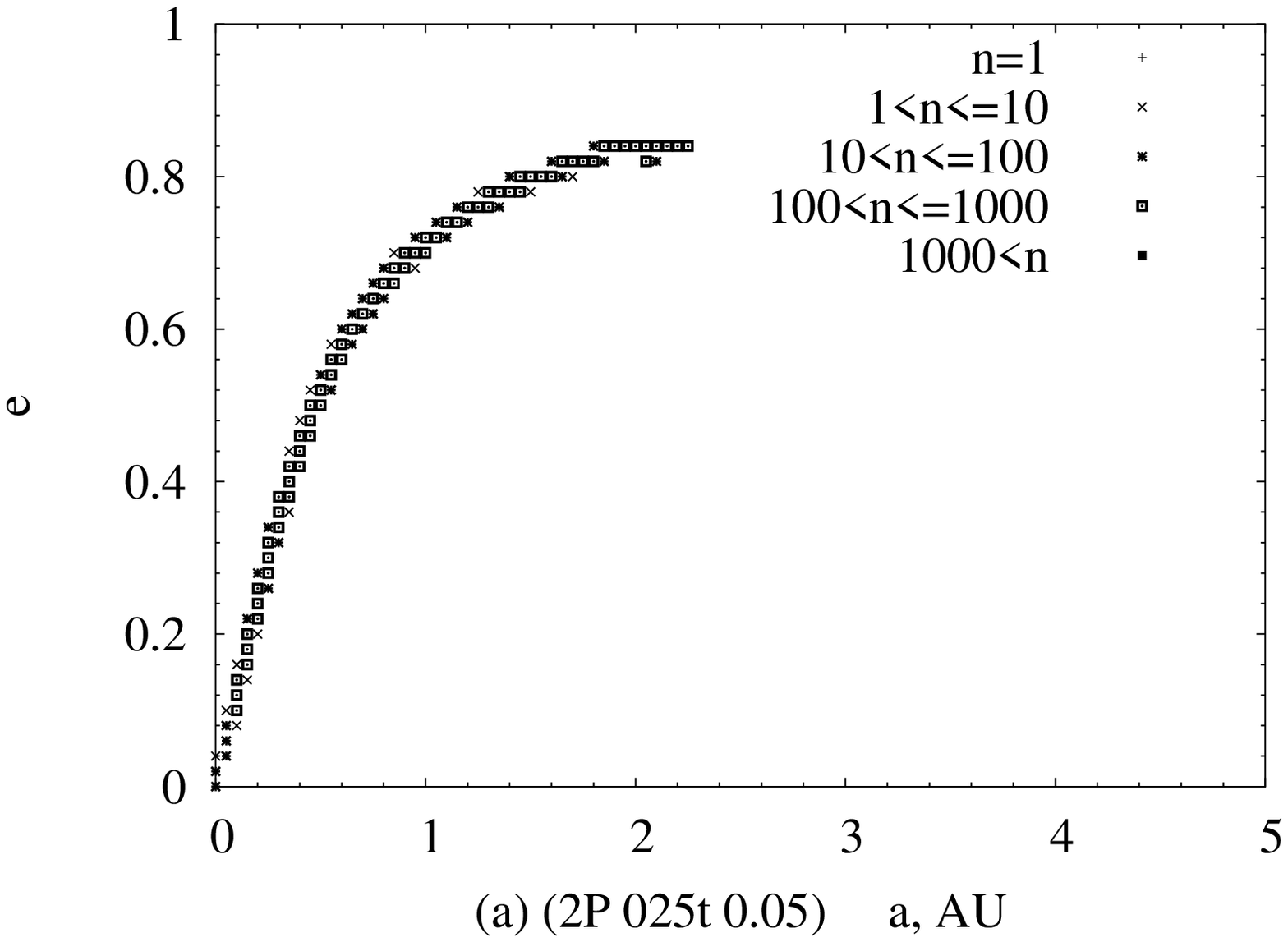} 
\includegraphics[width=81mm]{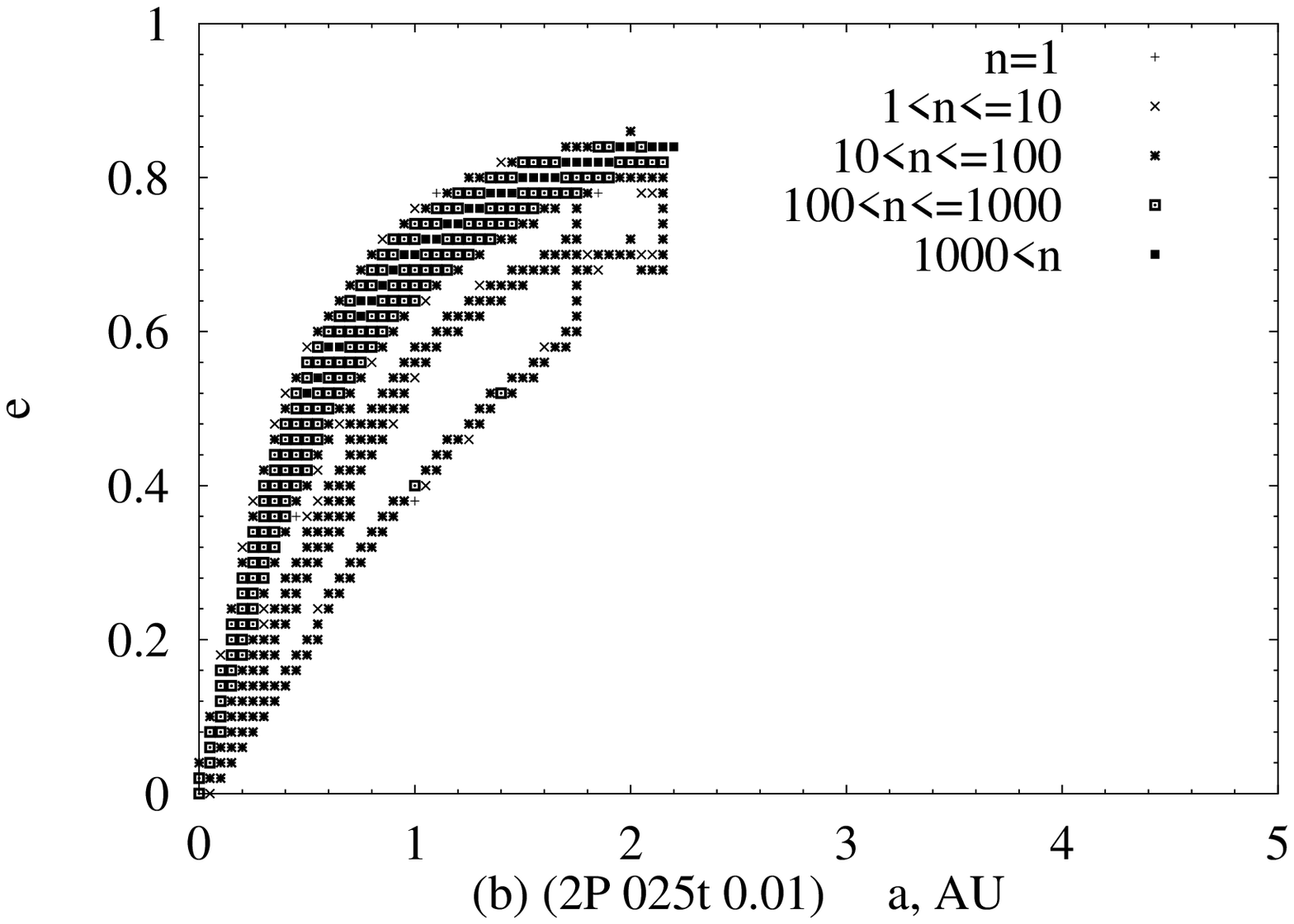} 
\includegraphics[width=81mm]{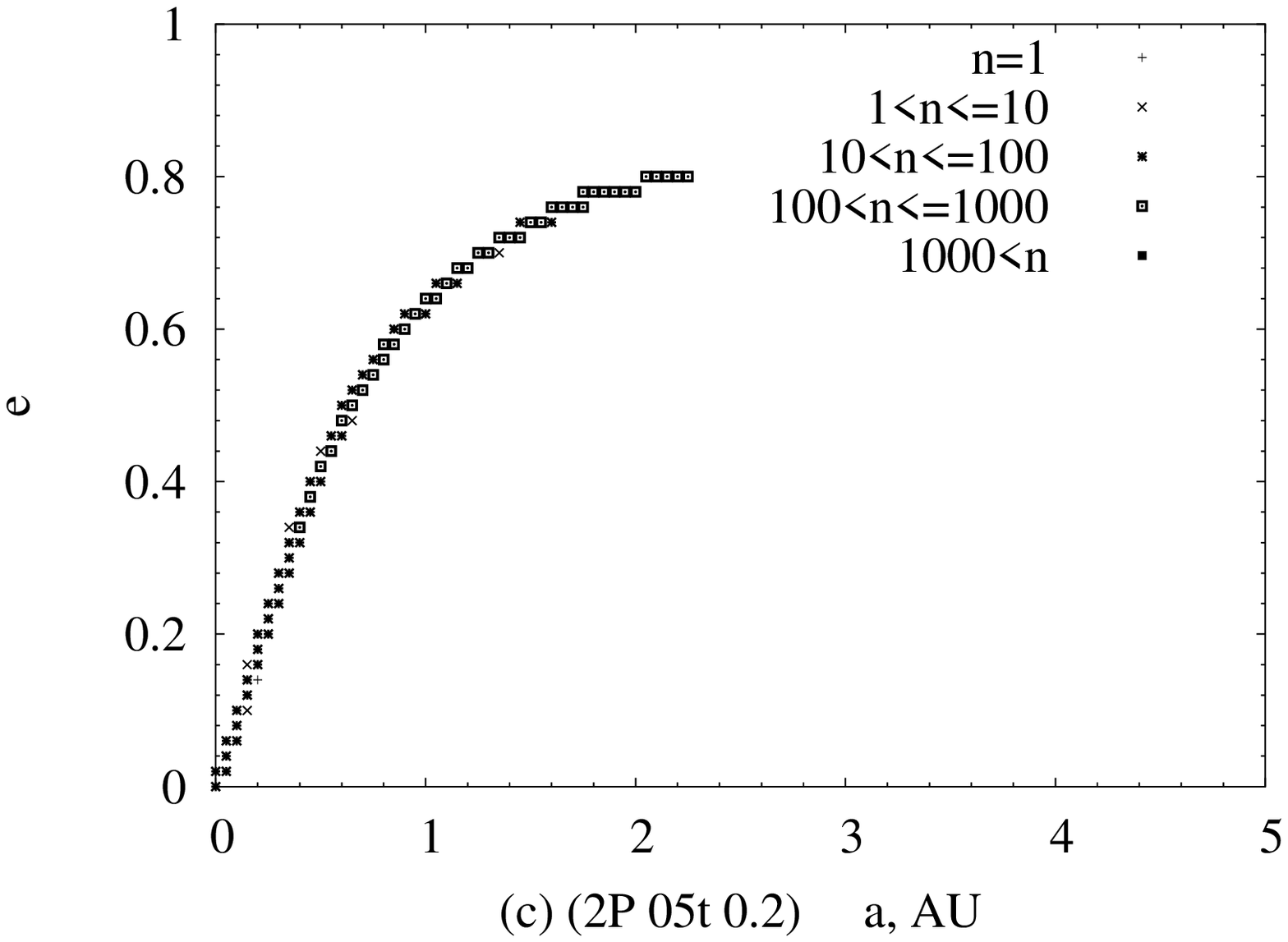} 
\includegraphics[width=81mm]{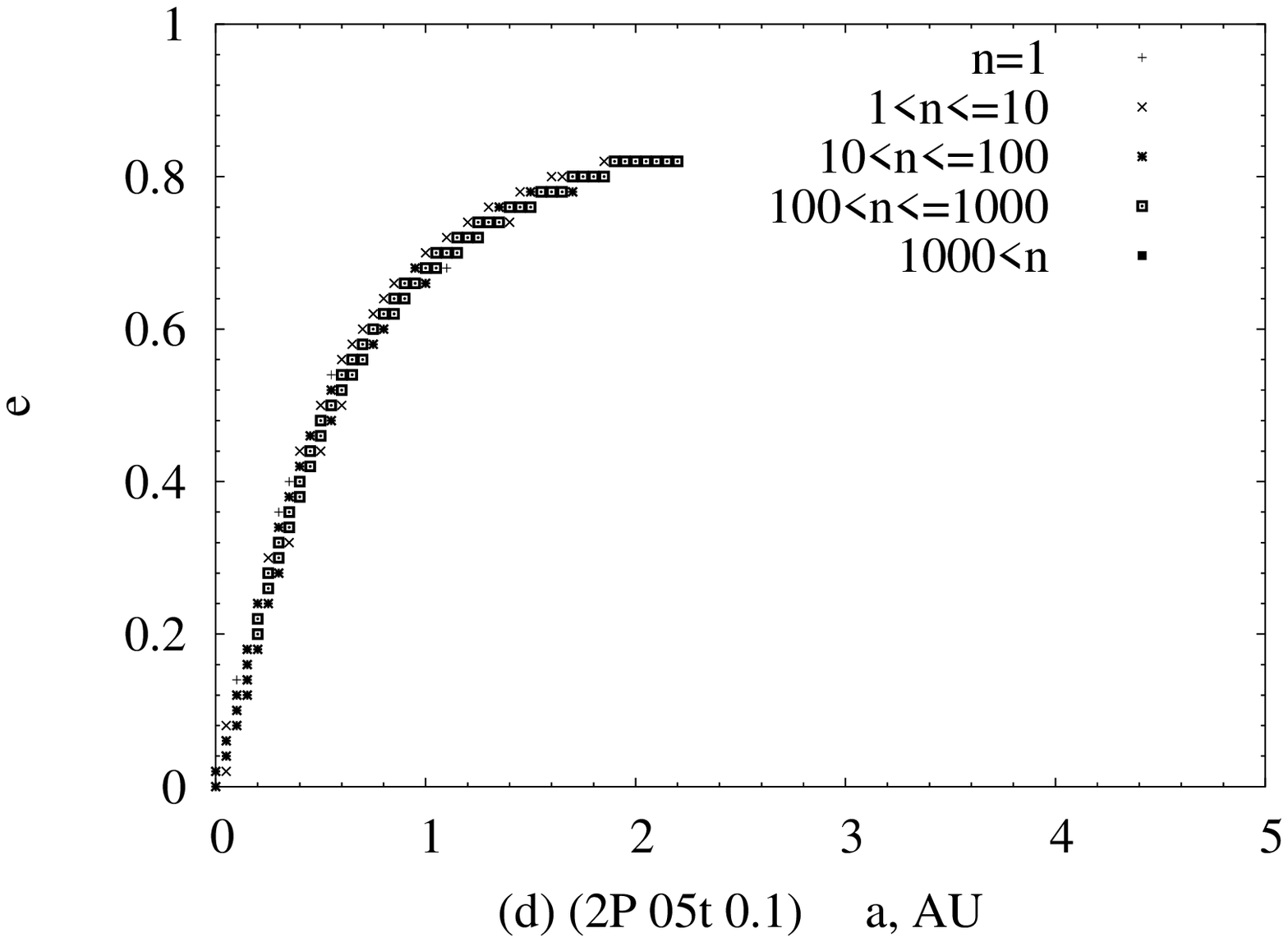} 
\includegraphics[width=81mm]{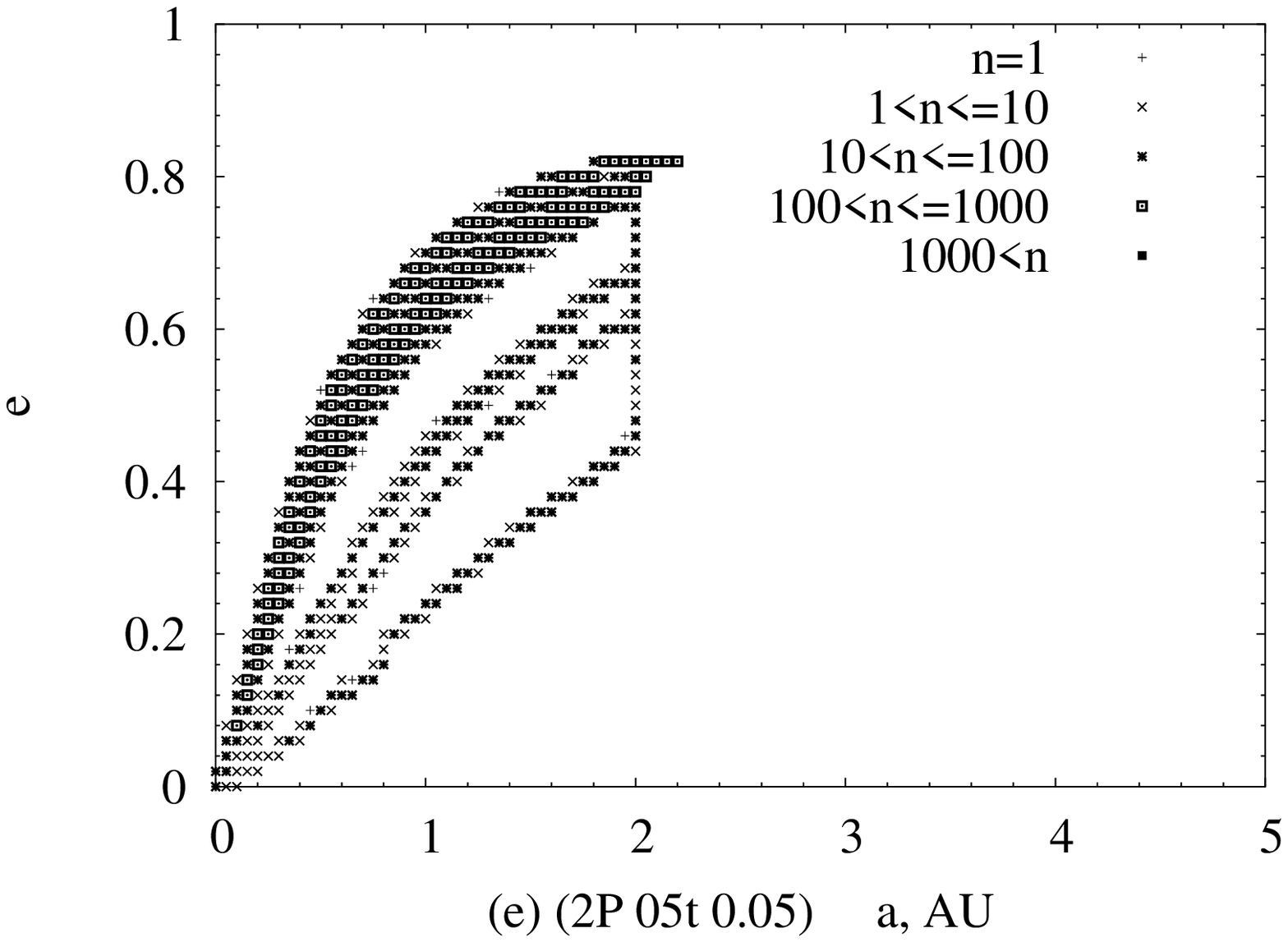} 
\includegraphics[width=81mm]{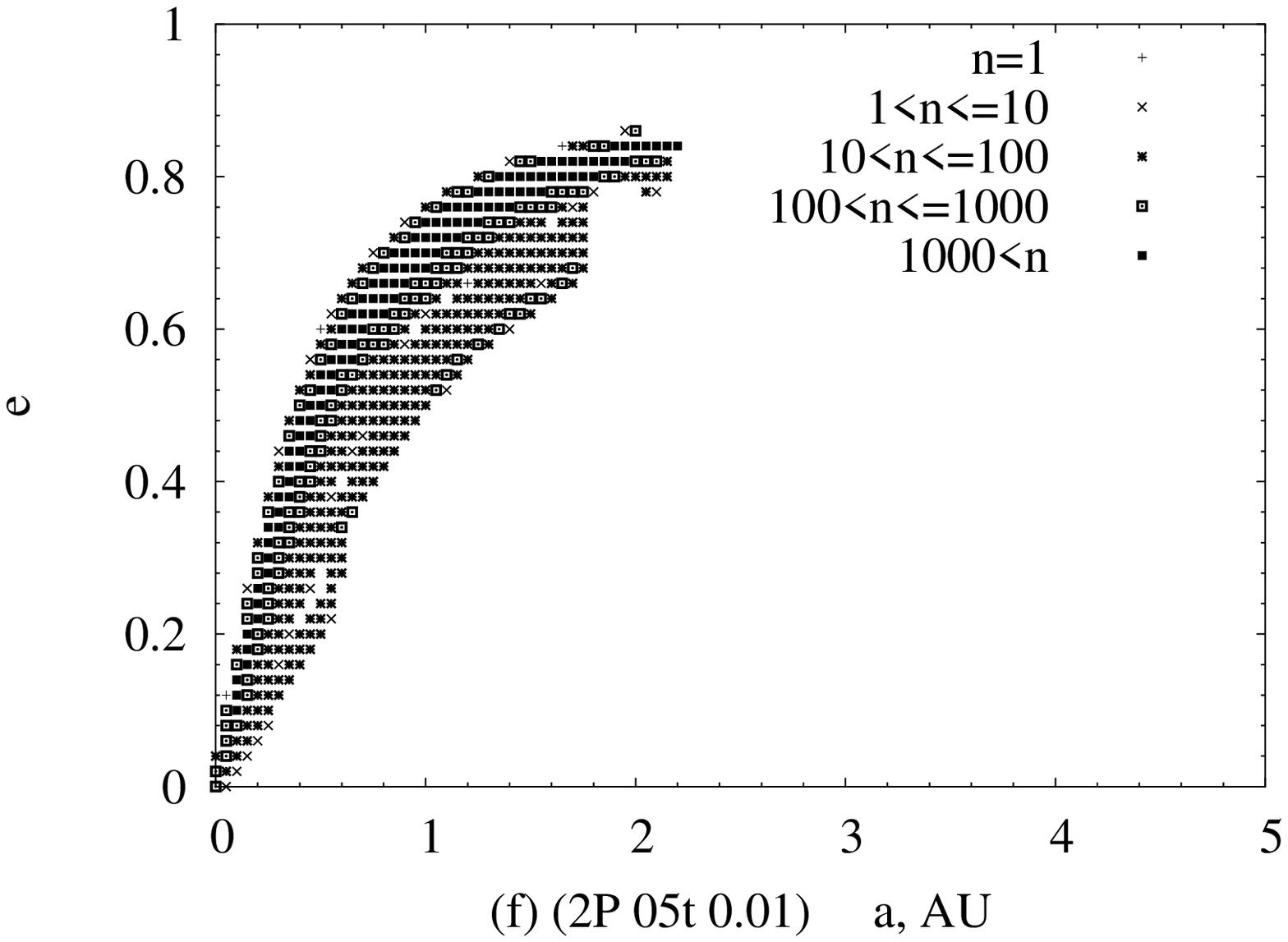} 

\caption{Distribution of dust particles with semimajor axis and eccentricity
at several values of $\beta$
(0.05 for (a),(e), 0.01 for (b),(f), 0.2 for (c), and 0.1 for (d))
for particles started from Comet 2P 
at the middle of the orbit (2P 0.25t) and at aphelion (2P 0.5t).
$n$ is the number of orbits with $e$ and $a$ in the interval of width 
of 0.02 and 0.05 AU, respectively. Orbital elements were stored 
with a step of 20 yr, and the number of particles in each run was 100 or 250.
}
	
\end{figure}%

\begin{figure}

\includegraphics[width=54mm]{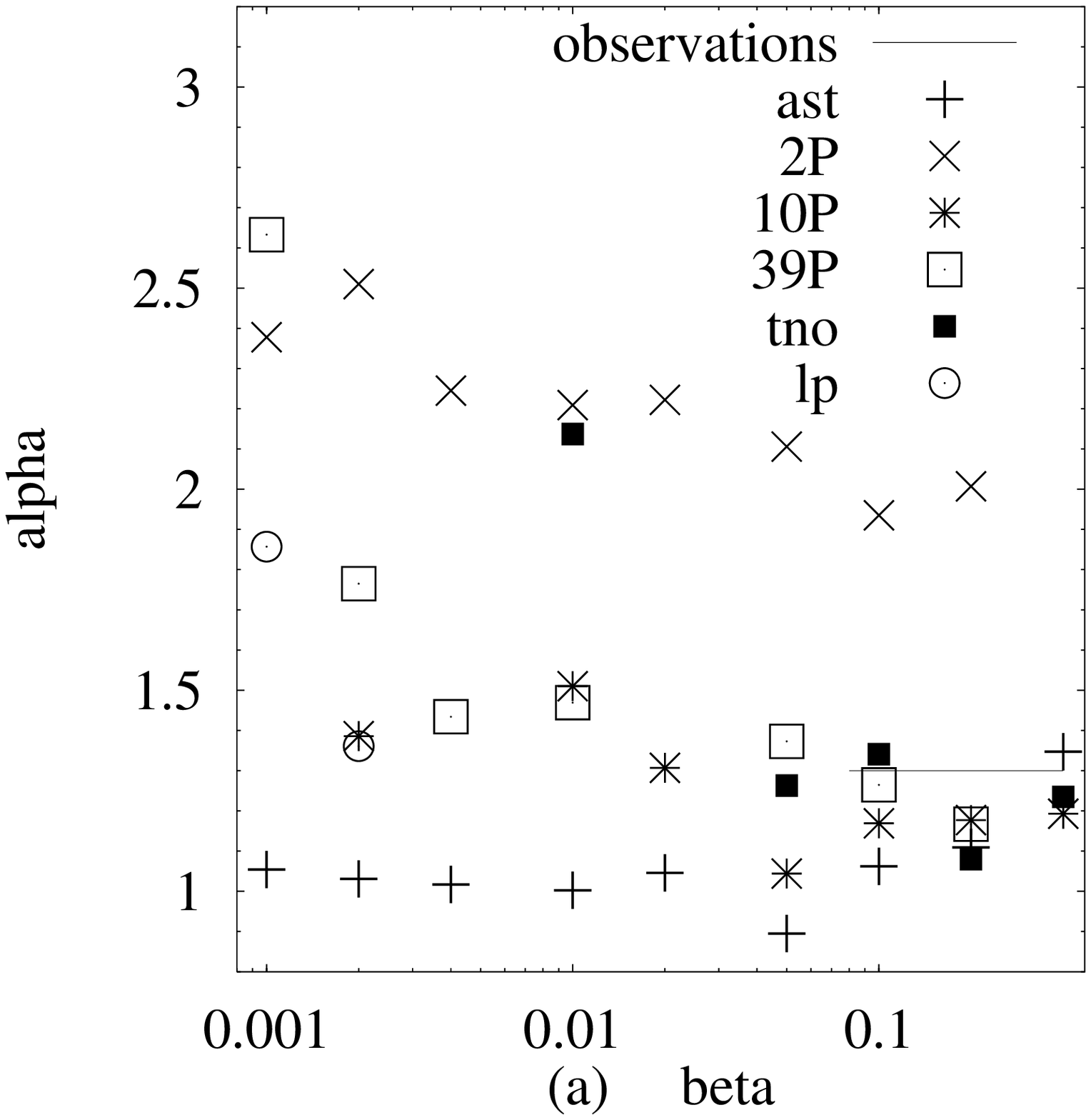} 
\includegraphics[width=54mm]{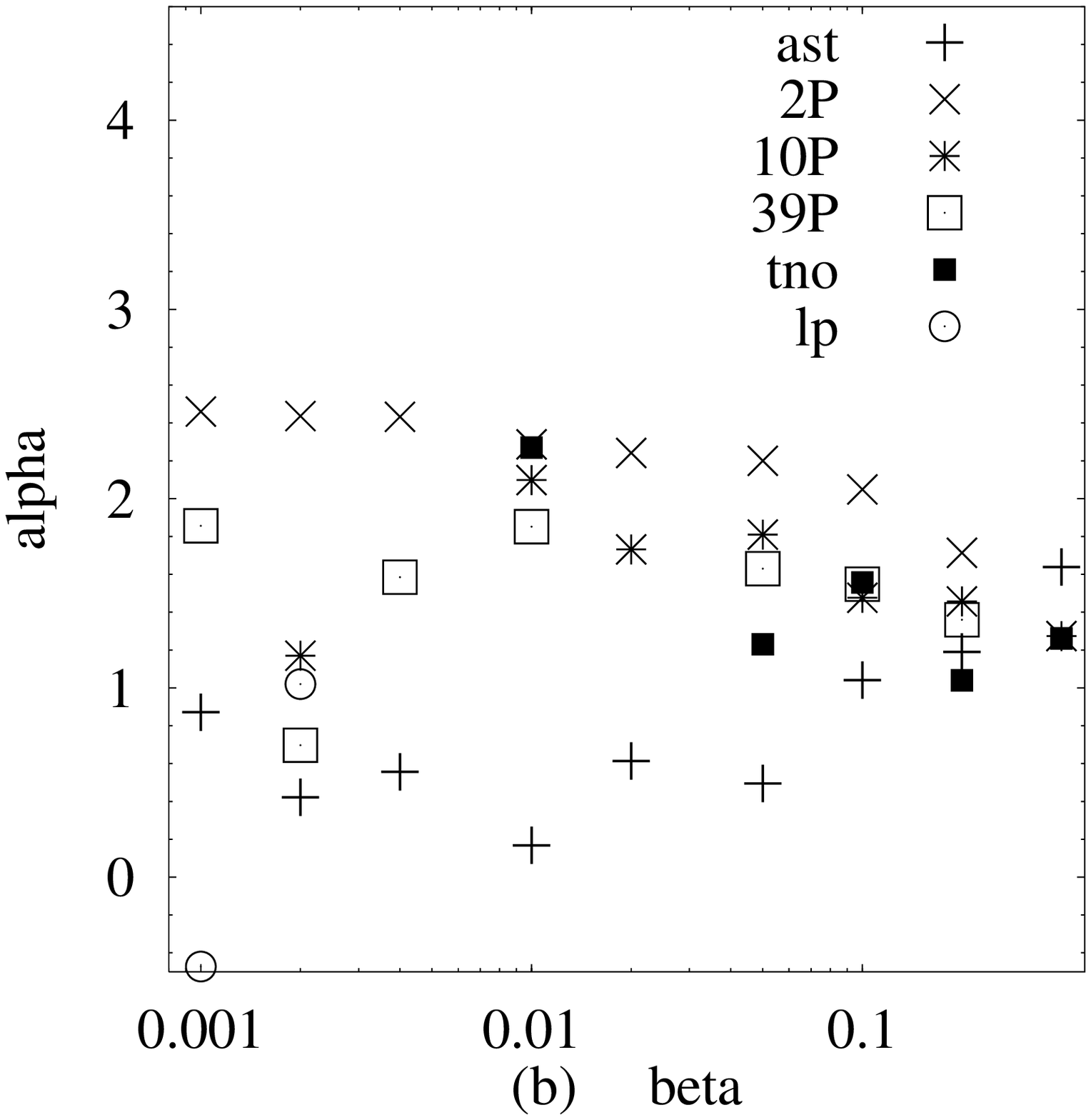} 
\includegraphics[width=54mm]{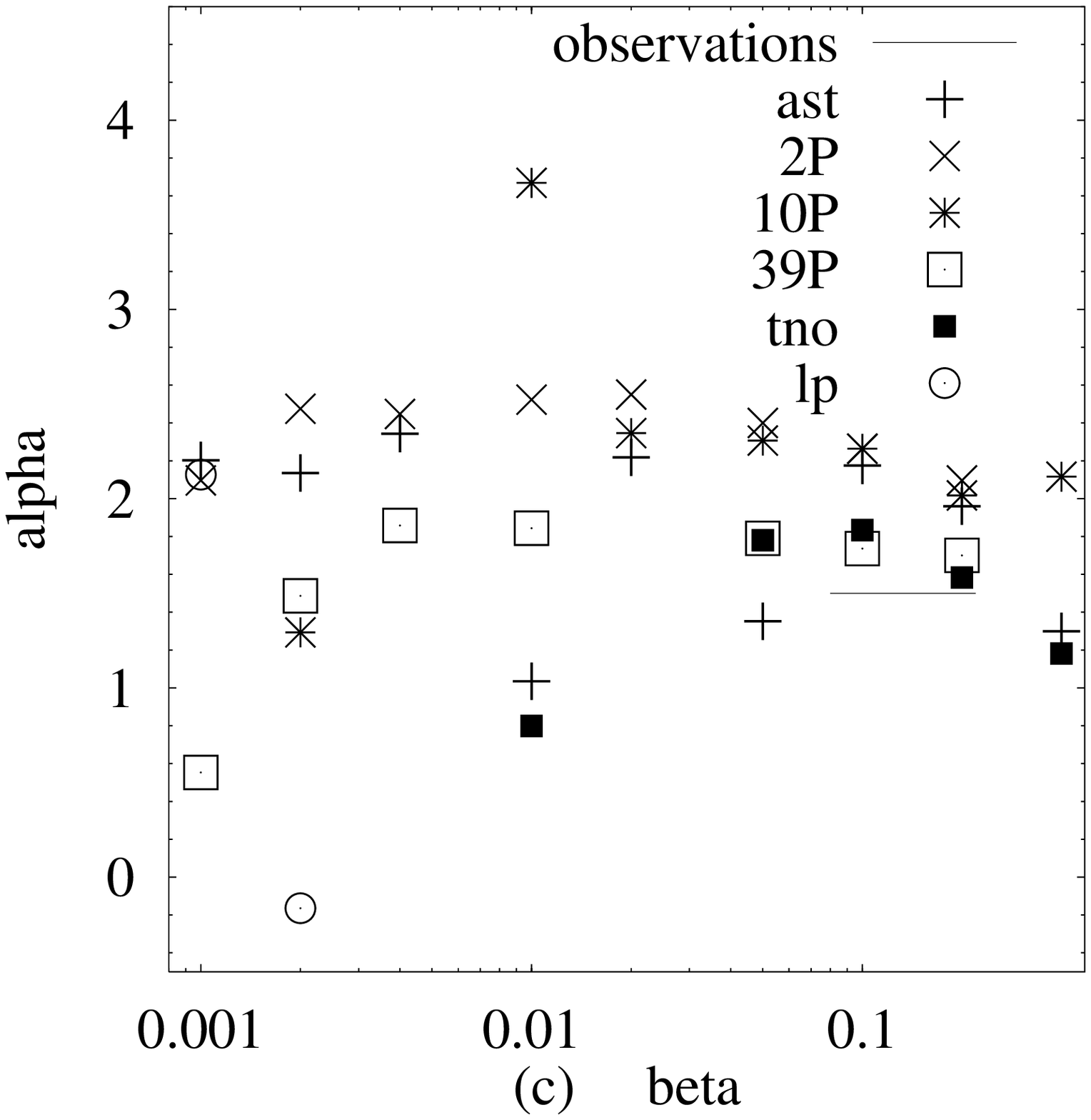} 

\caption{Values of $\alpha$ in $n(R) \propto R^{-\alpha}$ obtained by comparison 
of the values of the number density $n(R)$ at distance $R$ from the sun equal to 0.3 
and 1 AU (a), at $R$=0.8 and $R$=1.2 AU (b), 
and at $R$ equal to 1 and 3 AU (c).
The values are presented at several values of $\beta$
for particles started from Comets 10P and 39P (10P and 39P),
from trans-Neptunian objects (tno),
and from long-period comets (lp) at $e_o$=0.995, 
$q_o$=0.9 AU, and $i_o$ distributed between 0 and 180$^\circ$.
Horizontal bars correspond to observations.
}	
\end{figure}%

\clearpage

\begin{table}
\begin{center}
\caption{Characteristic velocity amplitude $v_a$ of `velocity-elongation'
 curves at 
90$^\circ$$<$$\epsilon$$<$270$^\circ$, minimum ($v_{\min}$) and 
maximum ($v_{\max}$) velocities at the above interval, mean eccentricities ($e_z$), and mean
inclinations ($i_z$) at distance from the sun 1$\le$$R$$\le$3 AU for particles 
from different sources at intervals of $\beta$}
\begin{tabular}{lllllll}
\tableline\tableline
Particles  & $\beta$ &  $e_z$  & $i_z$ & $v_a$  & $v_{\min}$ & $v_{\max}$  \\
               &             &           & (deg.) & (km s$^{-1}$) &(km s$^{-1}$) & (km s$^{-1}$) \\
\tableline
asteroidal & 0.0004-0.1 & $<$0.3 & 8-13 & 9   & (-13)-(-11) & 5-7 \\
10P        & 0.0001-0.2 & 0.2-0.5 & 7-18 & 8   & (-11)-(-9) & 3-9 \\
39P        & 0.01-0.2   & 0.2-0.4 & 12-15 & 8-9 & (-13)-(-9) & 4-8 \\
39P        & 0.0001-0.004 & 0.28-0.82 & 5-25 & 11-17 & (-20)-(-10) & 8-16 \\
2P         & 0.01-0.1  & 0.45-0.85 &0.45-0.8 & 13 & (-12)-(-7)&12-22 \\
2P         & 0.0001-0.004  & 0.6-0.9 & 0.6-0.9 & 14 & (-16)-(-12)&15-20 \\
2P 0.25t   & 0.002-0.2  & 0.6-0.85 & 4-11 & 16  &(-14)-4& 12-40 \\ 
2P 0.5t   & 0.002-0.4  & 0.55-0.85 & 4-11 & 14  & (-22)-(-2)& 2-29 \\
tno       & 0.05-0.4    & 0.15-0.45 & 13-22 & 16 & (-20)-(-14)&12-18 \\
tno       & 0.01        & 0.48-0.76 & 14-17 & 12 & -16 & 8 \\ 
lp, $q$=0.9 AU        & 0.002 & 0.1-0.25 & 154-160 & 41 & -44 & 38 \\
lp, $q$=0.9 AU        & 0.0001-0.001 & 0.28-0.76 & 105-132 & 33 & (-38)-(-32)&28-34 \\
lp, $q$=0.1 AU        & 0.0004 & 0.44-0.55 & 104-113 & 34 & -34 & 34 \\

\tableline
\end{tabular}
\end{center}
\end{table}

\newpage

\begin{thebibliography}{}


\item  Burns, J. A., Lamy, P. L., \& Soter, S. 1979, \icarus, 40, 1

\item Clarke, D., Matthews, S. A., Mundell, C. G., \& Weir, A. S.  1996, 
\aap, 308, 273



\item Dermott, S. F., Grogan, K., Durda, D. D., et al.  2001, 
in Interplanetary dust, ed.
B. A. S. Gustafson, S. Dermott, \& H. Fechtig 
(Berlin: Springer-Verlag), 569

\item Dermott, S. F., Durda, D. D., Grogan, K., \& Kehoe, T. J. J. 2002,
in Asteroids III, ed. W. F. Bottke, Jr., A. Cellino, P. Paolicchi, \& R. P. Binzel (Tucson: The University of Arizona Press),  423



\item East, I. R., \& Reay, N. K. 1984, \aap, 139, 512

\item Fried, J. M. 1978, \aap, 68, 259


\item Giese, R. H. 1963,
\ssr, 1, 589 

\item Giese, R. H., \& v. Dziembowski, C. 1969,
\ssr, 17, 949 

\item Gorkavyi, N. N., Ozernoy, L. M., Taidakova, T., \& Mather, J. C.
2000, astro-ph/0006435

\item Grogan, K., Dermott, S. F., \& Durda, D. D. 2001, \icarus, 152, 251


\item  Gr\"un, E., Fechtig, H.,  Kissel, J., \& Gammelin, P. 1977, J. Geophys., 42, 717

\item  Gr\"un, E., Zook, H.A., Fechtig, H.,  \& Giese, R.H. 1985, \icarus, 62, 244

\item  Gr\"un, E., Kruger, H., \& Landgraf, M. 2000, in Proc. 
ESO workshop (Nov. 2-5, 1998, Garching, Germany).
Springer, 99

\item Hicks, T. R., May, B. H., \& Reay, N. K. 1974,
\mnras, 166, 439 


\item Hong, S. S., 1985, 
\aap, 146, 67






\item Ipatov, S. I., \& Mather, J. C.  2003, Earth, Moon, \& Planets, 92, 89

\item Ipatov, S. I., \& Mather, J. C.  2004a, in 
Annals of the New York Acad. of Sci., v. 1017, 
Astrodynamics, Space Missions, and Chaos, 
ed. B. Belbruno, D. Folta, \& P. Gurfil (New York: NYAS), 46 

\item Ipatov, S. I., \& Mather, J. C. 2004b, Adv. Space Res., 33, 1524


\item   Ipatov, S. I. \& Mather, J. C. 2006a, Adv. Space Res., 37, 126 

\item Ipatov, S. I., \& Mather, J. C. 2006b, in
Dust in Planetary System, ed. H. Kruger \& A. Graps (ESA),
in press 


\item Ipatov, S. I., Mather, J. C., \& Taylor, P. A.  2004, in
Annals of the New York Acad. of Sci., v. 1017, 
Astrodynamics, Space Missions, and Chaos, 
ed. B. Belbruno, D. Folta, \& P. Gurfil (New York: NYAS), 66 







\item Ipatov, S. I., Kutyrev, A. S., Madsen, G. J., 
Mather, J. C., Moseley, S. H., \& Reynolds, R.J. 
2005, Lunar Planet. Sci. Conf., 36, 1266.

\item Ipatov, S. I., Kutyrev, A. S., Madsen, G. J., 
Mather, J. C., Moseley, S. H., \& Reynolds, R.J. 
2006, Lunar Planet. Sci. Conf., 37, 1471.


\item James, J. F. 1969, \mnras, 142, 45


\item  Kortenkamp, S. J. \& Dermott., S. F. 1998, \icarus, 135, 469 


\item Landgraf, M., Liou, J.-C., Zook, H. A., \& Gr\"un, E. 2002, \aj, 123, 2857


\item Lamy, P. L. \& Perrin, J.-M. 1986, 
\aap, 163, 269 


\item Levison, H.F. \& Duncan, M.J. 1994, Icarus, 108, 18 

\item Leinert, C. 1975,
Space Sci. Rev., 18, 281  

\item Leinbert, C., Link, H., Pitz, E., \& Giese, R. H. 1976,
\aap, 47, 221 

\item Leinbert, C., Richter, I., Pitz, E., \& Plank, B. 1981,
\aap, 103, 177


\item  Liou, J.-C. \& Zook, H. A.  1999, \aj, 118, 580


\item  Liou, J.-C., Dermott, S. F., \& Xu, Y. L. 1995, \planss,
43, 717

\item  Liou, J.-C.,  Zook, H. A., \&  Dermott, S. F. 1996,  
 \icarus, 124, 429

\item  Liou, J.-C., Zook, H. A., \& Jackson, A. A. 1999, \icarus,  141, 13


\item  Madsen, G. J., Reynolds, R. J., Ipatov, S. I., Kutyrev, A. S., 
Mather, J. C., \& Moseley, S. H. 2006,
in Dust in Planetary System, ed. H. Kruger \& A. Graps, (ESA),
in press 


\item  Moro-Martin, A. \& Malhotra, R.  2002, \aj, 124, 2305

\item  Moro-Martin, A. \& Malhotra, R.  2003, \aj, 125, 2255


\item Ozernoy, L. M. 2001, in 
IAU Colloq. 204: The Extragalactic Infrared Background 
and its Cosmological Implications, 
ed. M. Harwit \& M. G. Hauser, 17, 
astro-ph/0012033 

\item Reach, W.T. 1992, \apj, 392, 289

\item  Reach, W. T.,  Franz, B. A., \& Weiland, J. L. 1997, \icarus, 127, 461

\item Reynolds, R. J., Madsen, G. J., \& Moseley, S. H. 2004, \apj, 612, 1206

\item Weiss-Wrana, K. 
1983, 
\aap, 126, 240


\item Zook, H. A. 2001,
in Accretion of extraterrestrial matter throughout Earth's history, 
ed. B. Peucker-Ehrenbrink and B. Schmitz 
(NY: Kluwer), 75

\end{thebibliography}
\end{document}